\documentclass{article}

\usepackage[english]{babel}

\usepackage[letterpaper,top=2cm,bottom=2cm,left=3cm,right=3cm,marginparwidth=1.75cm]{geometry}
\usepackage{amsthm}
\usepackage{tikz}
\usepackage{centernot}
\usepackage{mathtools}
\usepackage{ stmaryrd }
\usetikzlibrary{matrix,arrows.meta}
\usepackage{amsmath}
\usepackage{tikz}
\usetikzlibrary{matrix,arrows,decorations.pathmorphing}
\usetikzlibrary{shapes,arrows,positioning}
\usepackage{graphicx}
\usepackage{amsmath}
\usepackage{amssymb}
\usepackage{latexsym}
\usepackage{amssymb,amsmath}
\usepackage{tikz-cd}
\usetikzlibrary{arrows}
\usepackage{amsfonts}
\usepackage{amssymb}
\usepackage{graphicx}
\usepackage{fancyhdr}
\usepackage{fancyvrb}
\usepackage{fancybox}
\usepackage{float}
\usepackage{geometry}
\usepackage{minitoc}
\usepackage{indentfirst}
\usepackage{multicol, color}
\usepackage[outerbars]{changebar}
\usepackage{pifont,textcomp}
\usepackage{amsfonts}
\usepackage{amsmath}
\usepackage{amscd}
\usepackage{array}
\usepackage{multirow}
\usepackage{mathrsfs}
\usepackage{amssymb}
\usepackage{amsthm}

\usepackage{amsmath}
\usepackage{amssymb}
\usepackage{amsfonts}
\usepackage{amsthm,amscd}

\newtheorem{theorem}{Theorem}[section]
\newtheorem{lemma}[theorem]{Lemma}
\newtheorem{corollary}[theorem]{Corollary}
\newtheorem{proposition}[theorem]{Proposition}
\theoremstyle{definition}
\newtheorem{definition}[theorem]{Definition}

\theoremstyle{remark}

\usepackage{amsmath}
\usepackage{graphicx}
\usepackage[colorlinks=true, allcolors=blue]{hyperref}

\title{Higher lattice gauge theory from representations of 2-groups and 3+1D topological phases}
\author{You}
\author{  Latévi M. Lawson$^{1,2,a}$ and Prince K. Osei$^{1,2,b}$   
	\space\\\\
  ${}^{1}$African Institute for Mathematical Sciences (AIMS) Ghana,\\
	 1 Shoppers Street, Manet, Spintex, Accra, Ghana\\\\
	 ${}^{2}$Quantum Leap Africa (QLA), AIMS Research Innovation Centre,\\ KN 3 Rd, Kigali, Rwanda.\\\\
	Latevi@aims.edu.gh$^a$ and pkosei@aims.edu.gh$^b$,   }
\begin{document}
\maketitle

\begin{abstract}
%A higher-lattice gauge theory model is based on a generalization of ordinary lattice gauge theory,
%where in addition to the ordinary 1-gauge field, there is a second 2-gauge field which describes a
%gauge symmetry on the first gauge field itself. An ordinary gauge field is based on group theory,
%however a higher lattice field is based on higher dimensional category theory. 
We construct a higher lattice
gauge theory based on the representation of 2-groups described by a
category of crossed modules on a lattice model described by path 2-groupoids. 
%in a 2-category presented as a lattice
%model described by path 2-groupoids.
Using these lattice gauge representations, an exactly solvable Hamiltonian for topological phases in 3+1 dimensions is  constructed. We show that the ground states of this model are topological observables.
\end{abstract}
\tableofcontents

\section{Introduction}

%The modern computing world is facing a revolutionary wave of change with the advent of quantum computing. At the forefront of this evolution is the topological quantum computer. The future for topological quantum computers appears promising due
%to their potential to revolutionize quantum computing by addressing challenges faced
%by traditional qubits. In fact, conventional quantum computers are prone to frequent
%errors because qubits are very sensitive to interference from the external environment.
%In contrast, topological quantum computers incorporate quantum information into the
%topological phases of matter, thus greatly reducing the incidence of errors. This allows for longer computations and the execution of complex algorithms. These robust
%qubits, based on the  topological phases of matter, can push the boundaries of computational power, enabling breakthroughs in fields like cryptography, optimization, and
%drug discovery. Although the development of this technology is still in its infancy, the
%possibilities are endless and will revolutionize our lives, science, and industry.

%The  fault-tolerant quantum computing is due to Alexei Kitaev .  In his seminal paper, Kitaev  defined a lattice model, henceforth called the

The Kitaev quantum double model, or simply the Kitaev model, for 2+1D, was invented for fault-tolerant quantum computing by leveraging  
  topological phases of matter \cite{1}. This model provides a framework for understanding and constructing topologically ordered states of matter that can potentially serve as robust platforms for quantum information processing.  In Kitaev’s proposal,
information is encoded in nontrivial loops of the underlying surface \cite{2} and is processed by the creation, braiding, and annihilation of topological charges \cite{3,4}.  Hamiltonians of the Kitaev model as well as   Levin and Wen’s string-net models \cite{4,5}  exhibit topological quantum order (TQO) \cite{5}  and  have a remarkable property that their ground state degeneracy cannot be lifted by generic local perturbations \cite{6,7,8,9}. 
The Kitaev quantum double  model $D(G)$ has as input a finite group $G$,  and the well-known  models are the toric code   based on the abelian group  $\mathbb{Z}_2$ \cite{1,2,3} and  on the non-abelian  group $S_3$ \cite{10,11,12}. Furthermore, the quantum double model has
been extended in a number of ways. In \cite{13} and \cite{14}, twisted  quantum double  models in   2+1D  and  3+1D have been defined, respectively.  A generalized 2+1D model in the  Hopf algebra setting  called the Hopf-algebraic Kitaev model has been  studied in \cite{15,16,17,18,19,20,21}.  Tensor network representations for the ground states of these models have been  constructed \cite{22,23}.

  Kitaev   models can also be viewed as  discrete gauge theories \cite{1} \ where the gauge   fields that colorate the  edges of the lattice are  built on group elements.  To provide rigorous descriptions of  the Kitaev lattice model at higher-dimensions, one  is  required to  extend the traditional lattice gauge theory to higher  gauge theory \cite{24,25,26,27}.  Higher  lattice gauge theories  of Kitaev’s quantum double model called higher Kitaev model   has been developed in \cite{28,29,30,31,32,33,34}.  The model can be defined on an  arbitrary dimensional manifold. It  consists of generalizing the gauge groups  to  their categorified version usually called strict 2-groups \cite{35,36,37,38,39}. In higher  lattice gauge theory, in addition to 1D-holonomy along the edges, one  also has  2D-holonomies along  faces of  the lattice \cite{33,40,41}. Exactly solvable Hamiltonian models of  mutually commuting operators for higher lattice gauge theory  have been developed in \cite{42,43,41}. They exhibit properties  of topological phases  such as a topologically dependent ground state degeneracy as in the Kitaev quantum double models. %\cite{KITAEV20032,bravyi2010topological,bravyi2011short}
. In  particular this implies that the ground state degeneracy  does not depend on the lattice and it is a topological invariant of manifold. For example, it was demonstrated in \cite{40}  that the model in question corresponds to the Hamiltonian realisation
 of the Yetter homotopy 2-type TQFT \cite{36,44,45}  and  the ground state degeneracy  can be identified
 with the Yetter topological partition function \cite{36}.  Similar ideas were applied to the 3+1D Crane-Yetter TQFT \cite{46}, giving rise to the Walker-Wang model. 
 
 Analogously to previous works on higher lattice gauge theory based on 2-groups \cite{33,40,41,47},
 % and on crossed modules of  semisimple Hopf algebras \cite{koppen2021exactly},
   we construct in the present paper a higher lattice gauge theory based on the representation of  2-groups \cite{39,48,49,50,51}   on   a   lattice model embedded in 3+1D manifolds. 
 % Note that a 2-group is the two-dimensional version of a group. In particular,  if a symmetry  is described by the action of a group  in a vector space,  2-groups describe  two levels of symmetry action in an appropriate category. 
   There are several equivalent manifestations of 2-groups \cite{48}, namely as   2-categories, as
 categorical groups (a certain type of category), or as crossed modules (an algebraic definition with 
 explicit categorical content). In the present  context, we adopt the view of 2-groups as crossed modules.  The representation of 2-groups can be regarded as 2-functors  between two equivalent definitions of  small strict 2-categories. This gives 2-categories that have the structures of 2-groupoids i.e., 2-categories with invertible morphisms. 
 %To find a suitable 2-category with the same structure as given 2-groups, we   consider morphisms of 2-category  to invertible morphisms. 
 %This gives a  category  which has the  structure of 2-groupoids. 
 We then  interpret higher lattice gauge theory as a 2-group representation in a lattice described by a path 2-groupoid, which is a particular example of a 2-groupoid that describes all possible motions of a point particle through the lattice model
  (see more details on path 2-groupoids \cite{26,33,51,52}). In this  setting, we  associate  2-functors (2-group representations) to the higher gauge  configurations. Gauge transformations are associated with local operators based  on vertices and edges of the
 underlying lattice. However, gauge invariants  are assigned to 1,2-holonomies. Finally, we  define a higher Hamiltonian model for 3+1D topological phases based  on local projectors  on  these lattice  gauge representations. We show that the  ground states of this model are invariant under mutation transformations of the graph that preserve the spatial  topology but not necessarily the local graph structure.
 
 The organization of this paper is as follows:
% Section \ref{sec2} reviews some background materials on equivalent definitions of 2-groups as 2-categories and as crossed modules that are required for the remainder of the paper.
 In section \ref{sec3}, we define a representation theory of 2-groups as strict 2-functors into 2-groupoids. We formulate the corresponding pseudo-natural transformations  and pseudo-natural equivalences as a stepping stone to defining a higher lattice gauge theory.
  Section \ref{HLGT} contains the key ingredients to describe a higher gauge theory, such as: matter fields, gauge configurations, gauge transformations and gauge invariants. From these data,  we
   construct  an exactly solvable topological Hamiltonian schema that encodes the higher lattice gauge theory model and  its ground  states. We show in section \ref{sec40} that these ground states are  topological observables. In the last section  \ref{sec5}, we present our conclusion.
%%%%%%%%%%%%%%%%%%%%%%%%%%%%%%%%%%%%%%%%%%%%%%%%%%%%%%%%%%%%%%%%%%%%%%%%%%%%%%%%%%%%%%%%%%%%%%%%%%%%%%%%%%%%%%%%%%%%%%%%%%%%%%%%%%%%%%%%%%%%%%%%%%%%%%%%
%%%%%%%%%%%%%%%%%%%%%%%%%%%%%%%%%%%%%%%%%%%%%%%%%%%%%%%%%%%%%%%%%%%%%%%%%%%%
\section{Representation  of 2-groups}\label{sec3}
The goal of this  section is to describe   representations of a 2-group  as 2-functors into a 2-category. This  theory of 2-group representations is based on analogy with the classical theory of group representations. Through this work, we define $2$-groups as a strict (small) $2$-category. See Appendix $A$  for a review of $2$-groups and for its  equivalent definitions.

\subsection{2-groups}  
 Before defining   representations of 2-groups, we brieflly review some equivalent definitions of 2-groups.There are equivalent definitions of  2-groups as: 2-groupoids,  crossed modules of groups and crossed modules over 2-groupoids called a category of crossed modules \cite{35,53,54}:

\begin{itemize}

\item  2-groups as 2-groupoids: A 2-group $\mathcal{G}$ is a small 2-category with  a single object such that all morphisms and 2-morphisms  are invertible. This 2-category is called  2-groupoids $\mathbb{G}=\left(\mathbb{G}_0,\,\mathbb{G}_1,\,\mathbb{G}_2\right)$  where $\mathbb{G}_0$ is the set of objects, $\mathbb{G}_1$ is  the set of morphisms and  the set of 2-morphisms is given by $\mathbb{G}_2$ with 
$\circ:\mathbb{G}_1 \,\,\,{}_{s}\times_t  \mathbb{G}_1\,\,\rightarrow \mathbb{G}_1$, $\circ_v:\mathbb{G}_2 \,\,\,\,{}_{s_v}\times_{t_v}  \mathbb{G}_2\rightarrow \mathbb{G}_2$, and $\circ_h:\mathbb{G}_2 \,\,\,{}_{s_h}\times_{t_h}  \mathbb{G}_2\rightarrow \mathbb{G}_2$ are the  composition of 1-morphisms, the vertical composition and the horizontal composition of 2-morphisms respectively. The maps $s,t: \mathbb{G}_1\rightarrow \mathbb{G}_0$, $s_v,t_v: \mathbb{G}_2\rightarrow \mathbb{G}_1, $ and $s_h,t_h:\mathbb{G}_2\rightarrow \mathbb{G}_0$ are the  source and the target functions respectively. The maps $\varepsilon: \mathbb{G}_0\rightarrow \mathbb{G}_1$,   $\varepsilon_h: \mathbb{G}_2\rightarrow \mathbb{G}_0 $ and $\varepsilon_v:\mathbb{G}_1\rightarrow \mathbb{G}_2$ are the identity maps for the composition of 1-morphisms, the vertical composition and the horizontal composition of 2-morphisms respectively. Finally, the inverse maps $(\eta,\eta_v,\eta_h)$  for  each  composition laws are given by $\eta: \mathbb{G}_1\rightarrow \mathbb{G}_1$, $\eta_v: \mathbb{G}_2\rightarrow \mathbb{G}_2$ and $\eta_h: \mathbb{G}_2\rightarrow \mathbb{G}_2$. See Appendix A for more explicit detail.

\item 2-groups as a crossed module of groups:  A 2-group $\mathcal{G}$ is  
a crossed module  $\mathcal{X}=\left( E\rightarrow G,\partial,\, \rhd\right) $ consists of groups $G$ and $E$ together with   a homomorphism $\partial:E\rightarrow G$
and a left action $ \rhd: G\times E\rightarrow E$ by automorphisms
satisfying the Peiffer relations:    
\begin{eqnarray}
\partial(g\rhd e)&=&g\partial(e)g^{-1} \quad \forall\, g\in G \quad \mbox{and}\quad e\in E,\label{P1} \\
\partial(e)\rhd f&=&efe^{-1} \quad\quad \forall\, e,f\in E.  \label{P2}
\end{eqnarray}

\item  2-groups as a category of crossed modules: From  a crossed module  $\mathcal{X}=\left( E\rightarrow G,\partial,\, \rhd \right)$,
we can associate  a 2-groupoid to $\mathcal{X}$ defined by $  \mathcal{G}(\mathcal{X})=\left(\mathcal{G}_0=\{*\},\,\mathcal{G}_1= G,\,\mathcal{G}_2=G \ltimes_{\rhd} E\right)$
where the semi-direct product $ G \ltimes_{\rhd} E $ with the group multiplication $\circ_h$ given by
\begin{eqnarray}
(g,e)\circ_h(g',e')=(gg',e(g\rhd e')), \quad g,g'\in G \quad \mbox{and}\quad e,e'\in E,
\end{eqnarray}
where $\circ_h$ is a horizontal composition map of 2-morphisms. With these equivalent definitions of 2-groups at hand, we can now define the representations of 2-groups. Furthermore, we define the non-trivial source, target and identity maps
\begin{eqnarray}
    s : G \ltimes_{\rhd} E \rightarrow G,(g, e) \mapsto g;\,  t : G \ltimes_{\rhd} E \rightarrow G,(g, e) \mapsto \partial(e)g; \,  \varepsilon : G \rightarrow G \ltimes_{\rhd} E,(g,e) \mapsto (g,1_E)
\end{eqnarray}
The composition laws are more explicitly decribed in appendix A.
\end{itemize}

\subsection{Representation of 2-groups}

 In the context of categories, a representation $(\phi, \mathcal{V})$ of a group $G$ can be understood as a functor from the category of group $G$ or  groupoid to a suitable category  of vector  space $\mathcal{V}$   over a field $k$  such that
 \begin{eqnarray}
 \phi: G\rightarrow \mathcal{V}.
 \end{eqnarray}

  This  representation is  also known  as the  action of groupoid $G$ on the category of a vector space $ \mathcal{V}$ by automorphisms  denoted  by $  G\times \mathcal{V}\rightarrow \mathcal{V}.$
As mentionned,   $G$ is a category
with one object $\{*\}$  whose  morphisms are of the form  $ *\xrightarrow{g} *$ with $g\in G$, a such  functor $\phi$ can be defined as    
\begin{eqnarray}
\phi(*\xrightarrow{g} *):=\phi(*) \xrightarrow{\phi(g)} \phi(*)= V\xrightarrow{\phi(g)}V,
\end{eqnarray}
 where $ \phi(*)=V\in\mathcal{V}$ is the groupoid object. To every group element $g\in G$, $\phi$ assigns an invertible operator $\phi(g): V\rightarrow V$. This assignment satisfies  
  \begin{eqnarray}
  (V\xrightarrow{\phi(g_1)}V)\circ( V\xrightarrow{\phi(g_2)}V)&=& V\xrightarrow{\phi(g_1g_2)}V \iff \phi(g_1)\phi(g_2)=\phi(g_1g_2).
  \\
  V\xrightarrow{\phi(1_G) }V&=&V\xrightarrow{1_V}V\iff \phi(1_G)=1_V .
 \end{eqnarray}
 
 Given a pair of functors  $\phi_1,\phi_2:G\rightarrow \mathcal{V} $. A natural transformation $\vartheta: \phi_1\Rightarrow \phi_2$ is given by a map $ \vartheta: V_1\rightarrow V_2$ which associate $\phi_1(*)=V_1$ and $\phi_1(*)=V_2$.  Saying that the transformation is ‘natural’ then means
 that this square commutes:
 \begin{eqnarray*}
          \begin{tikzpicture}[thick]
          \matrix(m)[matrix of math nodes,
          row sep=2.6em, column sep=2.8em,
          text height=1.5ex, text depth=0.25ex]
          {V_1&V_1\\
          V_2&V_2\\};
          \path[->,font=\scriptsize,>=angle 90]
          (m-1-1) edge node[auto] {$\phi_1(g)$} (m-1-2)
          edge node[auto] {$ \vartheta $} (m-2-1)
          (m-1-2) edge node[auto] {$\vartheta  $} (m-2-2)
          (m-2-1) edge node[auto] {$\phi_2(g)$} (m-2-2);
          \end{tikzpicture}
 \end{eqnarray*}
for each group element $g$ such that
\begin{eqnarray}
\phi_2(g)\vartheta=\vartheta\phi_1(g), \quad \forall g \in G.
\end{eqnarray}
Such a $\vartheta$ is sometimes referred to in representation theory as an intertwining operator, so a representation $\phi$  may be thought of as the category whose objects are the $k$-linear
representations of $G$ and whose arrows are the intertwining operators between such representations.\\

 Similar to the representation of a group in the category of vector spaces,  2-groups  can  represented in some 2-categories of 2-vector spaces.

\begin{definition}{\it  
Let $\mathcal{G}$ be a 2-group and $\mathcal{C}$ any 2-category.  A representation $\Phi$ of $\mathcal{G}$ can be regarded as a 2-functor into a  2-category $\mathcal{C}$ over a field $k$  
$$\Phi: \mathcal{G}\rightarrow \mathcal{C}.$$}
\end{definition} 
To find a suitable 2-category with the same structure as given, 2-groups, one has to consider morphisms of  $\mathcal{C}$ to invertible morphisms \cite{48,49,50,51}.  This gives a 2-category $\mathbb{G}=(\mathbb{G}_0,\mathbb{G}_1,\mathbb{G}_2)$ 
 that has the structure of 2-groupoids such that  
\begin{itemize}
\item $\mathbb{G}_0$ contains a single element  $X$,   $0$-cell of    $\mathbb{G}$.
\item $\mathbb{G}_1$ contains  1-cells (1-morphisms)  which  are  isomorphisms from $X$ to  itself $\gamma:  X\rightarrow X  $  and its inverse is given by $\gamma^{-1}:  X\leftarrow X  $
\begin{eqnarray*}
\begin{tikzpicture}[thick]
\node (A) at (-2.5,0) {$ X$};
\node (B) at (2.5,0) {$ X $};
\path[->] (A) edge  node[above] {$\gamma$} (B);

\node (F)  at (4.5,0){and };
\node (G)  at (6.5,0){$X$};
\node (H)  at (11.5,0){$X$};

\path[<-] (G) edge  node[above] {$\gamma^{-1}$} (H);

\end{tikzpicture}      
\end{eqnarray*}
   
\item  $\mathbb{G}_2$ group contains 2-cells (2-morphisms) are homotopies between 1-cells   $\alpha:\gamma\Rightarrow \gamma'$ and its inverse is given by  
$\alpha^{-1}:\gamma\Leftarrow \gamma'$
\begin{eqnarray*}
\begin{tikzpicture}[thick]
\node (A) at (-1.5,0) {$X$};
\node (B) at (3.5,0) {$X$};
\node at (1,0) {\rotatebox{270}{$\implies$}};
\path[->, font=\scriptsize,>=angle 90]node[right=1.03cm]{$\alpha$}
(A) edge [bend left] node[above] {$\gamma$} (B)
edge [bend right] node[below] {$\gamma'$} (B);

\node (F)  at (4.9,0){and};
\node (G)  at (6.5,0){$X$};
\node (H)  at (11.5,0){$X$};
\node (I) at (9.4,0) {$\alpha^{-1}$};
\node at (9,0) {\rotatebox{-270}{$\implies$}}node[right=0.10cm]{$ $};
\path[->] (G) edge [bend left=25] node[above] {$\gamma$} (H);
\path[->] (G) edge [bend right=25] node[below] {$\gamma'$} (H);

\end{tikzpicture}      
\end{eqnarray*}
\end{itemize}

Thus, in complete  analogy with the action of  2-groups $\mathcal{G}$ on a $2$-category $\mathcal{C}$, the  2-group actions on a 2-groupoid  $\mathbb{G}$ are described by a 2-functor from the  2-category  $\mathcal{G}$  to the 2-category  $\mathbb{G}$
\begin{eqnarray}
\Phi:\mathcal{G}\rightarrow \mathbb{G}.
\end{eqnarray}
This could  be  understood as linear transformations of 2-groups  $\mathcal{G}$  over the 2-groupoids  $\mathbb{G}$ i.e, $\Phi(\mathcal{G}):\mathbb{G}\rightarrow  \mathbb{G}$.
   Object in $\mathbb{G}$ given  by $X=\Phi(\mathcal{G}_0)$ is the unique functorial image lies within the   2-groupoid  $\mathbb{G}$. The 1-representation of $\mathcal{G}$ on   $\mathbb{G}$ is    a 1-automorphism of  $ \Phi(\mathcal{G}_1):X\xrightarrow{\gamma} X$.   However,  the 2-representations of $\mathcal{G}$ on   $\mathbb{G}$ is    a 2-automorphism of  $ \Phi(\mathcal{G}_2):\gamma\xRightarrow{\alpha} \gamma'$   where $\gamma,\gamma'\in  \mathbb{G}_1$ and $  \alpha\in  \mathbb{G}_2$.
   
   To precisely describe the properties of the 2-functor, we consider the specific example of the 2-group as   a category of crossed modules $\mathcal{G}(\mathcal{X})= (*,G, G\ltimes_{\rhd} E,\partial,\rhd)$, where $\{*\}$ is the single object, $G$ is the group of   1-morphisms of $\mathcal{G}$ and $ G \ltimes_{\rhd} E$ the group of  2-morphisms between 1-morphisms.
 
\begin{definition}
{\it   A 2-functor $  \Phi: \mathcal{G}(\mathcal{X})\rightarrow \mathbb{G}$ can be defined as follows:
  \begin{itemize}
 \item For the  single object $\{*\}$ in $ \mathcal{G}(\mathcal{X})$,  then  $\Phi(*)=X \in \mathbb{G}_0$  is the 0-cell of  $\mathbb{G}$.
 
 \item  For any 1-morphism $*\xrightarrow{g}*$  with $g\in G$ and  for the 1-cell $ X \xrightarrow{\gamma} X \in  \mathbb{G}_1$, we have
 \label{key}\begin{eqnarray}
   X\xrightarrow{\gamma} X \quad \xmapsto{\Phi(*)\xrightarrow{\Phi(g)}\Phi(*)}\quad   \Phi_X(*) \xrightarrow{\Phi_\gamma(g)} \Phi_X(*) \equiv X\xrightarrow{\Phi_\gamma(g)} X,
  \end{eqnarray}
 where $\Phi_\gamma(g)= \Phi(g)\gamma$ is a linear transformation of 1-cell to itself  by $\Phi(g),$      
 \item For any 2-morphism $ g\xRightarrow{(g, e)} \partial(e)g$  with $ (g, e)\in G \ltimes_{\rhd} E$, and for the 2-cell $ \gamma\xRightarrow{\alpha} \gamma'\in  \mathbb{G}_2$, we have

\begin{eqnarray}
\begin{tikzpicture}[thick]
\node (A) at (-2.3,0) {$X$};
\node (B) at (2.3,0) {$ X$};
\node at (0,0) {\rotatebox{270}{$\implies$}};
\path[->, font=\scriptsize,>=angle 90]node[right=0.10cm]{$\alpha$}
(A) edge [bend left=35] node[above] {$\gamma $} (B)
edge [bend right=35] node[below] {$ \gamma'$} (B);

\node (F)  at (4.6,0){$\xmapsto{\Phi(g)\xRightarrow{\Phi(g, e)} \Phi(\partial(e)g)}$};
\node (G)  at (6.5,0){$ X$};
\node (H)  at (11.5,0){$ X$};
\node at (9,0) {\rotatebox{270}{$\implies$}} node  at (9.90,0){$\Phi_\alpha(g,e)$};
\path[->] (G) edge [bend left=25] node[above] {$\Phi_\gamma(g)$} (H);
\path[->] (G) edge [bend right=25] node[below] {$\Phi_{\gamma'}(\partial(e)g)$} (H);

\end{tikzpicture}      
\end{eqnarray}
where $ \Phi_{\alpha}(g,e)=\Phi(g,e)\alpha$  is a linear transformation of 2-cell to itself  by $\Phi(g,e)$.
 \end{itemize}
}
\end{definition}
These assignments must satisfy the conditions to be a strict 2-functors, i.e they preserve the composition laws  and identities, which here means:
\begin{itemize}
\item For 1-morphisms  
 \begin{eqnarray}
X\xrightarrow{\Phi_{\gamma_1}(g_1)}X\xrightarrow{\Phi_{\gamma_2}(g_2)}X&=& X\xrightarrow{\Phi_{\gamma_1\circ\gamma_2}(g_1g_2)}X,\\
X\xrightarrow{\Phi_\gamma(1_G)}X&=&X\xrightarrow{1_{\Phi_\gamma(X)}}X,
\end{eqnarray}
% where $\circ:\mathbb{G}_1\times_{\mathbb{G}_0}\mathbb{G}_1\rightarrow %\mathbb{G}_1$ is the  composition law of 1-morphisms.
\item For all vertically composable 2-morphisms 
\begin{eqnarray}
 \Phi_{\alpha_1}(\partial(e)g_1,e_1)\circ_v \Phi_{\alpha_2}(g_2,e_2)&=&\Phi_{\alpha_1\circ_v\alpha_2}(g_2,e_1e_2), \\
\Phi_\alpha(g,1_E)&=&1_{\Phi_\alpha(g)},
\end{eqnarray}
%where $\circ_v: \mathbb{G}_2 \times_{\mathbb{G}_1} %\mathbb{G}_2\rightarrow \mathbb{G}_2$ is the vertical composition map of 2-morphisms.
This vertically composable 2-morphisms  can be illustrated as follows   
\begin{eqnarray}
\begin{tikzpicture}[thick]
\node (A) at (-3.5,0) {$X$};
\node (B) at (3.5,0) {$X$};
\path[->] (A) edge [bend left=45,""{name=F}] node[above]{$\Phi_{\gamma_1}(g_1)$} (B);
\path[->] (A) edge[]  node[left=1.4, above] {$\Phi_{\gamma_2}(\partial(e_1)g_1) $} (B);
\path[->] (A) edge [bend right=45, ""{name=D,  }] node[below] {$\Phi_{\gamma_3}(\partial(e_1e_2)g_1)$} (B);

\node at (0,0.5) {\rotatebox{270}{$\implies$}} node at (1.3,0.5) {$\Phi_{\alpha_1}(g_1,e_1)$};
\node at (0,-0.5) {\rotatebox{270}{$\implies$}}  node at (1.3,-0.5) {$\Phi_{\alpha_2}(g_2,e_2)$};
\node (F)  at (4.,0){$=$};
\node (G)  at (4.5,0){$X$};
\node (H)  at (10.5,0){$X$};
%\node (I) at (9.70,0) {$\delta\circ_v \xi$};
\node at (7.5,-0.6) {\rotatebox{270}{$\implies$}} node  at (8.,0) {$\Phi_{\alpha_1\circ_v\alpha_2}(g_1, e_1 e_2)$};
\path[->] (G) edge [bend left=35] node[above] {$\Phi_{\gamma_1}(g_1)$} (H);
\path[->] (G) edge [bend right=35] node[below] {$\Phi_{\gamma_3}(\partial(e_1e_2)g_1)$} (H);
\end{tikzpicture}      
\end{eqnarray}
and
\begin{eqnarray}
\begin{tikzpicture}[thick]
\node (A) at (-2.3,0) {$X$};
\node (B) at (2.3,0) {$X$};
\node at (0,0) {\rotatebox{270}{$\implies$}};
\path[->, font=\scriptsize,>=angle 90]node[right=0.10cm]{$\Phi_{\alpha}(g,1_G)$}
(A) edge [bend left=35] node[above] {$\Phi_\gamma(g) $} (B)
edge [bend right=35] node[below] {$ \Phi_{\gamma'}(g)$} (B);
\node (F)  at (5.,0){$= $};
\node (G)  at (6.5,0){$X$};
\node (H)  at (11.,0){$X$};
\node at (9,0) {\rotatebox{270}{$\implies$}} node  at (9.70,0){$1_{\Phi_\alpha(g)}$};
\path[->] (G) edge [bend left=25] node[above] {$\Phi_\gamma(g)$} (H);
\path[->] (G) edge [bend right=25] node[below] {$\Phi_{\gamma'}(g)$} (H);

\end{tikzpicture}      
\end{eqnarray}

\item For all horizontally composable 2-morphisms
\begin{eqnarray}
 \Phi_{\alpha_1}(g_1,e_1)\circ_h \Phi_{\alpha_2}(g_2,e_2)&=& \Phi_{\alpha_1\circ_h\alpha_2}((g_1,e_1)\circ_v(g_2,e_2))= \Phi_{\alpha_1\circ_h\alpha_2}(g_1g_2,e_1(g_1\rhd e_2)),\\
\Phi_\alpha(1_G,1_{1_{X}})&=&1_{1_{\Phi_\alpha(X)}},
\end{eqnarray}
% where $\circ_h: \mathbb{G}_2 \times_{\mathbb{G}_1} \mathbb{G}_2\rightarrow \mathbb{G}_2$ is the horizontal composition map of 2-morphisms. 
 This horizontally composable 2-morphisms  can be illustrated as follows:
\begin{eqnarray}
\begin{tikzpicture}[thick]
\node (A) at (-1.5,0) {$ X$};
\node (B) at (3,0) {$ X$};
\node (C)  at (7.9,0){$ X$};
\node (E)  at (3,0){$ $};
\node at (0.5,0) {\rotatebox{270}{$\implies$}}node[right=0.48cm]
{$\Phi_{\alpha_1}(g_1,e_1)$};
\path[->] (A) edge [bend left=35] node[above] {$\Phi_{\gamma_1} (g_1)$} (B);
\path[->] (A) edge [bend right=35] node[below] {$\Phi_{\gamma_1'}(\partial(e_1)g_1)$} (B);
\path[->] (B) edge [bend right=35] node[below] {$\Phi_{\gamma_2'}(\partial (e_2)g_2)$} (C);
\path[->] (B) edge [bend left=35] node[above] {$\Phi_{\gamma_2}(g_2)$} (C);

\node (E) at (5.58,0) {$\Phi_{\alpha_2}(g_2,e_2)$};
\node at (4.5,0) {\rotatebox{270}{$\implies$}};

\node (F)  at (8.4,0){};
\node (G)  at (-1.80,-3.5){$ =X$};
\node (H)  at (7.5,-3.5){$X$};
\node (I) at (4.7,-3.50) {$\Phi_{\alpha_1\circ_h\alpha_2}(g_1g_2,e_1(g_1\rhd e_2))$};
\node at (2.5,-3.50) {\rotatebox{270}{$\implies$}}node[right=0.10cm]{$ $};
\path[->] (G) edge [bend left=25] node[above] {$\Phi_{\gamma_1\circ\gamma_2}(g_1g_2)$} (H);
\path[->] (G) edge [bend right=25] node[below] {$\Phi_{\gamma_1'\circ \gamma_2' }(\partial (e_1e_2)g_1g_2)
$} (H);

\end{tikzpicture}      
\end{eqnarray}
and
\begin{eqnarray}
\begin{tikzpicture}[thick]
\node (A) at (-2.3,0) {$X$};
\node (B) at (2.3,0) {$X$};
\node at (0,0) {\rotatebox{270}{$\implies$}};
\path[->, font=\scriptsize,>=angle 90]node[right=0.10cm]{$\Phi_{\alpha}(1_G,1_{1_X})$}
(A) edge [bend left=35] node[above] {$\Phi_\gamma(1_G) $} (B)
edge [bend right=35] node[below] {$ \Phi_{\gamma'}(1_G)$} (B);
\node (F)  at (3.4,0){$= $};
\node (G)  at (5.5,0){$X$};
\node (H)  at (10.5,0){$X$.};
\node at (8,0) {\rotatebox{270}{$\implies$}} node  at (9.0,0){$1_{1_{\Phi_\alpha(X)}}$};
\path[->] (G) edge [bend left=25] node[above] {$\Phi_\gamma(1_G)$} (H);
\path[->] (G) edge [bend right=25] node[below] {$\Phi_{\gamma'}(1_G)$} (H);

\end{tikzpicture}      
\end{eqnarray}

\item 
The right and left whickering are given by
\begin{eqnarray}
\begin{tikzpicture}[thick]
%\node (A0) at (-3,0) {$X$};
\node (A) at (-1.7,0) {$X$};
\node (B) at (2.5,0) {$ X$};
\node (C)  at (4.05,0){$X$};
\node (E)  at (3,0){$ $};
\node at (0.4,0) {\rotatebox{270}{$\implies$}}node[right=0.550cm]{$\Phi_\alpha(g,e)$};
\path[->] (A) edge [bend left=35] node[above] {$\Phi_{\gamma}(g)$} (B);
\path[->] (A) edge [bend right=35] node[below] {$\Phi_{\gamma'}(\partial(e)g)$} (B);
%\draw[->] (A0) --node[above=0.01cm] {$\Phi_{\gamma_1}(g_1)$} (A) ;
\draw[->] (B) --node[above=0.01cm] {$\Phi_{\gamma_2}(g_2)$} (C) (C);

\node (F)  at (4.5,0){=};
\node (G)  at (5.,0){$X$};
\node (H)  at (10.5,0){$X$.};
\node (I) at (8.5,0) {$\Phi_\alpha(g, e)$};
\node at (7,0) {\rotatebox{270}{$\implies$}}node[right=0.10cm]{$ $};
\path[->] (G) edge [bend left=25] node[above] {$\Phi_{\gamma\circ\gamma_2}(gg_2)$} (H);
\path[->] (G) edge [bend right=25] node[below] {$\Phi_{\gamma'\circ\gamma_2}(\partial (e)gg_2)
$} (H);

\end{tikzpicture}
\end{eqnarray}
\end{itemize}

\begin{eqnarray}
\begin{tikzpicture}[thick]
\node (A0) at (-3,0) {$X$};
\node (A) at (-1.7,0) {$X$};
\node (B) at (2.5,0) {$ X$};
%\node (C)  at (4.05,0){$X$};
\node (E)  at (3,0){$ $};
\node at (0.4,0) {\rotatebox{270}{$\implies$}}node[right=0.550cm]{$\Phi_\alpha(g,e)$};
\path[->] (A) edge [bend left=35] node[above] {$\Phi_{\gamma}(g)$} (B);
\path[->] (A) edge [bend right=35] node[below] {$\Phi_{\gamma'}(\partial(e)g)$} (B);
\draw[->] (A0) --node[above=0.01cm] {$\Phi_{\gamma_1}(g_1)$} (A) ;
%\draw[->] (B) --node[above=0.01cm] {$\Phi_{\gamma_2}(g_2)$} (C) (C);

\node (F)  at (3.5,0){=};
\node (G)  at (5.,0){$X$};
\node (H)  at (10.5,0){$X$.};
\node (I) at (8.5,0) {$\Phi_\alpha(g,\Phi_{\gamma_1}(g_1)\rhd e)$};
\node at (7,0) {\rotatebox{270}{$\implies$}}node[right=0.10cm]{$ $};
\path[->] (G) edge [bend left=25] node[above] {$\Phi_{\gamma_1\circ\gamma}(g_1g)$} (H);
\path[->] (G) edge [bend right=25] node[below] {$\Phi_{\gamma_1\circ\gamma'}(g_1\partial (e)g)
$} (H);

\end{tikzpicture}
\end{eqnarray}

%\end{itemize}

  As we can see, the strict 2-functor $\Phi$ between strict 2-categories   $\mathcal{G}(\mathcal{X})$  and $\mathbb{G}$ preserves  the structures of both 1-morphisms and 2-morphisms and    the composition between a 1-morphism and a
2-morphism which is an intermediate step  called a whisker.

  %It is not only ensures  the  composition of a 2-morphism (a left whiskering)  but   also  a right whiskering i.e, a composition of a 2-morphism and a right a 1-morphism.  In fact, the attachment of the left whiskering can be understood as a special case of the horizontal composition. Similarly, the right whiskering of a 1-morphism on a 2-morphism This terminology can be explained by the picture

As with group representations, we have intertwiners between 2-group representations, which in the language of 2-category are pseudonatural transformations.

\begin{definition}
     {\it  Let   $\Phi,\Phi':\mathcal{G}(\mathcal{X})\rightarrow \mathbb{G}$ be a two  representations of $\mathcal{G}$ in  $\mathbb{G}$.  A pseudo-intertwiner   $\vartheta: \Phi\Rightarrow \Phi'$  is a map  $\vartheta: X\rightarrow X'$  as such that
 the diagram
\begin{eqnarray}
\begin{tikzpicture}[node distance=3cm and 4cm]
\node (A) at (-1.5,0) {$X$};
\node (B) at (2.5,0) {$X$};
\node (C) at (2.5,0) [below=of A] {$X'$};
\node (D) at (2.5,0) at (2.5,0) [below=of B] {$X'$};  
\node at (0.3,0) {\rotatebox{270}{$\implies$}};
\path[->, font=\scriptsize,>=angle 90]node[right=0.40cm]{$\Phi_{\alpha_1}(g_1,e_1)$}
(A) edge [bend left] node[above] {$\Phi_{\gamma_1}(g_1)$} (B)
edge [bend right] node[below] {$\Phi_{\gamma_1'}(\partial(e_1)g_1) $}(B);

\draw[->](A)--node [left=0.15cm] {$\vartheta$}(C);
\draw[->](B)--node [right=0.15cm] {$\vartheta$}(D);
\node at (0.25,-3.50) {\rotatebox{270}{$\implies$}};
\path[->, font=\scriptsize,>=angle 90] (C) node[ below=7cm, right=1.81cm]{$\Phi_{\alpha_2}(g_2,e_2)$}
(C) edge [bend left] node[above] {$\Phi_{\gamma_2}'(g_2)$} (D)
edge [bend right] node[below] {$\Phi_{\gamma_2'}'(\partial(e_2)g_2)$} (D);

%\draw[->](B)--node [left=0.15cm] {$\vartheta$}(C);
\node at (1.6,-1.890)   {\rotatebox{-135}{$\Longrightarrow$}}node  at (1.9,-1.40){$\vartheta(g) $};

\node at (-0.9,-2.40)   {\rotatebox{-135}{$\implies$}}node  at (-0.5,-1.90){$\vartheta(g') $};

\end{tikzpicture}      
\end{eqnarray}
 Commute \cite{39} i.e
 \begin{eqnarray}
 \left[ 1_{\vartheta}\circ\Phi_{\alpha_1}(g_1,e_1)\right].\vartheta(g)=\vartheta(g').\left[\Phi_{\alpha_2}(g_2,e_2)\circ 1_{\vartheta} \right],
 \end{eqnarray}
where the 2-morphism $ \vartheta(g'): \Phi_{\gamma_1'}(\partial(e_1)g_1)\vartheta\Rightarrow  \vartheta \Phi_{\gamma_2'}'(\partial(e_2)g_2)$ is located at the front face of the diagram while $ \vartheta(g): \Phi_{\gamma_1}(g_1)\vartheta\Rightarrow  \vartheta \Phi_{\gamma_2}'(g_2)$ is at the back.
}
\end{definition} 
The axioms for a pseudo-natural transformations guarantee that given a pair of strict 2-functors
$\Phi,\Phi' :\mathcal{G}(\mathcal{X})\rightarrow \mathbb{G}$ and pseudo-natural transformation  $\vartheta: X\rightarrow X'$
the following relations hold \cite{39} 
\begin{eqnarray}
\vartheta(\Phi_{\gamma_1}(g_1))\circ\vartheta( \Phi_{\gamma_2}(g_2))&=&\vartheta\left(\Phi_{\gamma_1\circ\gamma_2}(g_1g_2)\right)\quad \mbox{and}\quad \vartheta(\Phi_{\gamma}(1_G))=\vartheta(1_X),\\
\vartheta(\Phi_{\alpha_1}(u_1))\circ_v\vartheta( \Phi_{\alpha_2}(u_2))&=&\vartheta\left(\Phi_{\alpha_1\circ_v \alpha_2}(u_1\circ_v u_2) \right)\quad \mbox{and}\quad \vartheta(\Phi_\alpha (g,1_{g}))= \vartheta( 1_{\Phi_\alpha (g)}),\\
\vartheta(\Phi_{\alpha_1}(u_1))\circ_h\vartheta( \Phi_{\alpha_2}(u_2))&=&\vartheta\left(\Phi_{\alpha_1\circ_h \alpha_2}(u_1\circ_h u_2) \right) \quad \mbox{and}\quad   \vartheta(\Phi_\alpha (g,1_{1_X}))= \vartheta( 1_{ 1_{\Phi_\alpha(X)}}).
\end{eqnarray}

 Now, if the  intertwining operator  $\vartheta$ between $\Phi$ and $\Phi'$ is  inversible one defines a pseudo-natural equivalence as follows\\

\begin{definition}   
 {\it A pseudo-natural equivalence is a pseudo-natural transformation $\vartheta: \Phi\rightarrow \Phi'$ such that there exists $\vartheta^{-1}: \Phi'\rightarrow \Phi$ where $\vartheta\vartheta^{-1}=1_\Phi: \Phi\rightarrow \Phi$ the identity natural transformation for $\Phi$ and  $\vartheta^{-1}\vartheta=1_{\Phi'}: \Phi'\rightarrow \Phi'$ the identity natural transformation for $\Phi'$.
   }
\end{definition}

%%%%%%%%%%%%%%%%%%%%%%%%%%%%%%%%%%%%%%%%%%%%%%%%%%%%%%%%%%%%%%%%%%%%%%%%%%%%%%%%%%%%%%%%%%%%%%%%%%%%%%%%%%%%%%%%%%%%%%%%%%%%%%%%%%%%%%%%%%%%%%%%%%%%%%%%
\section{ Higher lattice gauge theory and  Hamiltonian formulation }\label{HLGT}
Higher lattice gauge theories \cite{28,29,30,31,32,33,34,40,41} are a generalization of lattice gauge theory \cite{1}, where there is a second gauge field that describes the parallel transport of the ordinary 1-gauge field across the surface. In lattice gauge theory, the gauge  fields that decorate the edges of the lattice are built on group elements. However, in higher lattice gauge theory, the finite group elements are replaced by 2-group elements  that decorate the faces of the lattice.  
Analogously to  research papers in the frameworks  \cite{40,41,43}, we  interpret higher lattice gauge theory as  2-group representations in a path 2-groupoid. The lattice  model described by a  path 2-groupoid is decorated by 2-group (category of crossed modules) actions as linear transformations. In this  setting, the  1-holonomy generated by 1-morphism decorates the edges of the lattice, and the 2-holonomy  generated by 2-morphisms transport edge decorations through the face of the lattice. 

 To  describe a  continuum  gauge theory, the key ingredients are matter  fields,  gauge configurations,  gauge transformations, and  gauge invariants. In this discrete case,  matter fields are described by particles moving in  a lattice  that are coupled  to the gauge  configuartions  described by 2-group representations.  The gauge transformations will play the  role of   pseudo-natural transformations.  The gauge invariance will be assigned to the pseudo-natural equivalence. Finally, we define an exactly
		solvable model for a 3+1D topological Hamiltonian that  encodes  this lattice gauge theory.       

\subsection{Lattice representation}

 The first input data  for the construction is a lattice  embedded in a three-dimensional (3D) topological manifold.
 The lattice model $(M,L)$ is described by a topological oriented manifold $M$ equipped  with a discrete structure  in terms of a cellular decompotion denoted a lattice $L$. Here, a lattice $L$ consists of 
 subsets $L^i$ for $i=\{ 0,1,2,3\},$  where each $ L^i$ is a closed topological i-disk embedded in $M$, such that $M^i:=\cup_{j=0}^iL^j$ \cite{33,40,41,42}. This cellular decomposition can be interpreted as follows:
An element  $ v\in L^0$ is a point of M, called  a vertex, and $L^0$ defines sets of 0-cells. An element $\gamma\in L^1$ is called  edges or link and  $L^1$ defines sets of 1-cells.
 An element  $ \alpha\in L^2$ is called  a face or plaquette, and $L^2$ defines sets of 2-cells.  An element  $b\in L^3$ is called  a blob, and $L^3$ defines sets of 3-cells. To describe the lattice structures, we need    some additional structures  which consist of  dressing this lattice  and coloring it by associating a path 2-groupoid.

\subsubsection{Dressed Lattice }
 Dressing the lattice consists of  ordering  vertices and  orienting   plaquettes. We choose  an ordering of the  vertices of the lattice such that $v_1<v_2< v_3<...$.     We can represent this ordering by assigning arrows oriented from lower to higher ordered vertices on the edges of the lattice.  The 1-skeleton $(L^0,L^1)$ then has the structure of a directed graph. We then equip the lattice with additional orientation data for the boundaries of 2- and 3-cells.
For any given elements of the  plaquette $\alpha\in L^2$ and the blob $b\in L^3$, we distinguish    vertices $ v_\alpha\in L^0$ and $ v_b\in L^0$ in the plaquette and the blob base-points respectively. 
For the  sake of  simplicity, we consider both  base points at the same fix, point i.e., $v_\alpha=v_b$.
 Orientation for the plaquette  is also fixed from its base point $v_\alpha$ in clockwise or anticlockwise. We refer to $(M,L)$ with the above structure  as a dressed lattice or lattice model.

\subsubsection{Path 2-Groupoid }
 Given the dressed  lattice $(M,L),$ on can canonically associate a  path 2-groupoid $\mathcal{P}_2(M,L)$ to the lattice \cite{26,33,52}.  The path 2-groupoid $\mathcal{P}_2(M,L)=(P_0, P_1, P_2)$ is a small strict 2-groupoid
whose underlying 1-category is the path groupoid $\mathcal{P}_1(M,L)=(P_0, P_1)$ \cite{26,33,52}. In the present case, $P_0\simeq L^0$, the sets of $0$-cells.  $P_1\simeq L^1$, the sets of $1$-cells which are isomorphic from $v$ to itself $\gamma: v\rightarrow v$. Elements  $ \gamma \xRightarrow{\alpha} \gamma'\in  P_2\simeq L^2$ form the  set of 2-cells. Finally, elements $b\in P_3\simeq L^3$ of the  blobs form the set of 3-cells, which form the smallest volumes over the surfaces such as the smallest cubes in the  lattice.  

 The primary distinction between using the path 2-groupoid $\mathcal{P}_2(M,L)$ rather than the 2-groupoid $\mathbb{G}$ in this context is that the path 2-groupoid is a particular example of a 2-groupoid that describes all possible motions of a point particle through the lattice model $(M, L)$, whereas the 2-groupoid is a broad, abstract mathematical structure \cite{26}.

\subsection{Gauge  configurations and holonomies}

\subsubsection{Gauge  configurations}

%Let us now study the gauge  configurations of  our  lattice model $(M,L)$. 
The gauge configurations can be regarded as the  action of   2-groups $\mathcal{G}$ on the lattice path 2-groupoid $\mathcal{P}_2(M,L)$ by automorphisms. The 2-groups  described by crossed modules $\mathcal{G}(\mathcal{X})=(*,G, G\ltimes_{\rhd} E,\partial,\rhd)$  decorate the edges (paths)   and the plaquettes (surfaces) of $\mathcal{P}_2(M,L)$ by linear transformations of $\Phi(G)  $ and  $\Phi(G\ltimes_{\rhd} E)$ respectively.

\begin{definition}
     {\it The gauge configurations of the lattice path 2-groupoid $\mathcal{P}_2(M,L)$ is described by the map   
    \begin{eqnarray}\label{hi}
    	\Phi(\mathcal{G}(\mathcal{X})): \mathcal{P}_2(M,L)\rightarrow \mathcal{P}_2(M,L),	   
     \end{eqnarray}
 which described a linear transformation of the  lattice path 2-groupoid $\mathcal{P}_2(M,L)$ to itself. The image of  the  single object $\{*\} \in\mathcal{G}(\mathcal{X})$  is $\Phi(*)=v\in L^0$  and the linear transformation of 1-morphisms and 2-morphisms of $\mathcal{G}(\mathcal{X})$ are given by:   
 \begin{itemize}
 \item  for each  $ *\xrightarrow{g}*\in G$,  we have the endofunctor 
 \begin{eqnarray}
 	\Phi(*)\xrightarrow{\Phi(g)}\Phi(*)&:& L^1 \longrightarrow L^1\cr
 	&& (v\xrightarrow{\gamma}v)\mapsto \Phi_v(*)\xrightarrow{\Phi_\gamma(g)}\Phi_v(*)\equiv  v\xrightarrow{\Phi_\gamma(g)}v,
 \end{eqnarray}
  where $\Phi_\gamma(g)=\Phi(g)(\gamma)$ is the parallel transport of the  edge $\gamma\in L^1$ by $\Phi(g)$, 
  \item  for each $ g\xRightarrow{(g,e)} \partial(e)g\in G\ltimes_{\rhd} E $, we have endofunctor 
   \begin{eqnarray}
  	 \Phi(g)\xRightarrow{\Phi(g,e)}\Phi(\partial(e)g)        &:& L^2 \longrightarrow L^2\cr
  	&& (\gamma\xRightarrow{\alpha}\gamma')\mapsto \Phi_\gamma(g)\xRightarrow{\Phi_\alpha(g,e)}\Phi_{\gamma'}(\partial(e)g), 
  \end{eqnarray}
   \end{itemize}
  where $ \Phi_\alpha(g,e)= \Phi(g,e)(\alpha)$   is the parallel transport of the plaquette $\alpha\in L^2$ by $\Phi(g,e)$.
%    such that  $\Phi(*)\in L^0\subset \mathcal{C}(M,L)$ and  
%    \begin{itemize}
%    %	\item for   $\mathcal{G}_0=\{*\}$  then
%    	\item for $g \in G= \mathcal{G}_1$,   $x,y\in L^0\subset \mathcal{C}(M,L) $, and  $ \gamma \in L^1 \subset \mathcal{C}(M,L)$, then    the  1-morphism  representation is given by the map
%    	\begin{eqnarray}
%    	( x \xrightarrow{\gamma}y ) \xmapsto{\Phi(g)}          x \xrightarrow{\Phi_\gamma(g)} y, 
%    	\end{eqnarray}
%  where  $ \Phi_\gamma(g)= \Phi(g)\gamma$ represents the action of $\Phi(g)$ on $\gamma$.
%    	\item for each 2-morphisms  $u=(g,e)\in G\ltimes_{\rhd} E$, $\gamma_1,\gamma_2 \in L^1  $ and  $ \alpha \in L^2 \subset \mathcal{C}(M,L)$, then the  2-morphisms  representation is given by the map
%    		\begin{eqnarray}
%    		 \gamma_1 \xrightarrow{\alpha}\gamma_2  \xrightarrow{\Phi(u)}  \Phi_{\gamma_1}(g_1) \xrightarrow{\Phi_\alpha(u)} \Phi_{\gamma_2}(g_2), 
%    	\end{eqnarray}
%    where $	g_2=\partial(e)g_1$ and $\Phi_\alpha(u) = \Phi(u)\gamma $ is the  action of $ \Phi_\alpha(u)$ on $\alpha$. These representations can be illustrated as follows   	
   % \end{itemize}
These representations  can be illustrated as follows   
	
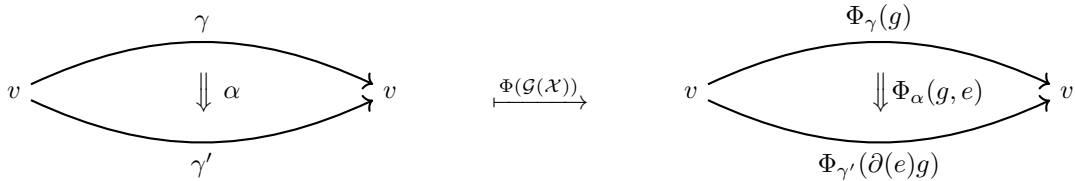
\begin{eqnarray*}
	\begin{tikzpicture}[thick]
		\node (A) at (-2.5,0) {$v$};
		\node (B) at (2.5,0) {$v$};
		\path[->] (A) edge [bend left=25,""{name=F,}] node[above] {$\gamma$} (B);
		\path[->] (A) edge [bend right=25, ""{name=D,  }] node[below] {$\gamma'$} (B);
		
		\node at (0,0) {\rotatebox{270}{$\implies$}} node at (0.4,0) {$\alpha$};

		\node (F)  at (4.5,0){$\xmapsto{\Phi(\mathcal{G}(\mathcal{X}))}  $};
		\node (G)  at (6.5,0){$v$};
		\node (H)  at (11.5,0){$v$};
		\node at (9,0) {\rotatebox{270}{$\implies$}} node  at (9.78,0) {$\Phi_\alpha(g,e)$};
		\path[->] (G) edge [bend left=25] node[above] {$\Phi_{\gamma}(g)$} (H);
		\path[->] (G) edge [bend right=25] node[below] {$\Phi_{\gamma'}(\partial(e)g)$} (H);
	\end{tikzpicture}	      
\end{eqnarray*}
}
\end{definition}

%To understand how field configurations look like in practice, consider the example of X
%taken to be the pentagon presented on figure
%%From  this definition, a 2-gauge configuration defines a triple of set maps 

\subsubsection{Holonomies}
In the lattice model represented by  crossed modules, there are  holonomies  generated by  1-gauge fields along the edges called 1-holonomies and the  2-holonomies along surfaces  generated by 2-gauge fields. Akin to 1-holonomy, which describes parallel transport of
point particles around the surface, 2-homolomies describe the parallel transport of paths over a surface. Similarly to the construction in \cite{33,40,43}  the 1-holonomies $ Hol_{v_{\alpha}}^1(M,L,\mathcal{G}(\mathcal{X}))$ and 2-holonomies $Hol_{v_{\alpha}}^2(M,L,\mathcal{G}(\mathcal{X}))$ are encoded by the representations $\Phi$ of $\mathcal{G}(\mathcal{\mathcal{X}})$ in the lattice model $\mathcal{P}_2(M,L)$ from the base point $v_\alpha$ to itself and must satisfy  certain flatness constraints.
\begin{definition}
{\it
Let $\alpha_i\in L^2$   be  plaquettes of the lattice $(M,L)$,  we choose   the base  points  $v_{\alpha_i}  \in bd(\alpha_i)$, to be  $0$-cells. Let $p^-(v_{\alpha_i}\rightarrow v_{\alpha_i})\in L^1$ be the oriented paths  anti-aligned with the orientations of  plaquettes $\alpha_i$ from the base-point $v_{\alpha_i}$ to itself.   Furthermore, we  fix     orientations of surfaces $\alpha_i$ inducing a cyclic  order of edges (paths) around each surface. Let $\theta_i=\pm 1$  be integers assigned to the orientations  of surfaces. $\theta_i=1$ corresponds to an anticlockwise orientation, while  the clockwise orientation of the plaquette  corresponds to $\theta_i=-1$.         The 1-holonomy $   Hol_{v_{\alpha}}^1(M,L,\mathcal{G}(\mathcal{X}))$  and  the 2-holonomy $   Hol_{v_{\alpha}}^2(M,L,\mathcal{G}(\mathcal{X}))$   are defined  as follows:
\begin{itemize}
\item  1-holonomy  also called the 1-fake-curvature \cite{33,40}  is associated with the direct
path representations   around the plaquette representations
\begin{eqnarray}
    Hol_{v_{\alpha_i}}^1=\partial \Phi_{\alpha_i} (u_i)p^-(v_{\alpha_i}\rightarrow v_{\alpha_i})= \Phi_{\partial \alpha_i} (g_i,\partial e_i)p^-(v_{\alpha_i}\rightarrow v_{\alpha_i})
\end{eqnarray}
where $\partial\alpha\in L^1$ is the oriented boundary of the plaquette $\alpha$ connecting  $v_\alpha$ to  $v_\alpha$. It is given by $\partial\alpha=1_v$.  
 Note that,  the boundary path around the palquette   oriented  from $v_\alpha$  to  
  $v_\alpha$   defined 1-fake flatness constraint of the face  given by
\begin{eqnarray}
\partial Hol_{v_{\alpha}}^1= \Phi_{\partial \alpha}(1_G).
\end{eqnarray}

    \item 2-holonomy, also called 2-fake flat configuration \cite{33,40}, is associated with   the  plaquette representations.  It is given by  
\begin{eqnarray}
    Hol_{v_{\alpha_i}}^2= \prod_{i}\Phi_{\gamma_i}(g_i)\rhd \left(\Phi_{\alpha_i}(u_i)\right)^{\theta_i},
\end{eqnarray}
The 2-fake-flatness constraint over the surface is given by  
\begin{eqnarray}
\partial  Hol_{v_{\alpha_i}}^2= \Phi_{\partial b}(1_G,1_E).
\end{eqnarray}

\end{itemize}}

\end{definition}

\begin{figure}
 \begin{eqnarray*}
  \begin{tikzpicture}[thick]
  \node (A0) at (-3,0) {$v$};
  \node (A) at (-1.5,0) {$v_\alpha$};
  \node (B) at (3.5,0) {$v_1$};
  % \node (C)  at (4.5,0){$*$};
  \node (E)  at (3,0){$ $};
  \node at (0.9,0) {\rotatebox{270}{$\implies$}}node[right=0.99cm]{$\Phi_\alpha(g,e)$};
  \draw[ thick, ->] (2.0,0) arc (-2:100:0.5);
  \path[->] (A) edge [bend left=28] node[above] {$\Phi_{\gamma_2}(g_2)$} (B);
  \path[->] (A) edge [bend right=28] node[below] {$\Phi_{\gamma_3}(g_3)$} (B);
  \draw[->] (A0) --node[above=0.01cm] {$\Phi_{\gamma_1}(g_1)$} (A) ;
 
  \end{tikzpicture}   
 \end{eqnarray*}
  \caption{ Here we consider a  left whiskering representation i.e., a  left composition of  a 1-morphism with  a 2-morphism}\label{en47}
 \end{figure}
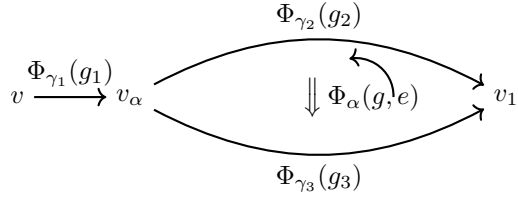

 To illustrate the latter definitions, we consider two examples of graphs. For  the first one (Fig \eqref {en47}), we combine an edge $\gamma$ that runs from the vertex $v$ to a base point $v_\alpha$ of a plaquette $\alpha$ that is anticlockwise oriented from its base point.
  The 1-holonomy $ Hol_{v_{\alpha}}^1$  and the  2-holonomy $ Hol_{v_{\alpha}}^2$  are  expressed as follows
  \begin{eqnarray}  
  Hol_{v_{\alpha}}^1&=&\partial \Phi_{\alpha}(u) \Phi_{\gamma_2}(g_2)\Phi_{\gamma_3^{-1}}(g_3^{-1}),\\
  Hol_{v_{\alpha}}^2&=&\Phi_{\gamma_1}(g_1)\rhd \Phi_{\alpha}(u).
 \end{eqnarray}
 The corresponding 1-fake-flatness and 2-fake-flatness  constraints around and over the surface are  given by
 \begin{eqnarray}
 \partial \Phi_{\alpha}(u)&=&\Phi_{\gamma_2}(g_2)\Phi_{\gamma_3^{-1}}(g_3^{-1}), \, p^-(v_{\alpha}\rightarrow v_{\alpha}) =\Phi_{\gamma_3}(g_3) \Phi_{\gamma_2^{-1}}(g_2^{-1})\implies  \partial Hol_{v_{\alpha}}^1=
 \Phi_{1_v}(1_G),\\
 \partial Hol_{\partial b }^2&=& \Phi_{\alpha}(1_G,1_E).
 \end{eqnarray}
 In the second example Fig \eqref{en46}, we consider  
 the  standard tetrahedron $(v_0v_1v_2v_3) $ displayed below. It  decomposed into four triangulated sublattices: $(v_0v_1v_2)$, $(v_0v_2v_3)$, $(v_1v_2v_3)$ and $(v_0v_1v_3)$. The granular boundary of each of the labeled plaquettes is indicated in the figure below.  Note that the  plaquettes $\Phi_{\alpha_{1}}(g_{023},e_1)$,   $\Phi_{\alpha_{2}}(g_{021},e_2)$ and  $\Phi_{\alpha_{4}}(g_{013},e_4)$ have a common base point based at $v_0=v_\alpha$ while the plaquette $\Phi_{\alpha_{3}}(g_{231},e_3),$  is based  at the vertex $v_2$. For the  sake of  simplicity, we consider the  base points of the plaquettes and the blobs of each surface at the same  fix as points.
 \begin{figure}
     \centering
\begin{eqnarray*}
\begin{tikzpicture}[thick]
\node (A) at (-4.5,0) {$v_0$};
\node (B) at (4.5,0) {$v_3$};
\node (C) at (0,4.16)  {$v_1$};
\node (D) at (0,1.3)  {$v_2$};

\draw[->](A)--node [left=0.15cm] {$\Phi_{\gamma_{0 1}}(g_{01}) $}(C);
\draw[->](A)--node [below=0.0015cm] {$\Phi_{\gamma_{03}}(g_{03})$}(B);
\draw[->](C)--node [right=-0.01cm]  {$\Phi_{\gamma_{12}}(g_{12})$} (D);
\draw[->](C)--node [right=0.15cm] {$\Phi_{\gamma_{1 3}(g_{13})}$}(B);
\draw[->](A)--node [left=0.09cm,above] {$\Phi_{\gamma_{02}}(g_{02})$}(D);
            \draw[ thick, ->] (-0.8,0.3) arc (-2:100:0.5);  
\draw[->](D)--node [right=0.17cm,above] {$\Phi_{\gamma_{23}(g_{23})}$}(B);

\node at (0,0.5) {\rotatebox{270}{$\implies$}};
\node at (1.5,2.0) {\rotatebox{45}{$\implies$}} node  at (1.3,1.7){$\Phi_{\alpha_{3}}(g_{231},e_3)$};
        \draw[ thick, ->] (3.,1.0) arc (-2:100:0.5);
\node at (-1.0,2.5) {\rotatebox{124}{$\implies$}}node  at (-1.,2.09){$\Phi_{\alpha_{2}}(g_{021},e_2)$};
     \draw[ thick, ->] (-0.3,2.6) arc (-2:100:0.5);
   
\node  at (1.2,0.42){$\Phi_{\alpha_{1}}(g_{023},e_1)$};

\end{tikzpicture}      
\end{eqnarray*}

\begin{eqnarray*}
\begin{tikzpicture}[thick]
\node (A) at (-4.5,0) {$v_0$};
\node (B) at (4.5,0) {$v_3$};
\node (C) at (0,4.16)  {$v_1$};
% \node (D) at (0,1.3)  {$v_2$};  

\draw[->](A)--node [left=0.15cm] {$\Phi_{\gamma_{0 1}}(g_{01}) $}(C);
\draw[->](A)--node [below=0.0015cm] {$\Phi_{\gamma_{03}}(g_{03})$}(B);
%\draw[->](C)--node [right=-0.01cm]  {$\Phi_{\gamma_{12}}(g_{12})$} (D);
\draw[->](C)--node [right=0.15cm] {$\Phi_{\gamma_{1 3}(g_{13})}$}(B);
%\draw[->](A)--node [left=0.09cm,above] {$\Phi_{\gamma_{02}}(g_{02})$}(D);
%\draw[ thick, <-] (-0.8,0.3) arc (-2:100:0.5);  
% \draw[->](D)--node [right=0.17cm,above] {$\Phi_{\gamma_{23}(g_{23})}$}(B);

%\node at (0,0.5) {\rotatebox{270}{$\implies$}};
\node at (0.1,1.21) {\rotatebox{-90}{$\implies$}} node  at (0.,1.7){$\Phi_{\alpha_{4}}(g_{013},e_4)$};
\draw[ thick, ->] (0.9,2.0) arc (-2:100:0.5);
% \node at (-1.0,2.5) {\rotatebox{124}{$\implies$}}node  at (-1.,2.09){$\Phi_{\alpha_{2}}(g_{021},e_2)$};
%\draw[ thick, ->] (-0.3,2.6) arc (-2:100:0.5);

%\node  at (1.2,0.42){$\Phi_{\alpha_{1}}(g_{023},e_1)$};

\end{tikzpicture}      
\end{eqnarray*}
     \caption{A configuration a  standard tetrahedron $(v_0v_1v_2v_3).$  }
     \label{en46}
 \end{figure}

Furthermore,  the plaquettes of all the  sublattices $(v_0v_1v_2)$, $(v_1v_2v_3)$, $(v_0v_1v_3),$ and  $(v_1v_2v_3)$ are anticlockwise oriented from their base points. The  2-targets  of each  2-morphism are defined as follows:
\begin{eqnarray}
\Phi_{\gamma_{02}} (g_{02})\Phi_{\gamma_{23}} (g_{23}) &\xRightarrow{\Phi_{\alpha_{1}}(g_{023},e_1)}& \Phi_{\gamma_{03}} (g_{03}),\\
\Phi_{\gamma_{02}} (g_{02})\Phi_{\gamma_{12}^{-1}} (g_{12}^{-1}) &\xRightarrow{\Phi_{\alpha_{2}}(g_{021},e_2)}& \Phi_{\gamma_{01}} (g_{01}),\\
\Phi_{\gamma_{12}} (g_{12})\Phi_{\gamma_{23}} (g_{23}) &\xRightarrow{\Phi_{\alpha_{3}}(g_{123},e_3)}& \Phi_{\gamma_{13}} (g_{13}),\\
\Phi_{\gamma_{01}} (g_{01})\Phi_{\gamma_{13}} (g_{13}) &\xRightarrow{\Phi_{\alpha_{3}}(g_{013},e_4)}& \Phi_{\gamma_{03}} (g_{03}).
\end{eqnarray}
It follows that the 1-holonomy and the 2-holonomy   of the tetrahedron are the composition of 1-holonomy  and  2-holonomy  of each sublattice
\begin{eqnarray}
Hol_{v_{0,2}}^1&=& \partial\Phi_{\alpha_1}(g_{023},e_1)\Phi_{\gamma_{02}} (g_{02})\Phi_{\gamma_{23}} (g_{23})\Phi_{\gamma_{03}^{-1}} (g_{03}^{-1})\cr&&
\partial\Phi_{\alpha_2}(g_{021},e_2)\Phi_{\gamma_{01}}(g_{01}) \Phi_{\gamma_{12}}(g_{12}) \Phi_{\gamma_{02}^{-1}}(g_{02}^{-1}) \cr&&
\partial\Phi_{\alpha_3}(g_{123},e_3) \Phi_{\gamma_{12}^{-1}} (g_{12}^{-1})\Phi_{\gamma_{13}}(g_{13}) \Phi_{\gamma_{23}^{-1}}(g_{23}^{-1})\cr&&
\partial\Phi_{\alpha_1}(g_{013},e_4)\Phi_{\gamma_{01}}(g_{01})\Phi_{\gamma_{13}}(g_{13})\Phi_{\gamma_{03}^{-1}}(g_{03}^{-1})
%&=&\partial \left(\Phi_{\alpha_1}(g_{023},e_1)\Phi_{\alpha_2}(g_{021},e_2)\Phi_{\alpha_3}(g_{123},e_3)\Phi_{\alpha_1}(g_{013},e_4)\right)\cr&&
%\Phi_{\gamma_{02}} (g_{02})\Phi_{\gamma_{23}} (g_{23})\Phi_{\gamma_{03}^{-1}} (g_{03}^{-1})\Phi_{\gamma_{01}}(g_{01}) \Phi_{\gamma_{12}}(g_{12}) \Phi_{\gamma_{02}^{-1}}(g_{02}^{-1})\Phi_{\gamma_{13}} (g_{13})\Phi_{\gamma_{23}^{-1}}(g_{23}^{-1}) \Phi_{\gamma_{12}^{-1}}(g_{12}^{-1}) \cr&&
%\Phi_{\gamma_{01}}(g_{01})\Phi_{\gamma_{13}}(g_{13})\Phi_{\gamma_{03}^{-1}}(g_{03}^{-1})
\\
Hol_{v_{0,2}}^2&=&   \Phi_{\alpha_2}(g_{021},e_2)\Phi_{\alpha_1}(g_{023},e_1)
\left(\Phi_{\gamma_{02}} (g_{02})\rhd\Phi_{\alpha_3}(g_{231},e_3)\right)\Phi_{\alpha_4}(g_{013},e_4)
\end{eqnarray}
The corresponding 1-fake-flatness and 2-fake-flatness  constraints around and over each surfaces are  given by
\begin{eqnarray}
\partial Hol_{v_{0,2}}^1&=& \partial\Phi_{\alpha_1}(g_{023},e_1)p_1^-(v_0\rightarrow v_0)\partial\Phi_{\alpha_2}(g_{021},e_2)p_2^-(v_0\rightarrow v_0)\cr&&\times\partial\Phi_{\alpha_3}(g_{123},e_3) p^-(v_2\rightarrow v_2)\partial\Phi_{\alpha_1}(g_{013},e_4)p_3^-(v_0\rightarrow v_0)\cr
&=& \Phi_{1_v}(1_G),\\
\partial Hol_{v_{0,2}}^2&=&\Phi_{\partial b_i}(1_G,1_E),
\end{eqnarray}
where the boundaries of each plaquette $\partial\Phi_{\alpha_i}(g_i,e_i)$ and their corresponding anti-oriented paths  $p_i^-(v_{\alpha_i}\rightarrow v_{\alpha_i})$ are given by
\begin{eqnarray}
\partial\Phi_{\alpha_1}(g_{023},e_1)&=& \Phi_{\gamma_{03}}(g_{03})\Phi_{\gamma_{23}^{-1}}(g_{23}^{-1})\Phi_{\gamma_{02}} (g_{02}^{-1}), \quad p_1^-(v_0\rightarrow v_0)=\Phi_{\gamma_{02}} (g_{02})\Phi_{\gamma_{23}} (g_{23})\cr&& \Phi_{\gamma_{03}^{-1}} (g_{03}^{-1}),\\
\partial\Phi_{\alpha_2}(g_{021},e_2)&=&\Phi_{\gamma_{02}}(g_{02})\Phi_{\gamma_{12}^{-1}}(g_{12}^{-1})\Phi_{\gamma_{01}}(g_{01})\quad   p_2^-(v_0\rightarrow v_0)= \Phi_{\gamma_{01}^{-1}}(g_{01}^{-1}) \Phi_{\gamma_{12}}(g_{12}) \cr&& \Phi_{\gamma_{02}^{-1}}(g_{02}^{-1}),\\
\partial\Phi_{\alpha_3}(g_{123},e_3)&=&\Phi_{\gamma_{23}} (g_{13})\Phi_{\gamma_{13}}(g_{12}) \Phi_{\gamma_{12}^{-1}}(g_{12}^{-1}),            \quad  p^-(v_2\rightarrow v_2)=   \Phi_{\gamma_{12}^{-1}} (g_{12}^{-1})\Phi_{\gamma_{13}}(g_{13})\cr&& \Phi_{\gamma_{23}^{-1}}(g_{23}^{-1}),\\
\partial\Phi_{\alpha_1}(g_{013},e_4)&=& \Phi_{\gamma_{13}}(g_{13})\Phi_{\gamma_{03}}(g_{03})\Phi_{\gamma_{13}^{-1}}(g_{13}^{-1}) \Phi_{\gamma_{01}^{-1}}(g_{01}^{-1}) \quad p_3^-(v_0\rightarrow v_0)=\Phi_{\gamma_{01}}(g_{01})\Phi_{\gamma_{13}}(g_{13})\cr&&\Phi_{\gamma_{03}^{-1}}(g_{03}^{-1}).
\end{eqnarray}

\subsection{Gauge transformation} \label{sec4}
Having described how the gauge field fits into the lattice picture by means of the action of 2-groups $\Phi(\mathcal{G}(\mathcal{X}))$ on the lattice path 2-groupoids $\mathcal{P}_2(M,L)$, we now look at the local
gauge transformations in analogy  with  pseudo-natural transformations. We start with the Hilbert  space  $  \mathcal{H}(M,L,\mathcal{G}(\mathcal{X})),$ which defines the  state space representations. We then define  a set of operators acting on  $\mathcal{H}(M,L,\mathcal{G}(\mathcal{X}))$, which we call higher gauge operators. We deduce from these actions  the commutation relations between  operators.

\subsubsection{Hilbert  space}

Associated with the representation category of the 2-group and the lattice model is a Hilbert Space  $\mathcal{H}(M,L,\mathcal{G}(\mathcal{X}))$  spanned by all possible configurations of the
 elements  $g\in G$ on the edges $\gamma_i$ and  the  elements $ u\in G\ltimes_{\rhd} E$ on the plaquettes $\alpha_i$. It is  defined  as the tensor product of the local representation spaces assigned to the  edges and the plaquettes, respectively
 \begin{equation}\label{HI}
 \mathcal{H}(M,L,\mathcal{G}(\mathcal{X})):= \mbox{span}\left \{\left|\bigotimes_{\gamma\in L^1} \Phi_{\gamma}(g) \bigotimes_{\alpha\in L^2} \Phi_{\alpha}(u)  \right \rangle\right\}.
 \end{equation} 
We equip $\mathcal{H}(M,L,\mathcal{G}(\mathcal{X}))$ with a complete Hilbert space structure by defining
a positive definite Hermitian inner product
\begin{eqnarray}
	\langle .|.\rangle:  \mathcal{H}(M,L,\mathcal{G}(\mathcal{X}))\otimes  \mathcal{H}(M,L,\mathcal{G}(\mathcal{X})) \rightarrow \mathbb{C},
\end{eqnarray}
in which the states $ |\Phi_{\gamma}(g)\Phi_{\alpha}(u) \rangle    $ form an orthonormal basis:
\begin{eqnarray}
\langle \Phi_{\gamma}(g)\Phi_{\alpha}(u)|\Phi_{\gamma}(g')\Phi_{\alpha}(u') \rangle=\delta_{gg'}\delta_{uu'}.	
\end{eqnarray}
\subsubsection{Gauge operators}
   Now that, we have considered the   gauge fields, we can describe gauge transforms. There are two   types of gauge transforms: those associated with the more familiar 1-gauge field and those associated with
the 2-gauge field. The 1-gauge transformation $\hat A_v^g \,(g\in G)$ associated with the  vertex $v$ and the 2-gauge transformation $ \hat A_\gamma^e\, (e\in E)  $ associated with the  edge are linear operators  defined on  the Hilbert  space   
\begin{eqnarray}
\hat A_v^g,\hat A_\gamma^e :\mathcal{H}(M,L,\mathcal{G}(\mathcal{X}))\rightarrow \mathcal{H}(M,L,\mathcal{G}(\mathcal{X})).
\end{eqnarray}  
Let us now give an explicit description of  how vertex and edge operators act on $ \mathcal{H}(M,L,\mathcal{G}(\mathcal{X}))$.\\\\
\begin{itemize}
\item {\bf  Vertex operators}

 Let $v\in L^0$ be  a vertex  and $\alpha\in L^2$  be an adjacent plaquette. Such a pair $(v, \alpha)$ is also called a site \cite{40,41,42}. Note that $v$ is
not necessarily the base-point  $v_\alpha$ of the plaquette.
We define now a family of vertex operators, $ \hat A_v^h$ with $ h \in G$ acting on $ \mathcal{H}(M,L,\mathcal{G}(\mathcal{X}))$, whose support is the
set of the edges and plaquettes incident to the vertex $v$. Now, consider the set of edges incident to  the vertex $v$.   This action depends on the orientation of the edge,
inwards or outwards $v$. On the other  hand, this  action on the  surfaces  whose base-point is
$v$  gives a $h\rhd$
action on the labelled plaquette,  and  surfaces not based at that vertex are left unaffected.  \\ 

\begin{definition}
{\it For any $h\in G$, the vertex operator $ \hat A_v^h$ with $ h \in G$ acting on $ \mathcal{H}(M,L,\mathcal{G}(\mathcal{X}))$ based at
the site $(v, \alpha)$, is defined as \cite{40,41,42} 

\begin{eqnarray}
\hat A_v^h \Phi_\gamma(g)&=&  \begin{cases} h \Phi_\gamma(g)  & \text{if $v$ is the start of $\gamma$,} \\
	\Phi_\gamma(g)h^{-1} & \text{if $v$ is the end of $\gamma$,}\\
	\Phi_\gamma(g) & \text{otherwise.}
 \end{cases}\label{t45}\\
 \hat A_v^h \Phi_\alpha(u)&=&  \begin{cases} h\rhd \Phi_\alpha(u)  & \text{if\quad $v=v_\alpha$,} \\ 
	\Phi_\gamma(g) & \text{if \quad $v\neq v_\alpha$}.
 \end{cases}\label{t46}
 \end{eqnarray}
 }
 \end{definition}
To illustrate equations \eqref{t45} and \eqref{t46} of  definition $8$, we consider  the dressing graph Fig \eqref{en200} with the plaquette base point at $v_\alpha$ and  anticlockwise oriented. 
\begin{figure}[H]
\centering
	\begin{tikzpicture}[thick]
		\node (A) at (-1.5,0) {$v_\alpha$};
		\node (B) at (1.5,0) {$v_2$};
		\node (C) at (4.5,0) [below=of A] {$v_1$}; 
		\node (D) at (4.5,0)   [below=of B] {$v_3$};	
		
		\draw[->](A)--node [left=0.1cm] {$\Phi_{\gamma_{\alpha 1}}(g_{\alpha1})$}(C);
		\draw[->](B)--node [right=0.1cm] {$\Phi_{\gamma_{23}}(g_{23})$}(D);
		\draw[->](A)--node [above=0.1cm] {$\Phi_{\gamma_{\alpha 2}}(g_{\alpha2})$}(B);
		\draw[->](C)--node [below=0.1cm] {$\Phi_{\gamma_{1 3}}(g_{13})$}(D); 
        \draw[ thick, ->] (0.9,-0.75) arc (-2:70:0.5);
		\node at (-0.1,-0.75){$\Phi_\alpha(u)$};	
	\end{tikzpicture}
     \caption{ A configuration graph with $v_\alpha$ its base point and the ordering vertices $v_\alpha< v_1<v_2<v_3$. 
}\label{en200}
\end{figure}
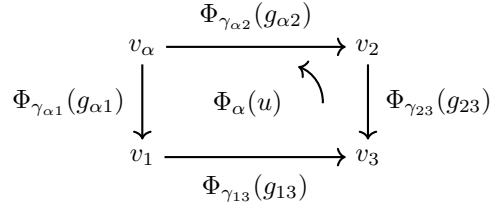
We have for  instance,
	
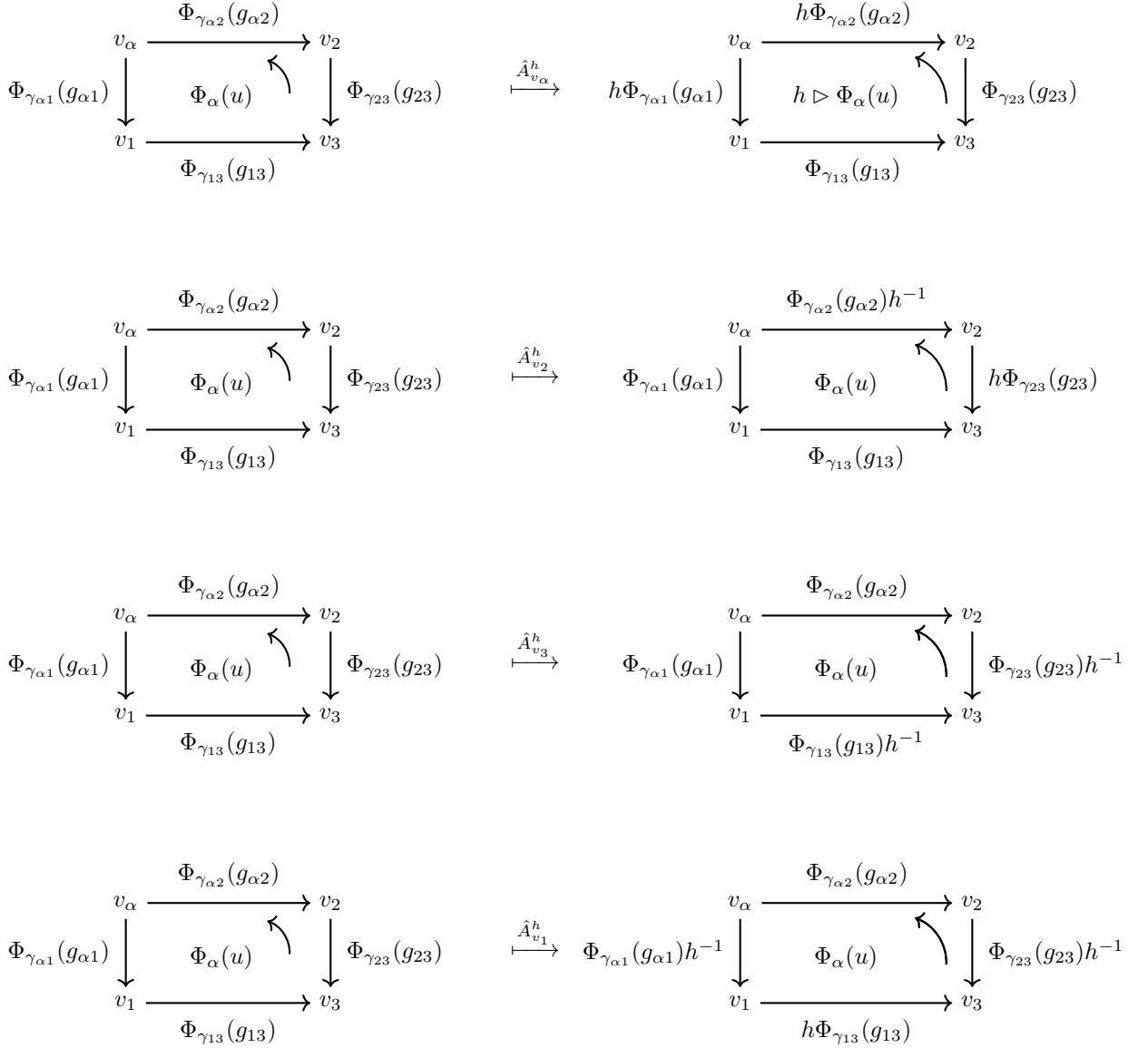
\begin{eqnarray*}
	\begin{tikzpicture}[thick]
	\node (A) at (-1.5,0) {$v_\alpha$};
	\node (B) at (1.5,0) {$v_2$};
	\node (C) at (2.5,0) [below=of A] {$v_1$}; 
	\node (D) at (2.5,0)   [below=of B] {$v_3$};	
		
\draw[->](A)--node [left=0.1cm] {$\Phi_{\gamma_{\alpha 1}}(g_{\alpha1})$}(C);
\draw[->](B)--node [right=0.1cm] {$\Phi_{\gamma_{23}}(g_{23})$}(D);
\draw[->](A)--node [above=0.1cm] {$\Phi_{\gamma_{\alpha 2}}(g_{\alpha2})$}(B);
\draw[->](C)--node [below=0.1cm] {$\Phi_{\gamma_{1 3}}(g_{13})$}(D);
 \draw[ thick, ->] (0.9,-0.75) arc (-2:70:0.5);

%\node at (0.5,-0.5) {\rotatebox{45}{$\Longrightarrow$}}
\node  at (-0.1,-0.8) {$\Phi_\alpha(u)$};
%\node at (-0.32,-1.2) {\rotatebox{45}{$=$}};	

\node (F)  at (4.5,-0.5){$\xmapsto{\hat A_{v_\alpha}^h }  $};

\node (G)  at (7.5,0){$v_\alpha$};
\node (H)  at (10.8,0){$v_2$};
\node (I)  at (6.4,1) [below=of G] {$v_1$};
\node (J)  at (10.5,1)[below=of H] {$v_3$};

\draw[->](G)-- node [above=0.1cm] {$h\Phi_{\gamma_{\alpha 2}}(g_{\alpha2})$}                    (H);

\draw[->](G)--  node [left=0.1cm] {$h\Phi_{\gamma_{\alpha 1}}(g_{\alpha1})$}          (I);
\draw[->](H)--  node [right=0.1cm] {$\Phi_{\gamma_{23}}(g_{23})$}                  (J);
\draw[->](I)-- node [below=0.1cm] {$\Phi_{\gamma_{1 3}}(g_{13})$}                 (J);

%\node at (9.76,-0.5) {\rotatebox{45}{$\Longrightarrow$}};
\node  at (9.05,-0.8) {$h\rhd\Phi_\alpha(u)$};
%\node at (9.0,-1.2) {\rotatebox{45}{$=$}};	
\draw[ thick, ->] (10.52,-0.9) arc (-2:70:0.7);

	\end{tikzpicture}	      
\end{eqnarray*}

	\begin{eqnarray*}
	\begin{tikzpicture}[ thick]
		\node (A) at (-1.5,0) {$v_\alpha$};
		\node (B) at (1.5,0) {$v_2$};
		\node (C) at (2.5,0) [below=of A] {$v_1$}; 
		\node (D) at (2.5,0)   [below=of B] {$v_3$};	
		
		\draw[->](A)--node [left=0.1cm] {$\Phi_{\gamma_{\alpha 1}}(g_{\alpha1})$}(C);
		\draw[->](B)--node [right=0.1cm] {$\Phi_{\gamma_{23}}(g_{23})$}(D);
		\draw[->](A)--node [above=0.1cm] {$\Phi_{\gamma_{\alpha 2}}(g_{\alpha2})$}(B);
		\draw[->](C)--node [below=0.1cm] {$\Phi_{\gamma_{1 3}}(g_{13})$}(D);

		%\node at (0.5,-0.5) {\rotatebox{45}{$\Longrightarrow$}}
		\node  at (-0.1,-0.8) {$\Phi_\alpha(u)$};
		%\node at (-0.32,-1.2) {\rotatebox{45}{$=$}};	
		 \draw[ thick, ->] (0.9,-0.75) arc (-2:70:0.5);

		\node (F)  at (4.5,-0.5){$\xmapsto{\hat A_{v_2}^h }  $};
		
		\node (G)  at (7.5,0){$v_\alpha$};
		\node (H)  at (10.9,0){$v_2$};
		\node (I)  at (6.4,1) [below=of G] {$v_1$};
		\node (J)  at (10.5,1)[below=of H] {$v_3$};
		
		\draw[->](G)-- node [above=0.1cm] {$\Phi_{\gamma_{\alpha 2}}(g_{\alpha2})h^{-1}$}                    (H);
		
		\draw[->](G)--  node [left=0.1cm] {$\Phi_{\gamma_{\alpha 1}}(g_{\alpha1})$}          (I);
		\draw[->](H)--  node [right=0.1cm] {$h\Phi_{\gamma_{23}}(g_{23})$}                  (J);
		\draw[->](I)-- node [below=0.1cm] {$\Phi_{\gamma_{1 3}}(g_{13})$}                 (J);

		%\node at (9.76,-0.5) {\rotatebox{45}{$\Longrightarrow$}};
		\node  at (9.05,-0.8) {$\Phi_\alpha(u)$};
		%\node at (9.0,-1.2) {\rotatebox{45}{$=$}};	
		\draw[ thick, ->] (10.52,-0.9) arc (-2:70:0.7);

	\end{tikzpicture}	      
\end{eqnarray*}

\begin{eqnarray*}
	\begin{tikzpicture}[ thick]
		\node (A) at (-1.5,0) {$v_\alpha$};
		\node (B) at (1.5,0) {$v_2$};
		\node (C) at (2.5,0) [below=of A] {$v_1$}; 
		\node (D) at (2.5,0)   [below=of B] {$v_3$};	
		
		\draw[->](A)--node [left=0.1cm] {$\Phi_{\gamma_{\alpha 1}}(g_{\alpha1})$}(C);
		\draw[->](B)--node [right=0.1cm] {$\Phi_{\gamma_{23}}(g_{23})$}(D);
		\draw[->](A)--node [above=0.1cm] {$\Phi_{\gamma_{\alpha 2}}(g_{\alpha2})$}(B);
		\draw[->](C)--node [below=0.1cm] {$\Phi_{\gamma_{1 3}}(g_{13})$}(D);
		\draw[ thick, ->] (0.9,-0.75) arc (-2:70:0.5);
		
		%\node at (0.5,-0.5) {\rotatebox{45}{$\Longrightarrow$}}
		\node  at (-0.1,-0.8) {$\Phi_\alpha(u)$};
		%\node at (-0.32,-1.2) {\rotatebox{45}{$=$}};	
		\draw[ thick, ->] (10.52,-0.9) arc (-2:70:0.7);

		\node (F)  at (4.5,-0.5){$\xmapsto{\hat A_{v_3}^h }  $};
		
		\node (G)  at (7.5,0){$v_\alpha$};
		\node (H)  at (10.9,0){$v_2$};
		\node (I)  at (6.4,1) [below=of G] {$v_1$};
		\node (J)  at (10.5,1)[below=of H] {$v_3$};
		
		\draw[->](G)-- node [above=0.1cm] {$\Phi_{\gamma_{\alpha 2}}(g_{\alpha2})$}                    (H);
		
		\draw[->](G)--  node [left=0.1cm] {$\Phi_{\gamma_{\alpha 1}}(g_{\alpha1})$}          (I);
		\draw[->](H)--  node [right=0.1cm] {$\Phi_{\gamma_{23}}(g_{23})h^{-1}$}                  (J);
		\draw[->](I)-- node [below=0.1cm] {$\Phi_{\gamma_{1 3}}(g_{13})h^{-1}$}                 (J);

		%\node at (9.76,-0.5) {\rotatebox{45}{$\Longrightarrow$}};
		\node  at (9.05,-0.8) {$\Phi_\alpha(u)$};
		%\node at (9.0,-1.2) {\rotatebox{45}{$=$}};	
		\draw[ thick, ->] (10.52,-0.9) arc (-2:70:0.7);

	\end{tikzpicture}	      
\end{eqnarray*}

\begin{eqnarray*}
	\begin{tikzpicture}[ thick]
		\node (A) at (-1.5,0) {$v_\alpha$};
		\node (B) at (1.5,0) {$v_2$};
		\node (C) at (2.5,0) [below=of A] {$v_1$}; 
		\node (D) at (2.5,0)   [below=of B] {$v_3$};	
		
		\draw[->](A)--node [left=0.1cm] {$\Phi_{\gamma_{\alpha 1}}(g_{\alpha1})$}(C);
		\draw[->](B)--node [right=0.1cm] {$\Phi_{\gamma_{23}}(g_{23})$}(D);
		\draw[->](A)--node [above=0.1cm] {$\Phi_{\gamma_{\alpha 2}}(g_{\alpha2})$}(B);
		\draw[->](C)--node [below=0.1cm] {$\Phi_{\gamma_{1 3}}(g_{13})$}(D);
		
		\draw[ thick, ->] (0.9,-0.75) arc (-2:70:0.5);
		%\node at (0.5,-0.5) {\rotatebox{45}{$\Longrightarrow$}}
		\node  at (-0.1,-0.8) {$\Phi_\alpha(u)$};
		%\node at (-0.32,-1.2) {\rotatebox{45}{$=$}};	
		\draw[ thick, ->] (10.52,-0.9) arc (-2:70:0.7);

		\node (F)  at (4.5,-0.5){$\xmapsto{\hat A_{v_1}^h }  $};
		
		\node (G)  at (7.5,0){$v_\alpha$};
		\node (H)  at (10.9,0){$v_2$};
		\node (I)  at (6.4,1) [below=of G] {$v_1$};
		\node (J)  at (10.5,1)[below=of H] {$v_3$};
		
		\draw[->](G)-- node [above=0.1cm] {$\Phi_{\gamma_{\alpha 2}}(g_{\alpha2})$}                    (H);
		
		\draw[->](G)--  node [left=0.1cm] {$\Phi_{\gamma_{\alpha 1}}(g_{\alpha1})h^{-1}$}          (I);
		\draw[->](H)--  node [right=0.1cm] {$\Phi_{\gamma_{23}}(g_{23})h^{-1}$}                  (J);
		\draw[->](I)-- node [below=0.1cm] {$h\Phi_{\gamma_{1 3}}(g_{13})$}                 (J);

		%\node at (9.76,-0.5) {\rotatebox{45}{$\Longrightarrow$}};
		\node  at (9.05,-0.8) {$\Phi_\alpha(u)$};
		%\node at (9.0,-1.2) {\rotatebox{45}{$=$}};	
		\draw[ thick, ->] (10.52,-0.9) arc (-2:70:0.7);

	\end{tikzpicture}	      
\end{eqnarray*}

\begin{lemma}\label{lem1}  
 {\it Let $v, v'\in L^0$ be  vertices  and $\alpha,\alpha'\in L^2$  be adjacent plaquettes; we have  
	\begin{eqnarray}\label{a40}
	\hat A_v^h \hat A_v^{h'}= \hat A_v^{hh'}\quad \mbox{and}\quad	[	\hat A_v^h, \hat A_{v'}^{h'}]= 0, \quad \forall\, h,h'\in G.
	\end{eqnarray}
    }
    \end{lemma}
\begin{proof}
Provided in Appendix B.
\end{proof}
   
%{\bf Lemma 4.6.1:} {\it Let $v,v'\in L^0$ be   two distinct vertices  and $\alpha\in L^2$  be an adjacent plaquette. Then the corresponding vertex operators commute with each other, that is:
%	\begin{eqnarray}
%	[	\hat A_v^h, \hat A_{v'}^{h'}]= 0, \quad \forall\, h,h'\in G
%	\end{eqnarray}}
%\begin{proof}
% Let $v,v'\in L^0$   and $\alpha,\alpha'\in L^2$ and $ \Phi_{\gamma}(g)\otimes \Phi_{\alpha}(u) \in \mathcal{H}(M,L,\mathcal{G})$	and  $  \Phi_{\gamma'}(g)\otimes \Phi_{\alpha'}(u) \in \mathcal{H}(M,L,\mathcal{G})$. Based on equation  (\ref{a}), we have  
% \begin{eqnarray}
%	\hat A_v^h \hat A_{v'}^{h'} \left(\Phi_{\gamma}(g)\otimes \Phi_{\alpha}(u)\otimes \Phi_{\gamma'}(g)\otimes \Phi_{\alpha'}(u) \right)&=& \left(h_{(1)}\Phi_{\gamma}(g)\otimes h_{(2)}\rhd \Phi_{\alpha}(u)\otimes h_{(1)}'\Phi_{\gamma'}(g)\otimes h_{(2)}\rhd\Phi_{\alpha'}(u) \right)\\
%	\hat A_v^h \hat A_{v'}^{h'} \left(\Phi_{\gamma}(g)\otimes \Phi_{\alpha}(u)\otimes \Phi_{\gamma'}(g)\otimes \Phi_{\alpha'}(u) \right)&=& \left(h_{(1)}\Phi_{\gamma}(g)\otimes h_{(2)}\rhd \Phi_{\alpha}(u)\otimes h_{(1)}'\Phi_{\gamma'}(g)\otimes h_{(2)}\rhd\Phi_{\alpha'}(u) \right)
% \end{eqnarray}
% 
%\end{proof}

%%That is, to find the action of a 2-gauge transform
%on a diagram we add a surface and combine the surface
%with the rest of the diagram, as shown in Figure 21. This
%fluctuates the plaquette labels surrounding an edge, as
%well as changing the edge label itself. This is similar
%to how the 1-gauge transform at a vertex fluctuates the
%edges around the vertex (along with any plaquettes based
%at that vertex)

\item {\bf Edge operators}

In addition to these 1-gauge transforms, we also have
2-gauge transforms, which act on an edge and the surface
that adjoins it. The 2-gauge transform denoted by $\hat A_\gamma^e (e\in E) $ on an edge $\gamma$
 labeled by   $\Phi_\gamma(g)$  acts
like parallel transport of the edge along a surface $\alpha$ labeled
by $\Phi_\alpha (u)$.

Moreover, let $\alpha \in L^2$ be the plaquette element, $v_\alpha \in L^0$ be the base-point of $\alpha$ and $\gamma=(\gamma_1,\cdots,\gamma_n)\in L^1$ be oriented  path. Consider the path around one of the plaquettes starting at the base point $v_\alpha$ of the plaquette and
traveling along its boundary, aligned or anti-aligned with its orientation. 
\begin{itemize}
    \item If the edge $\gamma$ is aligned with the orientation of the plaquette $\alpha$,
the  aligned path  up to the base point is denoted by
$p^+(v_\alpha \rightarrow s(\gamma)) $ where $s(\gamma),$ is the source of the edge $\gamma$ and the anti-aligned  path  is denoted by  $ p^-(v_\alpha \rightarrow t(\gamma) $ where $t(\gamma)$ is the target of the edge $\gamma$.

\item However, if the edge $\gamma$ anti-aligned with the orientation of the plaquette $\alpha$, we have $p^-(v_\alpha \rightarrow s(\gamma))$ which is the anti-aligned path starting at this base point of the plaquette  and circulating  against the orientation of
the plaquette,  and  $p^+(v_\alpha \rightarrow t(\gamma))$ is the path that is  aligned with the circulation of the plaquette.

\end{itemize}
For  example, we consider the configuration  graph illustrated by the  figure \eqref{e2}. 
\begin{figure}[H]
\centering
\begin{tikzpicture}[thick]
\node (A) at (-1.5,0) {$v_\alpha$};
\node (B) at (1.5,0) {$v_2$};
\node (C) at (4.5,0) [below=of A] {$v_1$};
\node (D) at (4.5,0)   [below=of B] {$v_3$};

\draw[->](A)--node [left=0.1cm] {$\Phi_{\gamma_{\alpha 1}}(g_{\alpha1})$}(C);
\draw[->](B)--node [right=0.1cm] {$\Phi_{\gamma_{23}}(g_{23})$}(D);
\draw[->](A)--node [above=0.1cm] {$\Phi_{\gamma_{\alpha 2}}(g_{\alpha2})$}(B);
\draw[->](C)--node [below=0.1cm] {$\Phi_{\gamma_{1 3}}(g_{13})$}(D);
        \draw[ thick, ->] (0.9,-0.75) arc (-2:70:0.5);
\node at (-0.1,-0.75){$\Phi_\alpha(u)$};
\end{tikzpicture}
     \caption{ A configuration graph with an anticlockwise orientation of the plaquette
}\label{e2}
\end{figure}
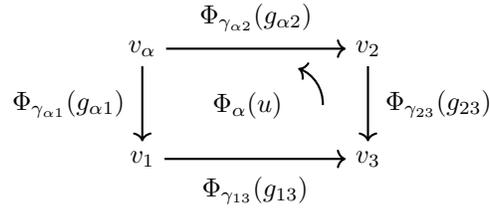
 The granular paths $p^\pm(v_\alpha \rightarrow s(\gamma))$ and $p^\pm(v_\alpha \rightarrow t(\gamma))$  are given by
  \begin{eqnarray}
  p^+(v_\alpha)&=&\varnothing,\quad \quad \quad \quad \quad \quad p^-(v_\alpha)=\varnothing\\
  p^+(v_\alpha\rightarrow s(\gamma_{\alpha1}))&=&\varnothing, \quad \quad p^-(v_\alpha\rightarrow
 t(\gamma_{\alpha1}))=\Phi_{\gamma_{\alpha2}}(g_{\alpha2})\Phi_{\gamma_{23}}(g_{23})\Phi_{\gamma_{13}^{-1}}(g_{13}^{-1}),\\
  p^+( v_\alpha\rightarrow s(\gamma_{13}))&=& \Phi_{\gamma_{\alpha1}}(g_{\alpha1}), \quad  p^-(v_\alpha\rightarrow t(\gamma_{13}))= \Phi_{\gamma_{\alpha2}}(g_{\alpha2})\Phi_{\gamma_{23}}(g_{23}),\\
  p^+(v_\alpha\rightarrow t(\gamma_{\alpha2}))&=&\Phi_{\gamma_{\alpha1}}(g_{\alpha1}) \Phi_{\gamma_{13}}(g_{13})\Phi_{\gamma_{23}^{-1}}(g_{23}^{-1}),\quad  p^-(v_\alpha\rightarrow
 s(\gamma_{\alpha2}))=\varnothing,\\
 p^+(v_\alpha\rightarrow t(\gamma_{23}))&=&\Phi_{\gamma_{\alpha1}}(g_{\alpha1}) \Phi_{\gamma_{13}}(g_{13}),\quad  \quad p^-(v_\alpha\rightarrow
 s(\gamma_{\alpha2}))=\Phi_{\gamma_{\alpha2}}(g_{\alpha2})   ,
     \end{eqnarray}

\begin{definition}
 {\it The action of the  operator $ A_{\gamma'}^e$ on the Hilbert $ \mathcal{H}(M,L,\mathcal{G}(\mathcal{X}))$  is defined as follows \cite{41,42}:
\begin{itemize}
\item Let $\gamma,\gamma'\in L^{1}$, for any $e\in  E $, the edge operator $ \hat A_{\gamma'}^e$  acting on $ \mathcal{H}(M,L,\mathcal{G}(\mathcal{X}))$ based on the edge
$ \gamma$, is given by
     \begin{eqnarray}\label{tae}
\hat A_{\gamma'}^{e} \Phi_\gamma(g)=  \begin{cases} \partial(e) \Phi_\gamma(g)  & \text{if $\gamma=\gamma'$,} \\
\Phi_\gamma(g) & \text{otherwise}.
\end{cases}
      \end{eqnarray}
\item Let $\gamma\in L^{1}$, for any $e\in  E$. We let ${p^+(v_\alpha\rightarrow s(\gamma))} \in G $ and ${p^-(v_\alpha\rightarrow s(\gamma))} \in G $  be the  paths that are aligned and anti-aligned respectively, with the orientation of the plaquette $\alpha$ from the base-point $v_\alpha$ to the edge. The edge operator $ \hat A_{\gamma}^e$  acting on $ \mathcal{H}(M,L,\mathcal{G}(\mathcal{X}))$ based on the plaquette
$ \alpha$, is given by
\begin{eqnarray}\label{ta}
    \hat A_{\gamma}^{e} \Phi_\alpha(u)=  \begin{cases}  \Phi_\alpha(u)\left(p^+(v_\alpha\rightarrow s(\gamma))\rhd e^{-1} \right)  & \text{ if $\gamma$ is on $\alpha$ and aligned with $\alpha$}, \\
\left({p^-(v_\alpha\rightarrow s(\gamma)}\rhd e\right) \Phi_\alpha(u)  & \text{if $\gamma$ is on $\alpha$ and aligned against $\alpha$},\\
\Phi_\alpha(u) & \text{otherwise}.\\
\end{cases}
\end{eqnarray}
\end{itemize}
}
\end{definition}
To illustrate equations \eqref{tae} and \eqref{ta} of  definition $9$, we consider  the dressing graph represented by  figure\eqref{e2}. We have, for  instance.
\begin{eqnarray*}
\begin{tikzpicture}[thick]
\node (A) at (-1.5,0) {$v_\alpha$};
\node (B) at (1.5,0) {$v_2$};
\node (C) at (4.5,0) [below=of A] {$v_1$};
\node (D) at (4.5,0)   [below=of B] {$v_3$};

\draw[->](A)--node [left=0.1cm] {$\Phi_{\gamma_{\alpha 1}}(g_{\alpha1})$}(C);
\draw[->](B)--node [right=0.1cm] {$\Phi_{\gamma_{23}}(g_{23})$}(D);
\draw[->](A)--node [above=0.1cm] {$\Phi_{\gamma_{\alpha 2}}(g_{\alpha2})$}(B);
\draw[->](C)--node [below=0.1cm] {$\Phi_{\gamma_{1 3}}(g_{13})$}(D);
        \draw[ thick, ->] (0.9,-0.75) arc (-2:70:0.5);
\node at (-0.1,-0.75){$\Phi_\alpha(u)$};

\node (F)  at (4.05,-0.5){$\xmapsto{\hat A_{\gamma_{\alpha 1}}^{e} }  $};

\node (G)  at (7.2,0){$v_\alpha$};
\node (H)  at (11.6,0){$v_2$};
\node (I)  at (6.4,1) [below=of G] {$v_1$};
\node (J)  at (11.5,1)[below=of H] {$v_3$};

\draw[->](G)-- node [above=0.1cm] {$\Phi_{\gamma_{\alpha 2}}(g_{\alpha2})$}                    (H);

\draw[->](G)--node [left=0.1cm] {$\partial(e)\Phi_{\gamma_{\alpha 1}}(g_{\alpha1})$}          (I);
\draw[->](H)-- node [right=0.1cm] {$\Phi_{\gamma_{23}}(g_{23})$}                  (J);
\draw[->](I)-- node [below=0.1cm] {$\Phi_{\gamma_{1 3}}(g_{13})$}                 (J);

\draw[ thick, ->] (10.54,-0.7) arc (-2:70:0.5);
\node  at (9.5,-0.8) {$\Phi_\alpha(u) e^{-1}$};
\end{tikzpicture}  
\end{eqnarray*}

\begin{eqnarray*}
\begin{tikzpicture}[thick]
\node (A) at (-1.5,0) {$v_\alpha$};
\node (B) at (1.5,0) {$v_2$};
\node (C) at (4.5,0) [below=of A] {$v_1$};
\node (D) at (4.5,0)   [below=of B] {$v_3$};

\draw[->](A)--node [left=0.1cm] {$\Phi_{\gamma_{\alpha 1}}(g_{\alpha1})$}(C);
\draw[->](B)--node [right=0.1cm] {$\Phi_{\gamma_{23}}(g_{23})$}(D);
\draw[->](A)--node [above=0.1cm] {$\Phi_{\gamma_{\alpha 2}}(g_{\alpha2})$}(B);
\draw[->](C)--node [below=0.1cm] {$\Phi_{\gamma_{1 3}}(g_{13})$}(D);
        \draw[ thick, ->] (0.9,-0.75) arc (-2:70:0.5);
\node at (-0.1,-0.75){$\Phi_\alpha(u)$};

\node (F)  at (4.05,-0.5){$\xmapsto{\hat A_{\gamma_{13}}^{e} }  $};

\node (G)  at (6.5,0){$v_\alpha$};
\node (H)  at (11.5,0){$v_2$};
\node (I)  at (6.4,1) [below=of G] {$v_1$};
\node (J)  at (11.6,1)[below=of H] {$v_3$};

\draw[->](G)-- node [above=0.1cm] {$\Phi_{\gamma_{\alpha 2}}(g_{\alpha2})$}                    (H);

\draw[->](G)--node [left=0.1cm] {$\Phi_{\gamma_{\alpha 1}}(g_{\alpha1})$}          (I);
\draw[->](H)--node [right=0.1cm] {$\Phi_{\gamma_{23}}(g_{23})$}                  (J);
\draw[->](I)-- node [below=0.1cm] {$\partial(e)\Phi_{\gamma_{1 3}}(g_{13})$}                 (J);

\draw[ thick, ->] (10.54,-0.7) arc (-2:70:0.5);
\node  at (9.05,-0.8) {$\Phi_\alpha(u)\left(p^+(v_\alpha\rightarrow s(\gamma_{13}))\rhd e^{-1}\right)$};

\end{tikzpicture}      
\end{eqnarray*}

\begin{eqnarray*}
\begin{tikzpicture}[thick]
\node (A) at (-1.5,0) {$v_\alpha$};
\node (B) at (1.5,0) {$v_2$};
\node (C) at (4.5,0) [below=of A] {$v_1$};
\node (D) at (4.5,0)   [below=of B] {$v_3$};

\draw[->](A)--node [left=0.1cm] {$\Phi_{\gamma_{\alpha 1}}(g_{\alpha1})$}(C);
\draw[->](B)--node [right=0.1cm] {$\Phi_{\gamma_{23}}(g_{23})$}(D);
\draw[->](A)--node [above=0.1cm] {$\Phi_{\gamma_{\alpha 2}}(g_{\alpha2})$}(B);
\draw[->](C)--node [below=0.1cm] {$\Phi_{\gamma_{1 3}}(g_{13})$}(D);
        \draw[ thick, ->] (0.9,-0.75) arc (-2:70:0.5);
\node at (-0.1,-0.75){$\Phi_\alpha(u)$};

\node (F)  at (4.05,-0.5){$\xmapsto{\hat A_{\gamma_{23}}^{e} }  $};

\node (G)  at (7.0,0){$v_\alpha$};
\node (H)  at (11.5,0){$v_2$};
\node (I)  at (6.4,1) [below=of G] {$v_1$};
\node (J)  at (11.5,1)[below=of H] {$v_3$};

\draw[->](G)-- node [above=0.1cm] {$\Phi_{\gamma_{\alpha 2}}(g_{\alpha2})$}                    (H);

\draw[->](G)--node [left=0.1cm] {$\Phi_{\gamma_{\alpha 1}}(g_{\alpha1})$}          (I);
\draw[->](H)--node [right=0.1cm] {$\partial(e)\Phi_{\gamma_{23}}(g_{23})$}                  (J);
\draw[->](I)-- node [below=0.1cm] {$\Phi_{\gamma_{1 3}}(g_{13})$}                 (J);

\draw[ thick, ->] (10.,-0.7) arc (-2:70:0.5);
\node  at (9.4,-0.8) {$\left(p^-(v_\alpha\rightarrow s(\gamma_{23}))\rhd e\right) \Phi_\alpha(u)$};

\end{tikzpicture}      
\end{eqnarray*}

\begin{eqnarray*}
\begin{tikzpicture}[thick]
\node (A) at (-1.5,0) {$v_\alpha$};
\node (B) at (1.5,0) {$v_2$};
\node (C) at (4.5,0) [below=of A] {$v_1$};
\node (D) at (4.5,0)   [below=of B] {$v_3$};

\draw[->](A)--node [left=0.1cm] {$\Phi_{\gamma_{\alpha 1}}(g_{\alpha1})$}(C);
\draw[->](B)--node [right=0.1cm] {$\Phi_{\gamma_{23}}(g_{23})$}(D);
\draw[->](A)--node [above=0.1cm] {$\Phi_{\gamma_{\alpha 2}}(g_{\alpha2})$}(B);
\draw[->](C)--node [below=0.1cm] {$\Phi_{\gamma_{1 3}}(g_{13})$}(D);
        \draw[ thick, ->] (0.9,-0.75) arc (-2:70:0.5);
\node at (-0.1,-0.75){$\Phi_\alpha(u)$};

\node (F)  at (4.05,-0.5){$\xmapsto{\hat A_{\gamma_{\alpha2}}^{e} }  $};

\node (G)  at (7.5,0){$v_\alpha$};
\node (H)  at (11.5,0){$v_2$};
\node (I)  at (6.4,1) [below=of G] {$v_1$};
\node (J)  at (11.5,1)[below=of H] {$v_3$};

\draw[->](G)-- node [above=0.1cm] {$\partial(e)\Phi_{\gamma_{\alpha 2}}(g_{\alpha2})$}                    (H);

\draw[->](G)--node [left=0.1cm] {$\Phi_{\gamma_{\alpha 1}}(g_{\alpha1})$}          (I);
\draw[->](H)--node [right=0.1cm] {$\Phi_{\gamma_{23}}(g_{23})$}                  (J);
\draw[->](I)-- node [below=0.1cm] {$\Phi_{\gamma_{1 3}}(g_{13})$}                 (J);

\draw[ thick, ->] (10.54,-0.7) arc (-2:70:0.5);
\node  at (9.5,-0.8) {$ e \Phi_\alpha(u)$};

\end{tikzpicture}      
\end{eqnarray*}

The edge gauge operator $ \hat A_{\gamma}^{e}$ defines a representation of $  E$ on  $ \mathcal{H}(M,L,\mathcal{G}(\mathcal{X}))$ and  commutes for   two distinct edges $\gamma,\gamma'\in L^1$. These are shown by the  following Lemma. \\

\begin{lemma}\label{l2}  
    {\it Let $\gamma, \gamma'\in L^1$ be distinct edges  and $\alpha,\alpha'\in L^2$  be adjacent plaquettes, we have  
\begin{eqnarray}\label{bz}
\hat A_\gamma^e \hat A_\gamma^{e'}= \hat A_\gamma^{ee'}\quad \mbox{and}\quad [ \hat A_\gamma^e, \hat A_{\gamma'}^{e'}]= 0, \quad \forall\, e,e'\in   E.
\end{eqnarray}}
\end{lemma}
        \begin{proof}
          Provided in Appendix  
        \end{proof}
The vertex gauge operator  $ \hat A_v^h$ and the edge operator $ \hat A_{\gamma}^e$ commute in the   conditions given by the following Lemma.

\begin{lemma} \label{l3}  
{\it Let $v\in L^0$ be a vertex, with adjacent plaquette $\alpha\in L^2$, and let $\gamma\in L^1$ be an edge. Then for any $h\in G$ and $e\in  E$ the following hold:
\begin{eqnarray}\label{a}
\hat A_v^h \hat A_\gamma^{e}&=& \hat A_\gamma^{h\rhd e} \hat A_v^h,\quad  \text{if v is the starting vertex of $\gamma$}, \label{a1z}\\
\hat A_v^h \hat A_\gamma^{e}&=& \hat A_\gamma^{e} \hat A_v^h,\quad \quad \, \text{if v is not the starting vertex of $\gamma$.} \label{a2z}
\end{eqnarray}
}
\end{lemma}

\begin{proof}
          Provided in Appendix  B.
        \end{proof}
    
\end{itemize}

\subsection{ Gauge invariance }
In standard lattice gauge theory, gauge invariant quantities can be constructed from the
holonomy around closed loops.  In higher lattice gauge theory, one can build gauge-invariants from the 1-holonomy along the closed loop around the palquette and the 2-holonomy over the  closed surface. By gauge invariants, we mean the invariance of  1,2-holonomies under     gauge transforms $\hat A_v^h$ and  $\hat A_\gamma^e$.

 Now, we  assign operators associated with the 1-holonomy and 2-holonomy called the plaquette operator $ \hat B_\alpha$ and the blob operator $\hat B_b$,\label{key} respectively.  The plaquette operator $ \hat B_\alpha$ acts locally on  edges bounding the plaquette $\alpha$ and  checks that the  1-flux (1-holonomy) through a plaquette is equal to the identity of the group $G$. Hovever the blob operator  $ \hat B_b$ acts on the closed surface  and checks the 2-fake flatness  constraint (trivial 2-holonomy) of the surface.

\begin{definition}
{\it Let $\mathcal{P}_2(M,L)$ be a lattice model. Let $v_\alpha,v_b\in L^0$ be the base points of the plaquette $\alpha\in L^2$ and the blob $b\in L^3$ respectively. We have \cite{40,41,42}:
\begin{itemize}
\item The plaquette operator $\hat B_\alpha :\mathcal{H}(M,L,\mathcal{G}(\mathcal{X}))\rightarrow \mathcal{H}(M,L,\mathcal{G}(\mathcal{X}))$ is defined  as follows
\begin{eqnarray}
\hat B_\alpha  \left| \Phi_{\gamma}(g)  \Phi_{\alpha}(u)  \right \rangle &=&\delta\left( Hol_{v_{\alpha}}^1(M,L,\mathcal{G}(\mathcal{X})) ,\Phi_{\partial\alpha}(1_G)\right) \left| \Phi_{\gamma}(g)  \Phi_{\alpha}(u)  \right \rangle. %&=&\delta\left(\Phi_{\gamma}(g)\partial(\Phi_{\alpha}(g,e)) (\Phi_{\gamma'}(g'))^{-1}, \Phi_{\partial\alpha}(1_G)\right)     \left| \Phi_{\gamma}(g)  \Phi_{\alpha}(u)  \right \rangle.
\end{eqnarray}
\item The blob operator $\hat B_b :\mathcal{H}(M,L,\mathcal{G}(\mathcal{X}))\rightarrow \mathcal{H}(M,L,\mathcal{G}(\mathcal{X}))$ is defined  as follows
\begin{eqnarray}
\hat B_b  \left| \Phi_{\gamma}(g)  \Phi_{\alpha}(u)  \right \rangle  &=&\delta\left( Hol_{v_{\alpha}}^2(M,L,\mathcal{G}(\mathcal{X})) ,\Phi_{\partial b}(1_G,1_E)\right) \left| \Phi_{\gamma}(g)  \Phi_{\alpha}(u)  \right \rangle.
%,\cr &=& \delta\left(\Phi_{\gamma}(g)\rhd \Phi_{\alpha}(g,e),\Phi_{\partial b}(1_G,1_E)\right)     \left| \Phi_{\gamma}(g)  \Phi_{\alpha}(u)  \right \rangle,
\end{eqnarray}
where $\delta$ is the Kronecker and $\partial b$ the quantised 2-boundary of the blob.
\end{itemize} }
\end{definition}

Now we can  check how these holonomy operators  are also  invariant   under gauge transformations.
\begin{lemma}\label{l4}
    {\it Let $\alpha \in L^2$ be a plaquette element  and $b\in L^3$  be an adjacent blob element, we have the following  commutation relations  
\begin{eqnarray}
[\hat B_{\alpha}, \hat A_v^h]&=&  0  \quad \mbox{and}\quad [\hat B_{\alpha}, \hat A_\gamma^e]=0,\label{b10} \\
{ [\hat B_b, \hat A_v^h] }&=&  0  \quad   \mbox{and}\quad [\hat B_b, \hat A_\gamma^e]=0,\label{b100}
\end{eqnarray}
for all $h\in G$ and $e\in  E$.}
\end{lemma} 
\begin{proof}
    Provided in appendic B
\end{proof}

\subsection{ Hamiltonian formulation}\label{Hamiltonian}
Having defined the Hilbert space and local operators, we now define in this section   an exactly solvable  Hamiltonian model in   3+1D     topological higher  lattice gauge theory  while the temporal dimension is continuous. We will then demonstrate that the ground-states are  topological observable, i.e.,  they are invariant under the changes of the orientations of the edges of the lattice as well as the orientation and base-point of each plaquette.
\begin{definition}
    {\it
The Hamiltonian for higher lattice gauge theory is given by the sum of  projector terms \cite{40,41,42} 
\begin{eqnarray}\label{ha}
\hat H=-\sum_{Vertices,v} \hat{\mathcal{A}}_v - \sum_{edges,\gamma} \hat{\mathcal{A}}_\gamma- \sum_{plaquettes,\alpha} \hat{\mathcal{B}}_\alpha-\sum_{blobs,b} \hat{\mathcal{B}}_b,
\end{eqnarray}
where the sums run over all the vertices, edges, plaquettes, and blobs projectors.
 Here $\hat{\mathcal{A}}_v$ and $\hat{\mathcal{A}}_\gamma$ are the average, over gauge transforms at the vertex $v$ and at the edges $\gamma$ respectively. $\hat{\mathcal{B}}_\alpha$ and $\hat{\mathcal{B}}_b$ are the
 plaquette and the blob projector terms.  The sum of these projector terms represents the total energy of the system 
 and they  encode the topological realizations  of the Hamiltonian model. They are expressed as  follows. 
\begin{eqnarray}
 \hat{\mathcal{A}}_v =\frac{1}{|G|}\sum_{h\in G}\hat A_v^h,\quad  \hat{\mathcal{A}}_\gamma =\frac{1}{|E|}\sum_{e\in  E}\hat A_\gamma^e,\quad \hat{\mathcal{B}}_\alpha= \hat B_\alpha\quad \mbox{and}\quad \hat{\mathcal{B}}_b=\hat B_b.
 \end{eqnarray}}
\end{definition} 
%Projector operators $\hat{\mathcal{P}}$ defined ​ onto specific
Projector operators $\hat{\mathcal{P}}$ defined in
 subspaces of  the Hilbert space  $\mathcal{H}(M,L,\mathcal{G}(\mathcal{X}))$ satisfy $\hat{\mathcal{P}}^2=\hat{\mathcal{P}}$ and $\hat{\mathcal{P}}^\dag=\hat{\mathcal{P}},$ which  are  the idempotency property and the Hermiticity property, respectively.  \\

\begin{proposition}\label{p1}
     {\it Operators $ \hat{\mathcal{A}}_v, \hat{\mathcal{A}}_\gamma, \hat{\mathcal{B}}_\alpha, \hat{\mathcal{B}}_b: \mathcal{H}(M,L,\mathcal{G}(\mathcal{X}))\rightarrow \mathcal{H}(M,L,\mathcal{G}(\mathcal{X}))$  satisfy 
\begin{eqnarray}
\hat{\mathcal{A}}_v^2 &=& \hat{\mathcal{A}}_v,\quad  \hat{\mathcal{A}}_v^\dag = \hat{\mathcal{A}}_v,\quad \mbox{and}\quad
\hat{\mathcal{A}}_\gamma^2 = \hat{\mathcal{A}}_\gamma,\quad  \hat{\mathcal{A}}_\gamma^\dag = \hat{\mathcal{A}}_\gamma,\\
\hat{\mathcal{B}}_\alpha^2 &=& \hat{\mathcal{B}}_\alpha,\quad  \hat{\mathcal{B}}_\alpha^\dag = \hat{\mathcal{B}}_\alpha\quad \mbox{and}\quad
\hat{\mathcal{B}}_\alpha^2 = \hat{\mathcal{B}}_b,\quad \hat{\mathcal{B}}_b^\dag = \hat{\mathcal{B}}_b.
\end{eqnarray} }
\end{proposition}

\begin{proof}
          Provided in Appendix  B
        \end{proof}

As projectors,  $ \hat{\mathcal{A}}_v, \hat{\mathcal{A}}_\gamma, \hat{\mathcal{B}}_\alpha,$ and $\hat{\mathcal{B}}_b$ have eigenvalues of  zero and one. Now,  we call a Hamiltonian term \eqref{ha} exactly solvable, if all of its  projector operator terms mutually commute. Referring to the results in proposition \ref{p1} and in lemmas \ref{l5},\ref{l4}, \ref{l3}, \ref{l2}, the exactly solvable model is  shown by the following results:

\begin{lemma}  \label{l5}
    {\it  Let $v,v'\in L^0$ , $\gamma,\gamma'\in L^1$, $\alpha,\alpha'\in L^2,$ and $b,b'\in L^3$. A Hamiltonian \eqref{ha} is exactly  solvable if its  projector operator terms satisfy the following  commutation rules
	\begin{eqnarray}
		\hat{\mathcal{A}}_v\hat{\mathcal{A}}_v&=&	\hat{\mathcal{A}}_v,\quad \hat{\mathcal{A}}_\gamma\hat{\mathcal{A}}_\gamma=\hat{\mathcal{A}}_\gamma,\quad \hat{\mathcal{B}}_\alpha\hat{\mathcal{B}}_\alpha=\hat{\mathcal{B}}_\alpha,\quad  \hat{\mathcal{B}}_b \hat{\mathcal{B}}_b=\hat{\mathcal{B}}_b \label{z1}\\
		{[\hat{\mathcal{A}}_v,\hat{\mathcal{A}}_{v'}] }&=&0,\quad {[\hat{\mathcal{A}}_\gamma,\hat{\mathcal{A}}_{\gamma'}] }=0,\quad {[\hat{\mathcal{B}}_\alpha,\hat{\mathcal{B}}_{\alpha'}] }=0,\quad {[\hat{\mathcal{B}}_b,\hat{\mathcal{B}}_{b'}] }=0,\label{z2}\\
			{[\hat{\mathcal{A}}_v,\hat{\mathcal{A}}_{\gamma}] }&=&0,\quad {[\hat{\mathcal{A}}_v,\hat{\mathcal{B}}_{\alpha}] }=0,\quad {[\hat{\mathcal{A}}_v,\hat{\mathcal{B}}_{b}] }=0,\quad {[\hat{\mathcal{A}}_\gamma,\hat{\mathcal{B}}_{\alpha}] }=0,\label{z3}\\
			{[\hat{\mathcal{A}}_\gamma,\hat{\mathcal{B}}_{b}] }&=&0,\quad {[\hat{\mathcal{B}}_\alpha,\hat{\mathcal{B}}_{b}] }=0.\label{z4}
		\end{eqnarray}}
     \end{lemma}   
\begin{proof}
 The proofs of equations \eqref{z1} follow from the proofs of equations $\hat{\mathcal{A}}_v^2 = \hat{\mathcal{A}}_v$, $\hat{\mathcal{A}}_\gamma^2 = \hat{\mathcal{A}}_\gamma$,\, $\hat{\mathcal{B}}_\alpha^2 = \hat{\mathcal{B}}_\alpha$ and $\hat{\mathcal{B}}_b^2 = \hat{\mathcal{B}}_b$ provided in the appendix B.
    The proofs of equations \eqref{z2},\,\eqref{z3}, \eqref{z4} follow from the proofs of equations \eqref{lem1}, \eqref{a2z},\eqref{b10} and \eqref{b100} provided in the appendix \ref{b}.
\end{proof}

Having considered the various energy terms of the Hamiltonian model \eqref{ha}, we wish to use them to examine the ground states (GS). The degeneracy of GS will generally depend on the topology of the manifold. This means that one can have GS degeneracy on the torus but not the sphere for example, because topological models on the sphere have a unique ground state in the absence of symmetry. In the present case, we  keep the
discussion general, but it may be useful to have the sphere
in mind.

The GS  subspace  $ \mathcal{H}^0(M,L,\mathcal{G}(\mathcal{X})) $  of the lattice Hamiltonian defined in \eqref{ha} satisfies the stabiliser constraints defined by  \cite{40,41,42} 
\begin{eqnarray}
\mathcal{H}^0(M,L,\mathcal{G}(\mathcal{X}))=\{ |\psi\rangle \in \mathcal{H}(M,L,\mathcal{G}(\mathcal{X})):	\hat{\mathcal{A}}_v|\psi\rangle =|\psi\rangle,\,  	\hat{\mathcal{A}}_\gamma|\psi\rangle =|\psi\rangle,\, 	\hat{\mathcal{B}}_\alpha|\psi\rangle =|\psi\rangle, \, 	\hat{\mathcal{B}}_b|\psi\rangle =|\psi\rangle   \}.
\end{eqnarray}
It is direct to check that
\begin{eqnarray}
\hat{A}_v^h\hat{\mathcal{A}}_v=\hat{\mathcal{A}}_v,\quad \hat{A}_\gamma^e\hat{\mathcal{A}}_\gamma=\hat{\mathcal{A}}_\gamma, \quad \hat{B}_\alpha\hat{\mathcal{B}}_\alpha=\hat{\mathcal{B}}_\alpha, \quad \hat{B}_b\hat{\mathcal{B}}_b=\hat{\mathcal{B}}_b.
\end{eqnarray}
Hence, one can show that  the GS $|\psi\rangle $ are invariant under   gauge transforms  and holonomy operators 
\begin{eqnarray}
\hat{A}_v^h|\psi\rangle&=&  |\psi\rangle,\quad
\hat{A}_\gamma^e|\psi\rangle=|\psi\rangle,\\
\hat{B}_\alpha|\psi\rangle&=&  |\psi\rangle,\quad \hat{B}_b|\psi\rangle=  |\psi\rangle.
\end{eqnarray}
In the following, we will need the corresponding
ground state projector, namely
\begin{eqnarray}
	\hat P= \prod_{Vertices,v}\hat{\mathcal{A}}_v\prod_{edges,\gamma}\hat{\mathcal{A}}_\gamma\prod_{plaquette,\alpha}\hat{\mathcal{B}}_\alpha\prod_{blob,b}\hat{\mathcal{B}}_b,
\end{eqnarray}
and then the subspace of the ground states is
\begin{eqnarray}
	\mathcal{H}^0(M,L,\mathcal{G}(\mathcal{X}))=\{ |\psi\rangle \in \mathcal{H}(M,L,\mathcal{G}(\mathcal{X})):	\hat P|\psi\rangle =|\psi\rangle  \}.
\end{eqnarray}
 It was demonstrated in \cite{40}  that  the ground-state subspace of the topological higher
lattice gauge theory Hamiltonian schema corresponds to the state space defined by the Yetter homotopy 2-type TQFT.

\section{Topological invariance and unitarity of the ground states} \label{sec40}
In mathematics, topology is the study of the global properties of manifolds that are insensitive to local smooth deformations. The overused, but still illustrative example is the topological equivalence between a donut and a coffee cup.   Small smooth deformations, such as taking a bite on the side of the donut or chipping away a piece of the cup will change the object locally, but the topology remains unchanged.
Only global violent deformations, such as cutting the donut in half or breaking the cup handle, will change the topology by removing the hole. Analogously, these topological properties can be systematically recovered for a discrete model of
topological phases which are called 
the mutation symmetry transformations \cite{1,13,55,56}. As  discussed in \cite{40,41,42}, for higher-lattice gauge theory models, the dressing of the lattice based on the orientations of the edges as well as the orientation and base point of each plaquette, must somehow be resilient to changes to these details.

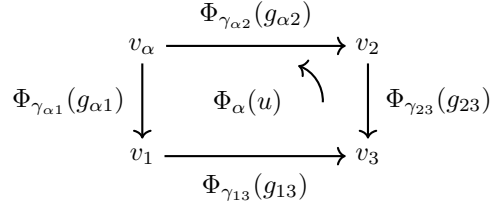
\begin{figure}[H]
\centering
	\begin{tikzpicture}[thick]
		\node (A) at (-1.5,0) {$v_\alpha$};
		\node (B) at (1.5,0) {$v_2$};
		\node (C) at (4.5,0) [below=of A] {$v_1$}; 
		\node (D) at (4.5,0)   [below=of B] {$v_3$};	
		
		\draw[->](A)--node [left=0.1cm] {$\Phi_{\gamma_{\alpha 1}}(g_{\alpha1})$}(C);
		\draw[->](B)--node [right=0.1cm] {$\Phi_{\gamma_{23}}(g_{23})$}(D);
		\draw[->](A)--node [above=0.1cm] {$\Phi_{\gamma_{\alpha 2}}(g_{\alpha2})$}(B);
		\draw[->](C)--node [below=0.1cm] {$\Phi_{\gamma_{1 3}}(g_{13})$}(D); 
        \draw[ thick, ->] (0.9,-0.75) arc (-2:70:0.5);
		\node at (-0.1,-0.75){$\Phi_\alpha(u)$};	
	\end{tikzpicture}
     \caption{Configuration graph: $v_\alpha$ is the plaquette base point, $v_i\,(i=1,2,3)$ are the vertices, $\Phi_{\gamma_j}(g_j)$ are the colored edges $\gamma_i$ and   $\Phi_\alpha(u)$ is  the colored plaquette $\alpha$ anti-clockwise rotation. }\label{en49}
\end{figure}

In this subsection, we will show that in  a certain sense, the GS of the Hamiltonian defined by means of the ground state projector is invariant under the change of  the branching structure.
Consider $ \hat P_1|\psi\rangle\in \mathcal{H}_1^0$ and $\hat P_2|\psi\rangle\in \mathcal{H}_2^0$ as two GS projectors and  $\hat T_i:\mathcal{H}_1^0\rightarrow \mathcal{H}_2^0$ as given transformations of the lattice dressing. We show that these transformations let the GS projector $\hat P$ topologically observable and are unitary on the ground states
\begin{eqnarray}\label{top1}
 \hat T_i^{-1} \hat P_1\hat T_i=\hat P_2\quad \mbox{and}\quad  \langle \hat T_i\phi |\hat T_i\psi\rangle=\langle \phi |\psi\rangle. 
\end{eqnarray}
Let $\Gamma_1\in \mathcal{H}_1^0$ and $\Gamma_2\in\mathcal{H}_2^0 $ be two graphs that are transformed from $\Gamma_1$ into $\Gamma_2$ by the action of operator $\hat T_i$. We also distinguish three tranformation operators $\hat T_1,\hat T_2,\hat T_3\in \hat T,$ that flip the orientation of edges as well as  the   orientation of plaquettes and 
   move the base point of  plaquettes, respectively. To demonstrate how these operator transforms change one graph into another, we take a look at a configuration graph  provided by the figure \eqref{en49}

  \begin{itemize}
      \item The edge-flipping transformation $\hat T_1$ reverses the orientation of the  edge  labeled by $\Phi_{\gamma}(g)$ into $\Phi_{\gamma^{-1}}(g^{-1})$ \cite{42} i.e., 
  \begin{eqnarray}
    \hat T_1 |\Phi_{\gamma}(g)\rangle&=&|\Phi_{\gamma^{-1}}(g^{-1})\rangle.
\end{eqnarray}

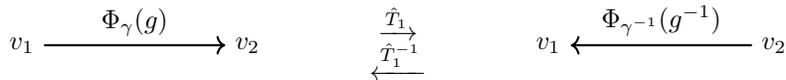
\begin{figure}[H]
\centering
\begin{tikzpicture}[thick]
\node (A) at (-1.5,0) {$v_1$};
\node (B) at (1.5,0) {$v_2$};
\path[->] (A) edge  node[above] {$\Phi_{\gamma}(g)$} (B);

\node (F)  at (3.5,-0.2){$\xleftarrow{\hat T_1^{-1}}$};
\node (F1)  at (3.5,0.3){$\xrightarrow{\hat T_1}$};
\node (G)  at (5.5,0){$v_1$};
\node (H)  at (8.5,0){$v_2$};

\path[<-] (G) edge  node[above] {$\Phi_{\gamma^{-1}}(g^{-1})$} (H);

\end{tikzpicture}      
  \caption{Flipping  edge transformation and its inverse. 
}\label{en2}
\end{figure}
The edge flipping transformation $\hat T_1$ on a certain edge $\Phi_{\gamma_{\alpha 1}}(g_{\alpha 1})$  of the graph \eqref{en49} is illustrated as follows;
 
	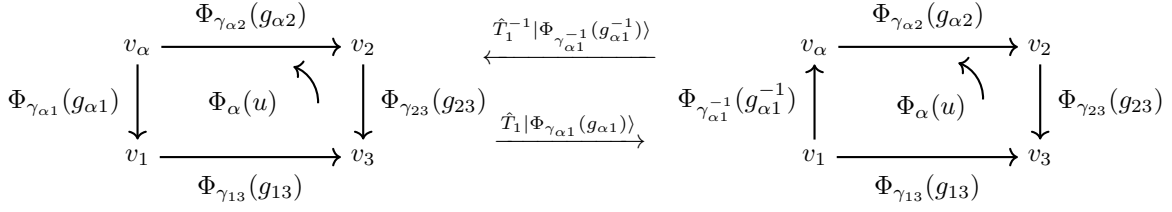
\begin{figure}[H]
\centering
	\begin{tikzpicture}[thick]
		\node (A) at (-1.5,0) {$v_\alpha$};
		\node (B) at (1.5,0) {$v_2$};
		\node (C) at (4.5,0) [below=of A] {$v_1$}; 
		\node (D) at (4.5,0)   [below=of B] {$v_3$};	
		
		\draw[->](A)--node [left=0.1cm] {$\Phi_{\gamma_{\alpha 1}}(g_{\alpha1})$}(C);
		\draw[->](B)--node [right=0.1cm] {$\Phi_{\gamma_{23}}(g_{23})$}(D);
		\draw[->](A)--node [above=0.1cm] {$\Phi_{\gamma_{\alpha 2}}(g_{\alpha2})$}(B);
		\draw[->](C)--node [below=0.1cm] {$\Phi_{\gamma_{1 3}}(g_{13})$}(D); 
        \draw[ thick, ->] (0.9,-0.75) arc (-2:70:0.5);
	\node at (-0.1,-0.75){$\Phi_\alpha(u)$};

		\node (F)  at (4.25,0.1){$\xleftarrow{\hat T_1^{-1}|\Phi_{\gamma_{\alpha1}^{-1}}(g_{\alpha 1}^{-1})\rangle } $};
        \node (F1)  at (4.25,-1.09){$\xrightarrow{\hat T_1|\Phi_{\gamma_{\alpha1}}(g_{\alpha 1})\rangle } $};
		
		\node (G)  at (7.5,0){$v_\alpha$};
		\node (H)  at (10.5,0){$v_2$};
		\node (I)  at (6.4,1) [below=of G] {$v_1$};
		\node (J)  at (10.5,1)[below=of H] {$v_3$};
		
		\draw[->](G)-- node [above=0.1cm] {$\Phi_{\gamma_{\alpha 2}}(g_{\alpha2})$}                    (H);
		
		\draw[<-](G)--node [left=0.1cm] {$\Phi_{\gamma_{\alpha 1}^{-1}}(g_{\alpha1}^{-1})$}          (I);
		\draw[->](H)-- node [right=0.1cm] {$\Phi_{\gamma_{23}}(g_{23})$}                  (J);
		\draw[->](I)-- node [below=0.1cm] {$\Phi_{\gamma_{1 3}}(g_{13})$}                 (J);
		
		\draw[ thick, ->] (9.74,-0.7) arc (-2:70:0.5);
		\node  at (9.05,-0.8) {$\Phi_\alpha(u)$};	
	\end{tikzpicture}
   \caption{Flipping  edge transformation and its inverse
}\label{en3} 
\end{figure}
As we can see, this transform reverses the orientation of the given edge \eqref{en49} of the graph and  lets  others edges unaffected.

\item We now consider the analogous procedure where we reverse the orientation of a plaquette $\alpha,$ and  its invert is labeled by  $\alpha^{-1} $ . We denote this operation by $\hat T_2$ for a plaquette labeled by $\Phi_\alpha(u),$ and its representation is  given by   
\begin{eqnarray}
    \hat T_2 |\Phi_\alpha(u)\rangle= |\left(\Phi_\alpha(u)\right)^{-1}\rangle =|\Phi_{\alpha^{-1}}(u^{-1})\rangle.
\end{eqnarray} 
This transformation  can only  affect a given plaquette orientation  itself (it is not affected by other plaquettes). In the present example \eqref{en49}, this transformation can be illustrated as follows;

	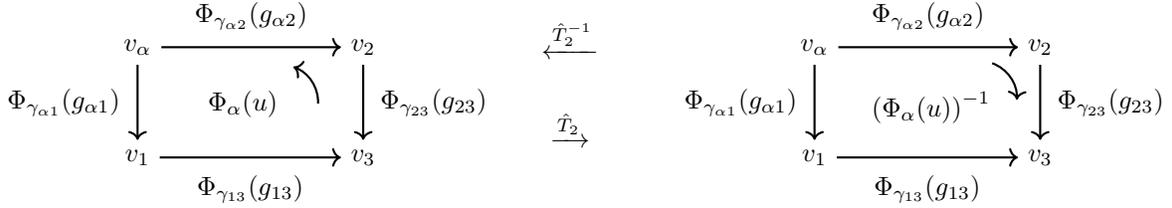
\begin{figure}[H]
\centering
	\begin{tikzpicture}[thick]
		\node (A) at (-1.5,0) {$v_\alpha$};
		\node (B) at (1.5,0) {$v_2$};
		\node (C) at (4.5,0) [below=of A] {$v_1$}; 
		\node (D) at (4.5,0)   [below=of B] {$v_3$};	
		
		\draw[->](A)--node [left=0.1cm] {$\Phi_{\gamma_{\alpha 1}}(g_{\alpha1})$}(C);
		\draw[->](B)--node [right=0.1cm] {$\Phi_{\gamma_{23}}(g_{23})$}(D);
		\draw[->](A)--node [above=0.1cm] {$\Phi_{\gamma_{\alpha 2}}(g_{\alpha2})$}(B);
		\draw[->](C)--node [below=0.1cm] {$\Phi_{\gamma_{1 3}}(g_{13})$}(D); 
        \draw[ thick, ->] (0.9,-0.75) arc (-2:70:0.5);
		\node at (-0.1,-0.75){$\Phi_{\alpha}(u)$};

		\node (F)  at (4.25,0.1){$\xleftarrow{\hat T_2^{-1} } $};
        \node (F1)  at (4.25,-1.09){$\xrightarrow{\hat T_2 } $};
		
		\node (G)  at (7.5,0){$v_\alpha$};
		\node (H)  at (10.5,0){$v_2$};
		\node (I)  at (6.4,1) [below=of G] {$v_1$};
		\node (J)  at (10.5,1)[below=of H] {$v_3$};
		
		\draw[->](G)-- node [above=0.1cm] {$\Phi_{\gamma_{\alpha 2}}(g_{\alpha2})$}                    (H);
		
		\draw[->](G)--node [left=0.1cm] {$\Phi_{\gamma_{\alpha 1}}(g_{\alpha1})$}          (I);
		\draw[->](H)-- node [right=0.1cm] {$\Phi_{\gamma_{23}}(g_{23})$}                  (J);
		\draw[->](I)-- node [below=0.1cm] {$\Phi_{\gamma_{1 3}}(g_{13})$}                 (J);
		
		\draw[ thick, <-] (10.19,-0.7) arc (-2:70:0.5);
		\node  at (9.05,-0.8) {$\left(\Phi_{\alpha}(u)\right)^{-1}$};
			
	\end{tikzpicture}	      
\caption{Flipping  plaquette orientation transformation and its inverse.}\label{en4} 
\end{figure}

As we can see from this representation \eqref{en4}, reversing the orientation of the plaquette $\alpha$ does not change the orientations of edges $\gamma_i$ around this plaquette.

\item The final procedure to consider is changing the base point of a plaquette. We denote the procedure that moves the base point
of plaquette $\alpha$ from a vertex $v_{\alpha_1}$ to a vertex  $v_{\alpha_2}$ by $\hat T_3 $. This operation changes the labeled plaquette $\Phi_\alpha(u)$ to $p(v_{\alpha_1}\rightarrow  v_{\alpha_2})^{-1}\rhd \Phi_\alpha(u)$,  where $ p(v_{\alpha_1}\rightarrow  v_{\alpha_2})$ is   the path along which we move the base point
\begin{eqnarray}\label{qa}
   \hat T_3\Phi_\alpha(u)= p(v_{\alpha_1}\rightarrow  v_{\alpha_2})^{-1}\rhd \Phi_\alpha(u),
\end{eqnarray}

 We finally  consider the  changing of the base point of a plaquette. Let's  consider $ v_{\alpha_1}$ and $ v_{\alpha_2}$, we denote the procedure that moves the base point
of plaquette $\alpha $ from a vertex $ v_{\alpha_1}$ to a vertex  $ v_{\alpha_2}$ by $\hat T_3 $. This operation changes the labeled plaquette $\Phi_\alpha(u)$ to $p(v_{\alpha_1}\rightarrow  v_{\alpha_2})^{-1}\rhd \Phi_\alpha(u)$  
\begin{eqnarray}\label{qa}
   \hat T_3\Phi_\alpha(u)= p(v_{\alpha_1}\rightarrow  v_{\alpha_2})^{-1}\rhd \Phi_\alpha(u),
\end{eqnarray}
where $ p(v_{\alpha_1}\rightarrow  v_{\alpha_2})$ is   the path along which we move the base point.
The process of shifting the base point $v_{\alpha}=v_{\alpha_1}$ to the base point $v_{\alpha}=v_{\alpha_2}$ may be shown using the graph \eqref{en49} above as an example 
	\begin{figure}[H]
\centering
	\begin{tikzpicture}[thick]
		\node (A) at (-1.5,0) {$v_{\alpha}$};
		\node (B) at (1.5,0) {$v_2$};
		\node (C) at (4.5,0) [below=of A] {$v_1$}; 
		\node (D) at (4.5,0)   [below=of B] {$v_3$};	
		
		\draw[->](A)--node [left=0.1cm] {$\Phi_{\gamma_{\alpha 1}}(g_{\alpha1})$}(C);
		\draw[->](B)--node [right=0.1cm] {$\Phi_{\gamma_{23}}(g_{23})$}(D);
		\draw[->](A)--node [above=0.1cm] {$\Phi_{\gamma_{\alpha 2}}(g_{\alpha2})$}(B);
		\draw[->](C)--node [below=0.1cm] {$\Phi_{\gamma_{1 3}}(g_{13})$}(D); 
        \draw[ thick, ->] (0.9,-0.75) arc (-2:70:0.5);
		\node at (-0.1,-0.75){$\Phi_\alpha(u)$};

	\node (F)  at (4.05,-0.1){$\xleftarrow{\hat T_3^{-1} } $};
        \node (F1)  at (4.05,-1.09){$\xrightarrow{\hat T_3 } $};
		
		\node (G)  at (6.5,0){$v_{\alpha_1}$};
		\node (H)  at (10.9,0){$v_2$};
		\node (I)  at (6.4,1) [below=of G] {$v_{\alpha_2}$};
		\node (J)  at (10.5,1)[below=of H] {$v_3$};
		
		\draw[->](G)-- node [above=0.1cm] {$\Phi_{\gamma_{\alpha 2}}(g_{\alpha2})$}                    (H);
		
		\draw[->](G)--node [left=0.1cm] {$\Phi_{\gamma_{\alpha_1\alpha_2}}(g_{\alpha_1\alpha_2})$}          (I);
		\draw[->](H)-- node [right=0.1cm] {$\Phi_{\gamma_{23}}(g_{23})$}                  (J);
		\draw[->](I)-- node [below=0.1cm] {$\Phi_{\gamma_{1 3}}(g_{13})$}                 (J);
		
		\draw[ thick, ->] (9.74,-0.7) arc (-2:70:0.5);
		\node  at (8.9,-0.8) {$ p(v_{\alpha_1}\rightarrow  v_{\alpha_2})^{-1}\rhd \Phi_\alpha(u)$};	
	\end{tikzpicture}
   \caption{Modifying a plaquette's base point: The expression $p(v_{\alpha_1}\rightarrow v_{\alpha_2})=\Phi_{\gamma_{\alpha_1\alpha_2}} (g_{\alpha_1 \alpha_2})$ represents the moving plaquette base point's path.  
}\label{en5} 
\end{figure}
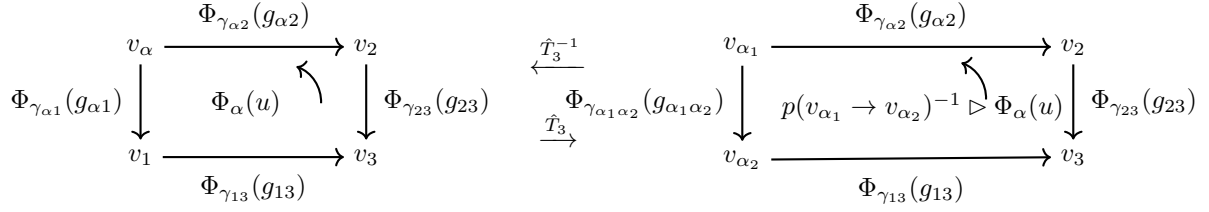
Based on the latter transformations, we have the following properties.

\begin{corollary} {\it  For given transformations  $\hat T_i:\mathcal{H}_1^0\rightarrow \mathcal{H}_2^0$, we have: 
		\begin{eqnarray}\label{t211}
		\text{i)}\quad \hat T_i=	\hat T_i^\dag	\quad\mbox{and}\quad   \text{ii)}\quad \hat T_i\hat T_j=	\hat T_j\hat T_i \quad \mbox{with}\quad  i,j=1,2,3.
		\end{eqnarray}}
      \end{corollary}  
\begin{proof}

$i)$ 
\begin{eqnarray*}
    \hat T_1 |\Phi_\gamma(g)\Phi_\alpha(u)\rangle &=&|\Phi_{\gamma^{-1}}(g^{-1})\Phi_\alpha(u)\rangle\implies 
    \hat T_1^2 |\Phi_\gamma(g)\Phi_\alpha(u)\rangle = |\Phi_\gamma(g)\Phi_\alpha(u)\rangle,\\
     \hat T_2 |\Phi_\gamma(g)\Phi_\alpha(u)\rangle &=&|\Phi_{\gamma}(g)(\Phi_\alpha(u))^{-1}\rangle\implies 
    \hat T_2^2 |\Phi_\gamma(g)\Phi_\alpha(u)\rangle = |\Phi_\gamma(g)\Phi_\alpha(u)\rangle,\\
    \hat T_2 |\Phi_\gamma(g)\Phi_\alpha(u)\rangle &=&|\Phi_{\gamma(g)}(p(v_{\alpha_1}\rightarrow  v_{\alpha_2})^{-1}\rhd \Phi_\alpha(u))\rangle\cr
    \hat T_2^2 |\Phi_\gamma(g)\Phi_\alpha(u)\rangle &=&|\Phi_\gamma(g)\Phi_\alpha(u)\rangle.
\end{eqnarray*}
This demonstrates that 
$$ \hat T_i^2 =\mathbb{I}\implies \hat T_i=\hat T_i^{-1}. $$
A transformation is unitary if  $\hat T_i^{-1}=\hat T_i^\dag$, hence we have 
\begin{eqnarray*}
    \hat T_i=\hat T_i^{-1}=\hat T_i^\dag
\end{eqnarray*}
$ii)$
\begin{eqnarray*}
  \hat T_2 \hat T_1 |\Phi_\gamma(g)\Phi_\alpha(u)\rangle &=&|\Phi_{\gamma^{-1}}(g^{-1})(\Phi_\alpha(u))^{-1}\rangle= \hat T_1 \hat T_2 |\Phi_\gamma(g)\Phi_\alpha(u)\rangle\cr
 \hat T_3 \hat T_2 |\Phi_\gamma(g)\Phi_\alpha(u)\rangle &=& |\Phi_\gamma(g)((p(v_{\alpha_1}\rightarrow  v_{\alpha_2})^{-1}\rhd \Phi_\alpha(u)))^{-1}\rangle=\hat T_2 \hat T_3 |\Phi_\gamma(g)\Phi_\alpha(u)\rangle,\\
 \hat T_3 \hat T_1 |\Phi_\gamma(g)\Phi_\alpha(u)\rangle &=&  |\Phi_{\gamma^{-1}}(g^{-1})(p(v_{\alpha_1}\rightarrow  v_{\alpha_2})^{-1}\rhd \Phi_\alpha(u))\rangle=\hat T_1 \hat T_3 |\Phi_\gamma(g)\Phi_\alpha(u)\rangle
\end{eqnarray*}

This demonstrates that 
	\begin{eqnarray*}\label{t21}
	 \hat T_i\hat T_j =	\hat T_j\hat T_i \quad \mbox{with}\quad  i,j=1,2,3.
		\end{eqnarray*}
    
\end{proof}

Now we need to show that all the above transformations let the GS projector $\hat P$ be topologically invariant and the GS unitary equivalent. To do so, we need to show  that  each energy terms ($\hat{\mathcal{A}}_v, \hat{\mathcal{A}}_\gamma, \hat{\mathcal{B}}_\alpha,\hat{\mathcal{B}}_b $) of the GS projector $\hat P$  individually   preserved under the edge-flipping transformation $ \hat T_1$, the plaquette-orientation flipping transformation $ \hat T_2,$ and the moving base point transformation $ \hat T_3$ (see Appendix C).

\begin{theorem}
     {\it  Let $v\in L^{0}$, $\gamma\in L^{1}$,   $\alpha\in L^{2}$, and   $b\in L^{3}$. For  given transformations $\hat T_i:\mathcal{H}_1^0\rightarrow \mathcal{H}_2^0$ of the dressing lattice, we have
\begin{eqnarray}
    \hat T_i^{-1} \hat P \hat T_i =  \hat P 
\end{eqnarray}
}
\end{theorem}
\begin{proof}
    Let us consider $ \hat P= \prod_{v,\gamma,\alpha,b}\hat{\mathcal{A}}_v\hat{\mathcal{A}}_\gamma\hat{\mathcal{B}}_\alpha\hat{\mathcal{B}}_b$ and with the  condition $\hat T_i\hat T_i^{-1}=\mathbb{I}=\hat T_i^{-1}\hat T_i$ we have 
    \begin{eqnarray*}
    \hat T_1^{-1} \left(\prod_{v,\gamma,\alpha,b}\hat{\mathcal{A}}_v\hat{\mathcal{A}}_\gamma\hat{\mathcal{B}}_\alpha\hat{\mathcal{B}}_b      \right) \hat T_1 &=& \prod_{v,\gamma,\alpha,b}\left(\hat T_1^{-1}\hat{\mathcal{A}}_v\hat{\mathcal{A}}_\gamma\hat{\mathcal{B}}_\alpha\hat{\mathcal{B}}_b   \hat T_1\right) 
    \cr
    &=& \prod_{v,\gamma,\alpha,b}\left(\hat T_1^{-1}\hat{\mathcal{A}}_v T_1T_1^{-1}\hat{\mathcal{A}}_\gamma T_1T_1^{-1}\,\hat{\mathcal{B}}_\alpha T_1 T_1^{-1}\,\hat{\mathcal{B}}_b   \hat T_1\right)= \hat P \cr
  \hat T_2^{-1}   \hat T_1^{-1} \left(\prod_{v,\gamma,\alpha,b}\hat{\mathcal{A}}_v\hat{\mathcal{A}}_\gamma\hat{\mathcal{B}}_\alpha\hat{\mathcal{B}}_b      \right) \hat T_1\hat T_2  &=& \hat T_2^{-1} \hat P \hat T_2\cr
 &=& \prod_{v,\gamma,\alpha,b}\left(\hat T_2^{-1}\hat{\mathcal{A}}_v T_2T_2^{-1}\hat{\mathcal{A}}_\gamma T_2T_2^{-1}\,\hat{\mathcal{B}}_\alpha T_2 T_2^{-1}\,\hat{\mathcal{B}}_b   \hat T_2\right)\cr
   &=&\hat P,\cr
 \hat T_3^{-1} \hat T_2^{-1}   \hat T_1^{-1} \left(\prod_{v,\gamma,\alpha,b}\hat{\mathcal{A}}_v\hat{\mathcal{A}}_\gamma\hat{\mathcal{B}}_\alpha\hat{\mathcal{B}}_b      \right) \hat T_1\hat T_2 \hat T_3  &=&  \hat T_3^{-1} \hat P \hat T_3\cr
 &=& \prod_{v,\gamma,\alpha,b}\left(\hat T_3^{-1}\hat{\mathcal{A}}_v T_3T_3^{-1}\hat{\mathcal{A}}_\gamma T_3T_3^{-1}\,\hat{\mathcal{B}}_\alpha T_3 T_3^{-1}\,\hat{\mathcal{B}}_b \hat T_3\right)\cr
 &=& \hat P.
\end{eqnarray*}
This  demonstrate that $\hat T_i\, (i=1,2,3)$  let the GS projector $\hat P$ topologically invariant.
\end{proof}

\begin{theorem}{\it  Let $\phi,\psi \in \mathcal{H}^0 $ be two different ground states,  and   for  given transformations $\hat T_i (i=1,2,3) $ of the dressing lattice, we have
\begin{eqnarray}\label{top1}
 \langle \hat T_i\phi |\hat T_i\psi\rangle=\langle \phi |\psi\rangle. 
\end{eqnarray}
}
\end{theorem}
\begin{proof}
  \begin{eqnarray}\label{top1}
 \langle \hat T_i\phi |\hat T_i\psi\rangle=\langle \phi |\hat T_i^\dag\hat T_i\psi\rangle.
\end{eqnarray}
Using the first relation of equation \eqref{t211} such that $\hat T_i^\dag\hat T_i= \mathbb{I}$, we have 
\begin{eqnarray}
\langle \hat T_i\phi |\hat T_i\psi\rangle=\langle \phi |\hat T_i^\dag\hat T_i\psi\rangle =\langle \phi |\psi\rangle. 
\end{eqnarray}
This  demonstrates that $\hat T_i $ are unitary on the ground states.
\end{proof}

 Consequently, there is a bijection between the ground states on any two graphs related by the mutation moves. Since two  such graphs  have the same spatial topology, the GS of our model, is a topological invariant  and a well-defined defined topological observable.

 \end{itemize}

\section{Conclusion }\label{sec5}

%Isolating quantum system  to make it free of decoherence is the real challenge to build a quantum  computer that is exponentially faster than its classical counterpart. Topological   phase of matter is a promising  candidate for  circumventing the problem of decoherence. These phases have quasi particles excitation called anyons. 
%Even though the tech giant Microsoft has recently announced the creation of the first stable topological qubits based on the Majorana quasiparticle \cite{castelvecchi2025microsoft}, there are still many technical challenges to overcome before   the realization of  topological quantum computers. These challenges include the development of advancing mathematical  models  to  manipulate topological states.

 The fault-tolerant quantum computing is due to Alexei Kitaev who proposed  2+1D lattice gauge theory models based on finite 1-group \cite{1}.  In this paper we generalized this model  into 3+1D lattice gauge theory models based on the representation of finite 2-groups. As is well studied in \cite{39,48,49,50,51}, 2-groups can be equivalently defined as a category of crossed modules   or as  2-groupoids.
 In the present context, we  labelled the lattice by a path 2-groupoids which is a strict 2-groupoid. Then, we  gauged this lattice by the representations of  crossed modules over the 2-groupoids called a category of crossed modules. This  consists of labeling the edges described by   path 1-groupoids by the representation of  1-group $G$. The plaquettes described by   path 2-groupoids are labelled  by the representations of a semi-direct product of  $G \ltimes_{\rhd} E$. From this lattice gauge configuration, we described the gauge transforms associated   with the  1-gauge fields and the 2-gauge fields.  We have also
 built gauge invariants from the closed loops around the surfaces and on 
 from closed surfaces.  
 % We also studied the gauge invariants by constructing the plaquette operators and the blob operators associated the 1-holonomy  and the 2-holonomy respectively. 
 From these data,  we constructed  an exactly solvable topological Hamiltonian  that encodes the higher lattice gauge theory. We constructed its ground  states  that encoded the quantum informations and we have shown  that these ground states are consistent against the flips of
the orientations of the edges as well as the orientation and the move base point of each plaquette.

Overall, the result achieved in this study is no longer different from   the ones obtained in \cite{33,40,41}. In the latter references the authors gauged the 3+1D lattice by finite  2-groups interpreted as crossed modules of groups \eqref{def1} while here,  referring on the works \cite{39,48,49,50,51}, we gauged this higher lattice described by  path 2-groupoids by the representations of 2-groups defined as the representations of  a category of crossed modules.   In summary, the current paper’s finding provides an additional method for obtaining the previous in  \cite{33, 40,41}.

\section*{Acknowledgments}
LML is
supported by funding partners through AIMS Ghana and the AIMS
Research and Innovation Centre.
\clearpage
%\section{References}
%\renewcommand{\refname}{}
%\bibliographystyle{ieeetr}
%\bibliography{sourc.bib}

\appendix
\section{Appendix A:  Equivalent definitions of  2-groups }\nonumber\label{sec2}
In this appendix, we provide some (standard) definitions of  2-groups as 2-categories and as crossed modules required for the section \ref{sec3} of the paper. Many of the definitions that appear in this section are adapted from Ref. \cite{57,58,59,60,61,62}.

\subsection{Linear representation  of a group}
Let $k$ be a field and $V$ a $k$-vector space. A linear
representation of a group G is a homomorphism of groups $\rho$ from G to $GL(V)$, where $GL(V)$ is  linear
isomorphisms of $V$ onto itself.

\begin{definition}
{\it Let $G$ be a group and $V$ a $k$-vector space. A $k$-linear representation $(\rho,V)$
of $G$ with representation space $V$ is a group homomorphism
\begin{eqnarray*}
\rho: G\rightarrow GL(V),
\end{eqnarray*}
 and satisfies
\begin{eqnarray}
\rho(g_1g_2)&=&\rho(g_1)\rho(g_2), \quad g_1,g_2\in G,\\
\rho(e)&=&1_V,
\end{eqnarray}}
\end{definition}
where $1_V$ is the identity of the vector space 
Thus, the representation
$\rho$ assigns to any elements $g$ of $G$ a linear isomorphism $\rho(g) : V_k \rightarrow V_k$.\\
\begin{proposition}  
 {\it    Two representations $(\rho_1, V_1)$ and $(\rho_2, V_2)$ of $G$
are said to be equivalent if there exist an isomorphism map or an intertwiner $T : V_1\rightarrow V_2$ satisfies the relation
\begin{eqnarray}
T\rho_1(g)=\rho_2(g)T, \quad \forall g\in G.
\end{eqnarray}}
\end{proposition}

\subsection{ Strict (small) 2-categories}
Informally speaking, a category is a collection of points (objects) and arrows between these points (morphisms) with enough structure so that we can compose arrows and that we have identities, i.e. arrows that behave like neutral elements under composition. The precise
definition is as follows. In all this paper, we restrict ourselves to strict small categories. Strict in the sens  that  equalities of morphisms are satisfied exactly and are not weakened to hold only
up to 2-isomorphism and small in the sens of
the collections of objects  and morphisms form proper sets. See \cite{24,25,26,27,39,52} for a more  a detailed on 2-categories.

\begin{definition}
    {\it   A Strict (small) 2-category \, $\mathcal{C}=(\mathcal{C}_0, \mathcal{C}_1,\mathcal{C}_2,s_v,t_v,s_h,t_h,s,t,\varepsilon,\varepsilon_v,\varepsilon_h,\circ,\circ_v,\circ_h)$ consists of   a set of objects $\mathcal{C}_0$, a set of 1-morphisms $\mathcal{C}_1$, and a set of 2-morphisms $\mathcal{C}_2$ given as  follows
% \begin{center}
% \begin{figure}[htbp]
% \resizebox{0.25\textwidth}{!}{
% \includegraphics {3}
% }
% \label{fig4}      
% \end{figure}
% \end{center}
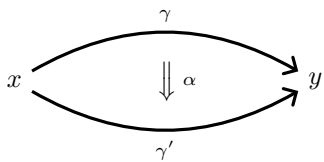
\begin{eqnarray*}
\begin{tikzpicture}[thick]
\node (A) at (-2,0) {$x$};
\node (B) at (2,0) {$y$};
\node at (0,0) {\rotatebox{270}{$\implies$}};
\path[->,very thick, font=\scriptsize,>=angle 90]node[right=0.10cm]{$\alpha$}
(A) edge [bend left] node[above] {$\gamma$} (B)
edge [bend right] node[below] {$\gamma'$} (B);
\end{tikzpicture}      
\end{eqnarray*}

with:
\begin{itemize}
\item The source and target maps
\begin{eqnarray*}
s,t&:& \mathcal{C}_1\rightarrow \mathcal{C}_0, \quad s(\gamma)=x,\quad t(\gamma)=y,         \\
s_h, t_h&:& \mathcal{C}_2 \rightarrow \mathcal{C}_0, \quad
s_h(\alpha)=x, \quad t_h (\alpha)=y,\\
s_v,t_v&:& \mathcal{C}_2\rightarrow \mathcal{C}_1, \quad s_v(\alpha)=\gamma,\quad t_v(\alpha)=\gamma'.
\end{eqnarray*}
\item The composition map  of  1-morphisms
\begin{eqnarray*}
\circ&:& \mathcal{C}_1 \times_{\mathcal{C}_0} \mathcal{C}_1\rightarrow \mathcal{C}_1,\\
(x \xrightarrow{\gamma_1} y)\circ (y\xrightarrow{\gamma_2} z)&=& x\xrightarrow{\gamma_2\circ \gamma_1} z,\quad \forall x,y,z\in \mathcal{C}_0,\,   \gamma_1,\gamma_2\in \mathcal{C}_1.
\end{eqnarray*}
\item Horizontal composition map  of  2-morphisms is given by
\begin{eqnarray*}
\circ_h&:& \mathcal{C}_2 \times_{\mathcal{C}_1} \mathcal{C}_2\rightarrow \mathcal{C}_2,
\end{eqnarray*}
with
\begin{eqnarray*}
\begin{tikzpicture}[ thick]
\node (A) at (-1.5,0) {$x$};
\node (B) at (1.5,0) {$y$};
\node (C)  at (4.5,0){$z$};
\node (E)  at (3,0){$ $};
\node at (0,0) {\rotatebox{270}{$\implies$}}node[right=0.10cm]{$\alpha$};
\path[->] (A) edge [bend left=35] node[above] {$\gamma$} (B);
\path[->] (A) edge [bend right=35] node[below] {$\gamma'$} (B);
\path[->] (B) edge [bend right=25] node[below] {$\chi'$} (C);
\path[->] (B) edge [bend left=25] node[above] {$\chi$} (C);

\node (E) at (3.3,0) {$\beta$};
\node at (3,0) {\rotatebox{270}{$\implies$}};

\node (F)  at (5.5,0){$=$};
\node (G)  at (6.5,0){$x$};
\node (H)  at (11.5,0){$z$};
\node (I) at (9.70,0) {$\beta\circ_h \alpha$};
\node at (9,0) {\rotatebox{270}{$\implies$}}node[right=0.10cm]{$\alpha$};
\path[->] (G) edge [bend left=20] node[above] {$\chi\circ \gamma$} (H);
\path[->] (G) edge [bend right=20] node[below] {$\chi'\circ \gamma'$} (H);

\end{tikzpicture}      
\end{eqnarray*}
such that the source $s_v(\beta\circ_h\alpha)=\chi\circ\gamma$ and the target $t_v(\beta\circ_h\alpha)=\chi'\circ\gamma'$
for all $ \gamma,\gamma',\chi,\chi'\in \mathcal{C}_1 $ and $\alpha,\beta\in \mathcal{C}_2$.  Vertical composition map  of  2-morphisms is  given by
\begin{eqnarray*}
\circ_v&:& \mathcal{C}_2 \times_{\mathcal{C}_1} \mathcal{C}_2\rightarrow \mathcal{C}_2,        
\end{eqnarray*}
\begin{eqnarray*}
\begin{tikzpicture}[thick]
\node (A) at (-2.5,0) {$x$};
\node (B) at (2.5,0) {$y$};
\path[->] (A) edge [bend left=45,""{name=F,}] node[above] {$\gamma$} (B);
\path[->] (A) edge [bend right=45, ""{name=D,  }] node[below] {$\gamma''$} (B);
\path[->] (A) edge[]  node[left=1, above] {$\gamma'$} (B);

\node at (0,0.5) {\rotatebox{270}{$\implies$}} node at (0.3,0.5) {$\alpha$};
\node at (0,-0.5) {\rotatebox{270}{$\implies$}}  node at (0.3,-0.5) {$\beta$};

\node (F)  at (4.5,0){$=$};
\node (G)  at (6.5,0){$x$};
\node (H)  at (11.5,0){$y$};
%\node (I) at (9.70,0) {$\delta\circ_v \xi$};
\node at (9,0) {\rotatebox{270}{$\implies$}} node  at (9.70,0) {$\beta\circ_v \alpha$};
\path[->] (G) edge [bend left=25] node[above] {$\gamma$} (H);
\path[->] (G) edge [bend right=25] node[below] {$\gamma''$} (H);
\end{tikzpicture}      
\end{eqnarray*}
such that the source $s_v(\beta\circ_v\alpha)=\gamma$ and the target $t_v(\beta\circ_v\alpha)=\gamma''$,
for all $ \gamma,\gamma',\gamma''\in \mathcal{C}_1 $ and $\alpha,\beta\in \mathcal{C}_2$.

\item The identity maps
\begin{eqnarray*}
\varepsilon&:& \mathcal{C}_0\rightarrow \mathcal{C}_1, \quad \varepsilon(x)=1_x, \\
\varepsilon_h&:& \mathcal{C}_0\rightarrow \mathcal{C}_2, \quad \varepsilon_h(x)=1_{1_x},\\
\varepsilon_v&:& \mathcal{C}_1\rightarrow \mathcal{C}_2, \quad \varepsilon_v(a)=1_{\gamma},
\end{eqnarray*}
\begin{eqnarray*}
\begin{tikzpicture}[thick]
\node (A) at (-2.3,0) {$x$};
\node (B) at (2.3,0) {$x$};
\node at (0,0) {\rotatebox{270}{$\implies$}};
\path[->, font=\scriptsize,>=angle 90]node[right=0.10cm]{$1_{1_x}$}
(A) edge [bend left=35] node[above] {$1_x$} (B)
edge [bend right=35] node[below] {$1_x$} (B);

\node (F)  at (4.5,0){$and $};
\node (G)  at (6.5,0){$x$};
\node (H)  at (11.5,0){$y$};
\node at (9,0) {\rotatebox{270}{$\implies$}} node  at (9.70,0){$1_\gamma$};
\path[->] (G) edge [bend left=25] node[above] {$\gamma$} (H);
\path[->] (G) edge [bend right=25] node[below] {$\gamma$} (H);

\end{tikzpicture}      
\end{eqnarray*}

\item Vertical and horizontal composition of 2-morphisms obey the interchange
law
\begin{eqnarray}
(\alpha_1\circ_v \beta_1)\circ_h (\alpha_2\circ_v \beta_2)=(\alpha_1\circ_h \beta_1)\circ_v (\alpha_2\circ_h \beta_2)
\end{eqnarray}
so that diagrams of the form
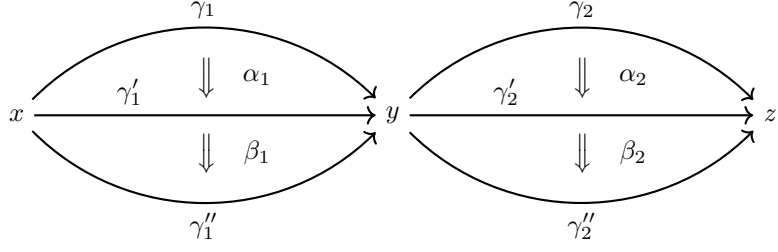
\begin{eqnarray*}
\begin{tikzpicture}[thick]
\node (A) at (-2.5,0) {$x$};
\node (B) at (2.5,0) {$y$};
\node (C) at (7.5,0) {z};
\path[->] (A) edge [bend left=45,""{name=F,}] node[above] {$\gamma_1$} (B);
\path[->] (A) edge [bend right=45, ""{name=D,  }] node[below] {$\gamma_1''$} (B);
\path[->] (A) edge[]  node[left=1, above] {$\gamma_1'$} (B);
\path[->] (B) edge [bend left=45, ""{name=D,  }] node[above] {$\gamma_2$} (C);
\path[->] (B) edge [bend right=45, ""{name=D,  }] node[below] {$\gamma_2''$} (C);
\path[->] (B) edge[]  node[left=1, above] {$\gamma_2'$} (C);

\node at (0,0.5) {\rotatebox{270}{$\implies$}} node at (0.7,0.5) {$\alpha_1$};
\node at (0,-0.5) {\rotatebox{270}{$\implies$}}  node at (0.7,-0.5) {$ \beta_1 $};

\node at (5,0.5) {\rotatebox{270}{$\implies$}} node at (5.7,0.5) {$\alpha_2$};
\node at (5,-0.5) {\rotatebox{270}{$\implies$}} node at (5.7,-0.5) {$\beta_2$};
\end{tikzpicture}      
\end{eqnarray*}
define unambiguous 2-morphisms.
\end{itemize}  

}
\end{definition}

A  2-category in which every  1-morphisms and 2-morphisms are   isomorphisms, is called 2-groupoids $\mathbb{G}=\left(\mathbb{G}_0,\mathbb{G}_1,\mathbb{G}_2\right)$ \cite{53,54,57}. In the next section, we will lay out with the definition of 2-groups.

\subsection{2-groups }
There are equivalent definitions of   2-groups as: 2-categories, categorical  groups and crossed modules \cite{53,54}. In this section, we will consider the definition of 2-groups as 2-categories  and  as crossed modules.
\subsubsection{2-groups as a 2-category}

{\bf Definition 2:} {\it  A 2-group $\mathcal{G}=\left(\mathcal{G}_0,\,\mathcal{G}_1,\,\mathcal{G}_2\right)$ is a small 2-category or a 2-groupoid with a single object $\mathcal{G}_0$,    and such that all morphisms $\mathcal{G}_1$ and 2-morphisms $\mathcal{G}_2$ are invertible.
For each 1-morphism $\gamma\in\mathcal{G}_1 $ there  is an inverse $ \gamma^{-1}\in \mathcal{G}_1$ and   for each 2-morphism $\alpha\in \mathcal{G}_2$ there are two inverses, the vertical inverse $\alpha_v^{-1}\in \mathcal{G}_2$  and the horizontal inverse $\alpha_h^{-1}\in \mathcal{G}_2$. The corresponding inverses maps $\eta_1,\eta_2$ are given by
\begin{eqnarray*}
\eta&:& \mathcal{G}_1\rightarrow\mathcal{G}_1,\quad \eta(\gamma)=\gamma^{-1},\\
\eta_v&:& \mathcal{G}_2\rightarrow \mathcal{G}_2,\quad \eta_v(\alpha)= \alpha_v^{-1},\\
\eta_h&:& \mathcal{G}_2\rightarrow \mathcal{G}_2,\quad \eta_h(\alpha)= \alpha_h^{-1},
\end{eqnarray*}
with:
\begin{eqnarray*}
\begin{tikzpicture}[thick]
\node (A) at (-2.5,0) {$x$};
\node (B) at (2.5,0) {$y$};
\path[->] (A) edge  node[above] {$\gamma$} (B);

\node (F)  at (4.5,0){$\xrightarrow{\eta}  $};
\node (G)  at (6.5,0){$x$};
\node (H)  at (11.5,0){$y$};

\path[<-] (G) edge  node[above] {$\gamma^{-1}$} (H);

\end{tikzpicture}      
\end{eqnarray*}
and
\begin{eqnarray*}
\begin{tikzpicture}[thick]
\node (A) at (-2.5,0) {$x$};
\node (B) at (2.5,0) {$y$};
\path[->] (A) edge [bend left=25,""{name=F,}] node[above] {$\gamma$} (B);
\path[->] (A) edge [bend right=25, ""{name=D,  }] node[below] {$\gamma'$} (B);

\node at (0,0) {\rotatebox{270}{$\implies$}} node at (0.4,0) {$\alpha_v$};

\node (F)  at (4.5,0){$\xrightarrow{\eta_v}  $};
\node (G)  at (6.5,0){$x$};
\node (H)  at (11.5,0){$y$};
\node at (9,0) {\rotatebox{-270}{$\implies$}} node  at (9.4,0) {$\alpha_v^{-1}$};
\path[->] (G) edge [bend left=25] node[above] {$\gamma$} (H);
\path[->] (G) edge [bend right=25] node[below] {$\gamma'$} (H);
\end{tikzpicture}      
\end{eqnarray*}
and

\begin{eqnarray*}
\begin{tikzpicture}[thick]
\node (A) at (-2.5,0) {$x$};
\node (B) at (2.5,0) {$y$};
\path[->] (A) edge [bend left=25,""{name=F,}] node[above] {$\gamma$} (B);
\path[->] (A) edge [bend right=25, ""{name=D,  }] node[below] {$\gamma'$} (B);

\node at (0,0) {\rotatebox{270}{$\implies$}} node at (0.4,0) {$\alpha_h$};

\node (F)  at (4.5,0){$\xrightarrow{\eta_h}  $};
\node (G)  at (6.5,0){$y$};
\node (H)  at (11.5,0){$x$};
\node at (9,0) {\rotatebox{270}{$\implies$}} node  at (9.4,0) {$\alpha_h^{-1}$};
\path[->] (G) edge [bend left=25] node[above] {$\gamma^{-1}$} (H);
\path[->] (G) edge [bend right=25] node[below] {${\gamma'}^{-1}$} (H);
\end{tikzpicture}      
\end{eqnarray*}
such that:
\begin{eqnarray*}
\gamma^{-1}\circ\gamma&=&1_x \quad \mbox{and}\quad \gamma\circ\gamma^{-1}= 1_y, \\
\alpha_v^{-1}\circ_v\alpha_v&=& 1_\gamma\quad\mbox{and}\quad  \alpha_v^{-1}\circ_v\alpha_v=  1_\gamma,\cr
\alpha_h^{-1}\circ_h\alpha_h&=& 1_{1_x}\quad\mbox{and}\quad \alpha_h\circ_h\alpha_h^{-1}= 1_{1_x}.
\end{eqnarray*}}

Given a 2-group $\mathcal{G}$, we can extract from it four pieces of information which form something called
a ‘crossed module’. Conversely, any crossed module gives a 2-group. In fact, 2-groups and crossed
modules are just different ways of describing the same concept. While less elegant than 2-groups,
crossed modules are good for computation, and also good for constructing examples.
\subsubsection{ 2-groups as crossed modules}
Crossed modules were first introduced by J. H. C. Whitehead \cite{60,61}  as a tool for homotopy theories. With Lane Mac  \cite{62}, they used crossed modules to generalize the fundamental  group of a space  to what it
might now call the fundamental 2-group.\\

\begin{definition}\label{def1}  {\it
A crossed module of groups  $\mathcal{X}=\left( E\rightarrow G,\partial,\, \rhd \right) $ consists of groups $G$ and $E$ with two maps $\partial$ and $\rhd$. The first map $\partial$ is a group homomorphism $\partial: E\rightarrow G$ i.e
\begin{eqnarray}
\partial(e_1.e_2)&=& \partial(e_1) \partial(e_2),\quad \quad \forall e_1,e_2\in E,\\
\partial(1_E)&=& 1_G, \quad \quad 1_E\in E, \mbox1_G\in G.
\end{eqnarray}
The  second map  $\rhd$ is    a left action of  G on E by automorphisms i.e  $\rhd: G\times E \rightarrow E $  and  satisfies the  following conditions
\begin{eqnarray}
(g_1g_2)\rhd e&=& g_1\rhd (g_2\rhd e) \quad  \mbox{and}\quad
1_G\rhd e   =  e,\quad \quad  g_1,g_2, 1_G\in G, \quad e\in E,  \\
g\rhd(e_1.e_2)&=&(g\rhd e_1).(g\rhd e_2),\quad \mbox{and} \quad  g\rhd 1_E= g, \quad g\in G, \quad  e_1,e_2,1_E\in E        
\end{eqnarray}
These maps are required to satisfy the following two compatibility conditions called  
the Peiffer relations      
\begin{eqnarray}
\partial(g\rhd e)&=&g\partial(e)g^{-1} \quad \forall\, g\in G \quad \mbox{and}\quad e\in E, \label{Pf1} \\
\partial(e)\rhd f&=&efe^{-1} \quad\quad \forall\, e,f\in E.  \label{Pf2} 
\end{eqnarray}}
\end{definition}

A structure with the same data as a crossed module and satisfying the first condition but not the second
Peiffer  relation is called a pre-crossed module.  Based on the latter two compatibility conditions, we have the following properties of crossed modules.\\

%{\bf Proposition 1:} {\it Let $e,f\in E$ and $g,h \in G$, we have:
%\begin{eqnarray}
%\partial((g\rhd e)( g\rhd f))&=&  \partial(g\rhd(ef))= g\partial(ef)g^{-1},\label{Pei1}\\
%\partial(g\rhd(h\rhd e))&=&\partial((gh)\rhd e) =gh\partial(e)h^{-1} g^{-1}.\label{Pei2}
%\end{eqnarray}}

From  a crossed module  $\mathcal{X}=\left( E\rightarrow G,\partial,\, \rhd \right)$,
we can recover the 2-group  $\mathcal{G}$ structure by  $ \mathcal{G}(\mathcal{X})=\left(\mathcal{G}_0=\{*\},\,\mathcal{G}_1= G,\,\mathcal{G}_2=G \ltimes_{\rhd} E\right)$ as 2-groupoids called category of crossed modules \cite{57,53,54} 
where the semi-direct product $ G \ltimes_{\rhd} E $ with the group multiplication $\circ_h$ given by
\begin{eqnarray*}
(g,e)\circ_h(g',e')=(gg',e(g\rhd e')), \quad g,g'\in G \quad \mbox{and}\quad e,e'\in E.
\end{eqnarray*}

The set of 1-morphisms of $\mathcal{G}(\mathcal{X})$ is given by all arrows of the form  $ *\xrightarrow{g} *$, with $g\in G$, the composition,  inverse composition  and the identity being
\begin{eqnarray}
(*\xrightarrow{g_1} *)\circ (*\xrightarrow{g_2} *)&=& *\xrightarrow{g_2g_1} *,\quad \forall g_1,g_2\in G,\\
\eta(*\xrightarrow{g_1} *\circ*\xrightarrow{g_2} *)&=& *\xrightarrow{g_2^{-1}g_1^{-1}} *,\quad \forall g_1,g_2\in G,\\
\varepsilon(*\xrightarrow{g} *)&=& *\xrightarrow{1_G} * ,\quad\quad\quad  \forall g\in G.
\end{eqnarray}
Let consider  a  pair $u = (g, e) \in G \ltimes_{\rhd} E$ as  a set of  2-morphisms of $\mathcal{G}(\mathcal{X})$ which is pictorially given as follows:
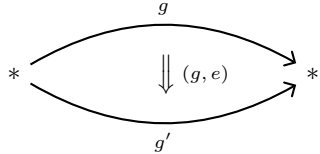
\begin{eqnarray*}
\begin{tikzpicture}[thick]
\node (A) at (-2,0) {$*$};
\node (B) at (2,0) {$*\,$};
\node at (0,0) {\rotatebox{270}{$\implies$}};
\path[->, font=\scriptsize,>=angle 90]node[right=0.10cm]{$(g,e)$}
(A) edge [bend left] node[above] {$g$} (B)
edge [bend right] node[below] {$g'$} (B);
\end{tikzpicture}      
\end{eqnarray*}
where  the source 1-morphism $g$ to the target 1-morphism   $g'=\partial(e)g\in G$ and  $e\in E$. There are horizontal and vertical composition operations for 2-morphisms. A  vertical composition of 2-morphism is given by
\begin{eqnarray*}
\begin{tikzpicture}[thick]
\node (A) at (-2.5,0) {$*$};
\node (B) at (2.5,0) {$*$};
\path[->] (A) edge [bend left=45,""{name=F,}] node[above] {$g$} (B);
\path[->] (A) edge [bend right=45, ""{name=D,  }] node[below] {$g''$} (B);
\path[->] (A) edge[]  node[left=1, above] {$g'$} (B);

\node at (0,0.5) {\rotatebox{270}{$\implies$}} node at (0.7,0.5) {$(g,e)$};
\node at (0,-0.5) {\rotatebox{270}{$\implies$}}  node at (0.7,-0.5) {$(g',e')$};

\node (F)  at (4.5,0){$=$};
\node (G)  at (6.5,0){$*$};
\node (H)  at (11.5,0){$*\,,$};
%\node (I) at (9.70,0) {$\delta\circ_v \xi$};
\node at (9,0) {\rotatebox{270}{$\implies$}} node  at (9.80,0) {$(g', e' e)$};
\path[->] (G) edge [bend left=25] node[above] {$g$} (H);
\path[->] (G) edge [bend right=25] node[below] {$g''$} (H);
\end{tikzpicture}      
\end{eqnarray*}
with $g'=\partial(e)g$ and $g''=\partial(e')\partial(e)g=\partial(e'e)g$. In other words, suppose that we have 2-morphisms
$u = (g, e)$ and $u' = (g', e')$. If $g' = \partial(e)g$, they are vertically composable, and their vertical
composite are given by
\begin{eqnarray}
u' \circ_v u=   (g', e')\circ_v(g, e)  = (g, e)\circ_v(\partial(e)g, e')=  (g',e'e).  
\end{eqnarray}                      
The  vertical identity $\varepsilon_v $ and  the  inverse  maps   are respectively given by
\begin{eqnarray}
\varepsilon_v(u)= (g,1_G)   \quad \mbox{and}\quad  \eta_v(u)=(\partial(e)^{-1}g, e^{-1})
\end{eqnarray}
\begin{eqnarray*}
\begin{tikzpicture}[thick]
\node (A) at (-2.5,0) {$*$};
\node (B) at (2.5,0) {$*$};
\path[->] (A) edge [bend left=45,""{name=F,}] node[above] {$g$} (B);
\path[->] (A) edge [bend right=45, ""{name=D,  }] node[below] {$g$} (B);
\path[->] (A) edge[]  node[left=1, above] {$g'$} (B);

\node at (0,0.5) {\rotatebox{270}{$\implies$}} node at (0.7,0.5) {$(g,e)$};
\node at (0,-0.5) {\rotatebox{270}{$\implies$}}  node at (0.97,-0.5) {$(g',e^{-1})$};

\node (F)  at (4.5,0){$=$};
\node (G)  at (6.5,0){$*$};
\node (H)  at (11.5,0){$* \,.$};
%\node (I) at (9.70,0) {$\delta\circ_v \xi$};
\node at (9,0) {\rotatebox{270}{$\implies$}} node  at (9.64,0) {$(g,1_G)$};
\path[->] (G) edge [bend left=25] node[above] {$g$} (H);
\path[->] (G) edge [bend right=25] node[below] {$g$} (H);
\end{tikzpicture}      
\end{eqnarray*}
Horizontal composition of 2-morphism is given by
\begin{eqnarray}\label{oh}
u'\circ_h u&=& (g',e')\circ_h(g,e)=(g'g,e(g\rhd e')), \quad g',g\in G \quad \mbox{and}\quad e,e'\in E,
\end{eqnarray}
\begin{eqnarray*}
\begin{tikzpicture}[thick]
\node (A) at (-1.5,0) {$*$};
\node (B) at (1.5,0) {$*$};
\node (C)  at (4.8,0){$*$};
\node (E)  at (3,0){$$};
\node at (0,0) {\rotatebox{270}{$\implies$}}node[right=0.10cm]
{$(g,e)$};
\path[->] (A) edge [bend left=35] node[above] {$g$} (B);
\path[->] (A) edge [bend right=35] node[below] {$\partial(e)g$} (B);
\path[->] (B) edge [bend right=35] node[below] {$\partial (e')g'$} (C);
\path[->] (B) edge [bend left=35] node[above] {$g'$} (C);

\node (E) at (3.68,0) {$(g',e')$};
\node at (3,0) {\rotatebox{270}{$\implies$}};

\node (F)  at (5.5,0){$=$};
\node (G)  at (6.5,0){$*$};
\node (H)  at (12.5,0){$*\,.$};
\node (I) at (10.2,0) {$(g,e(g\rhd e'))$};
\node at (9,0) {\rotatebox{270}{$\implies$}}node[right=0.10cm]{$$};
\path[->] (G) edge [bend left=25] node[above] {$g'g$} (H);
\path[->] (G) edge [bend right=25] node[below] {$\partial (e'e)g'g
$} (H);

\end{tikzpicture}      
\end{eqnarray*}
The horizontal inverse and the identity maps are  defined by
\begin{eqnarray}
\eta_h(u)&=&(g^{-1},g^{-1}\rhd e^{-1}),\quad g\in G \quad \mbox{and}\quad e\in E,\\
\varepsilon_h(u)&=&(1_G,1_{1_G}).
\end{eqnarray}
\begin{eqnarray*}
\begin{tikzpicture}[thick]
\node (A) at (-1.5,0) {$*$};
\node (B) at (3,0) {$*$};
\node (C)  at (8.5,0){$*$};
\node (E)  at (,0){$ $};
\node at (0.5,0) {\rotatebox{270}{$\implies$}}node[right=0.50cm]{$(g,e)$};
\path[->] (A) edge [bend left=45] node[above] {$g$} (B);
\path[->] (A) edge [bend right=45] node[below] {$\partial(e)g$} (B);
\path[->] (B) edge [bend right=40] node[below] {$g^{-1}\partial (e^{-1})$} (C);
\path[->] (B) edge [bend left=40] node[above] {$g^{-1}$} (C);

\node (E) at (6.7,0) {$(g^{-1},g^{-1}\rhd e^{-1})$};
\node at (5,0.0) {\rotatebox{270}{$\implies$}};

\node (F)  at (9.09,0){$=$};
\node (G)  at (9.5,0){$*$};
\node (H)  at (13.5,0){$*\, .$};
\node (I) at (12.40,0) {$(1_G,1_{1_G})$};
\node at (11.6,0) {\rotatebox{270}{$\implies$}}node[right=0.10cm]{$ $};
\path[->] (G) edge [bend left=45] node[above] {$1_G$} (H);
\path[->] (G) edge [bend right=45] node[below] {$1_G
$} (H);

\end{tikzpicture}      
\end{eqnarray*}

Vertical and horizontal compositions can be checked to satisfy the interchange law
\begin{eqnarray}
(u_1'\circ_v u_1)\circ_h(u_2'\circ_v u_2)= (u_1'\circ_h u_1)\circ_v(u_2'\circ_h u_2),
\end{eqnarray}
such that there is a well-defined 2-morphism associated with the diagram
\begin{eqnarray*}
\begin{tikzpicture}[thick]
\node (A) at (-2.5,0) {$*$};
\node (B) at (2.5,0) {$*$};
\node (C) at (7.5,0) {*\,,};
\path[->] (A) edge [bend left=45,""{name=F,}] node[above] {$g_1$} (B);
\path[->] (A) edge [bend right=45, ""{name=D,  }] node[below] {$g_1''$} (B);
\path[->] (A) edge[]  node[left=1, above] {$g_1'$} (B);
\path[->] (B) edge [bend left=45, ""{name=D,  }] node[above] {$g_2$} (C);
\path[->] (B) edge [bend right=45, ""{name=D,  }] node[below] {$g_2''$} (C);
\path[->] (B) edge[]  node[left=1, above] {$g_2'$} (C);

\node at (0,0.5) {\rotatebox{270}{$\implies$}} node at (0.7,0.5) {$(g_1,e_1)$};
\node at (0,-0.5) {\rotatebox{270}{$\implies$}}  node at (0.7,-0.5) {$(g_1',e_1')$};

\node at (5,0.5) {\rotatebox{270}{$\implies$}} node at (5.7,0.5) {$(g_2,e_2)$};
\node at (5,-0.5) {\rotatebox{270}{$\implies$}} node at (5.7,-0.5) {$(g_2',e_2')$};
\end{tikzpicture}      
\end{eqnarray*}
independent of the order of composition.
We can define a
composition between a 1-morphism and a 2-morphism which is  an intermediate step to defining horizontal composition for 2-morphism. This an intermediate step is called a whisker.  Let $ g \xRightarrow[]{(g,e)} \partial(e)g\in G \ltimes_{\rhd} E $ be the 2-morphism and $ *\xrightarrow{g'}*\in G$ be the 1-morphism, the
left whiskering of $g'$  on  $(g,e)$  is given by $(g,g'\rhd e):g'g\Rightarrow \partial(e)g'g$. This terminology can be explained by the picture

\begin{eqnarray*}
\begin{tikzpicture}[thick]
\node (A) at (-1.7,0) {$*$};
\node (B) at (2.5,0) {$*$};
\node (C)  at (4.05,0){$*$};
\node (E)  at (3,0){$$};
\node at (0.4,0) {\rotatebox{270}{$\implies$}}node[right=0.550cm]{$(g,e)$};
\path[->] (A) edge [bend left=35] node[above] {$g$} (B);
\path[->] (A) edge [bend right=35] node[below] {$\partial(e)g$} (B);
\draw[->] (B) --node[above=0.01cm] {$g'$} (C) (C);

\node (F)  at (5.5,0){=};
\node (G)  at (6.5,0){$*$};
\node (H)  at (11.5,0){$*\, .$};
\node (I) at (10.0,0) {$(g, e)$};
\node at (9,0) {\rotatebox{270}{$\implies$}}node[right=0.10cm]{$ $};
\path[->] (G) edge [bend left=25] node[above] {$gg'$} (H);
\path[->] (G) edge [bend right=25] node[below] {$\partial (e)gg'
$} (H);

\end{tikzpicture}
\end{eqnarray*}
The attachment of the left whisker can be understood as a special case of the horizontal
composition \eqref{oh} in which $g= g'$ and $e' = 1$ so that the left surface collapses to a line.
Similarly, the right whiskering of a 1-morphism  $ *\xrightarrow{g'}*\in G$ on 2-morphism $ g \xRightarrow[]{(g,e)} \partial(e)g\in G \ltimes_{\rhd} E $ is given by  $(g,{g'}\rhd e):gg'\Rightarrow \partial(e)gg'$. This terminology can be explained by the picture
\begin{eqnarray*}
\begin{tikzpicture}[thick]
\node (A0) at (-3,0) {$*$};
\node (A) at (-1.5,0) {$*$};
\node (B) at (3.5,0) {$*$};
% \node (C)  at (4.5,0){$*.$};
\node (E)  at (3,0){$ $};
\node at (0.9,0) {\rotatebox{270}{$\implies$}}node[right=0.99cm]{$(g,e)$};
\path[->] (A) edge [bend left=25] node[above] {$g$} (B);
\path[->] (A) edge [bend right=25] node[below] {$\partial(e)g$} (B);
\draw[->] (A0) --node[above=0.01cm] {$g'$} (A) ;

\node (F)  at (5.5,0){$=$};
\node (G)  at (6.5,0){$*$};
\node (H)  at (11.5,0){$*\,.$};
\node (I) at (10.1,0) {$(g,{g'}\rhd e)$};
\node at (9,0) {\rotatebox{270}{$\implies$}}node[right=0.10cm]{$$};
\path[->] (G) edge [bend left=25] node[above] {$gg'$} (H);
\path[->] (G) edge [bend right=25] node[below] {$\partial (e)gg'
$} (H);

\end{tikzpicture}      
\end{eqnarray*}

\section{ Appendix B: Proof of Lemmas }\label{b}

\subsection{Proof of Lemma \eqref{lem1}}
In this appendix, we provide the proofs of lemmas of section \eqref{sec4}. Let us recall the  Hilbert spaces in which are the defined  the lattice models such that
 \begin{equation*}\label{HI}
 \mathcal{H}(M,L,\mathcal{G}(\mathcal{X})):= \mbox{span}\left \{\left|\bigotimes_{\gamma\in L^1} \Phi_{\gamma}(g) \bigotimes_{\alpha\in L^2} \Phi_{\alpha}(u)  \right \rangle\right\}.
 \end{equation*}
 For the sake of simplicity, we define the Hilbert space from the following a configuration graph Fig\eqref{en61}. The vector basis of this graph is given by 
 \begin{eqnarray*}
   \mathcal{H}(M,L,\mathcal{G}(\mathcal{X}))= \mathcal{H}(M,L,\mathcal{G}(\mathcal{X})):= \{\mbox{span}\left|\Phi_{\gamma_{\alpha1}}(g_{\alpha1})\otimes \Phi_{\gamma_{\alpha2}}(g_{\alpha2})\otimes \Phi_{\gamma_{23}}(g_{23}) \otimes  \Phi_{\gamma_{13}}(g_{13})\otimes \Phi_\alpha(u)\right\rangle \}
 \end{eqnarray*}
 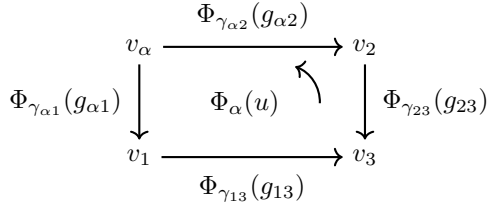
\begin{figure}
\centering
\begin{tikzpicture}[thick]
\node (A) at (-1.5,0) {$v_\alpha$};
\node (B) at (1.5,0) {$v_2$};
\node (C) at (4.5,0) [below=of A] {$v_1$};
\node (D) at (4.5,0)   [below=of B] {$v_3$};

\draw[->](A)--node [left=0.1cm] {$\Phi_{\gamma_{\alpha 1}}(g_{\alpha1})$}(C);
\draw[->](B)--node [right=0.1cm] {$\Phi_{\gamma_{23}}(g_{23})$}(D);
\draw[->](A)--node [above=0.1cm] {$\Phi_{\gamma_{\alpha 2}}(g_{\alpha2})$}(B);
\draw[->](C)--node [below=0.1cm] {$\Phi_{\gamma_{1 3}}(g_{13})$}(D);
        \draw[ thick, ->] (0.9,-0.75) arc (-2:70:0.5);
\node at (-0.1,-0.75){$\Phi_\alpha(u)$};
%\node (F)  at (3.9,-0.5){$\equiv $};

%node (G)  at (8.5,-0.5)
\end{tikzpicture}
 \caption{A configurative graph}\label{en61} 
\end{figure}
The actions of vertex transforms $\hat A_{v_\alpha}^h$ and $\hat A_{v_\alpha}^{h’}$ at the base point of the plaquette $v_{\alpha}$ read as follows
\begin{eqnarray*}
 \hat A_{v_\alpha}^h \left|\Phi_{\gamma_{\alpha1}}(g_{\alpha1}) \Phi_{\gamma_{\alpha2}}(g_{\alpha2}) \Phi_{\gamma_{23}}(g_{23})   \Phi_{\gamma_{13}}(g_{13}) \Phi_\alpha(u)\right\rangle &=& \left|h\Phi_{\gamma_{\alpha1}}(g_{\alpha1}) h\Phi_{\gamma_{\alpha2}}(g_{\alpha2}) \Phi_{\gamma_{23}}(g_{23})   \Phi_{\gamma_{13}}(g_{13}) h\rhd\Phi_\alpha(u)\right\rangle \\
  \hat A_{v_\alpha}^{h’}\hat A_{v_\alpha}^h \left|\Phi_{\gamma_{\alpha1}}(g_{\alpha1}) \Phi_{\gamma_{\alpha2}}(g_{\alpha2}) \Phi_{\gamma_{23}}(g_{23})   \Phi_{\gamma_{13}}(g_{13}) \Phi_\alpha(u)\right\rangle &=& |h’h\Phi_{\gamma_{\alpha1}}(g_{\alpha1}) h’h\Phi_{\gamma_{\alpha2}}(g_{\alpha2}) \Phi_{\gamma_{23}}(g_{23})  \cr&& \Phi_{\gamma_{13}}(g_{13}) (h’h)\rhd\Phi_\alpha(u)\rangle\\
  &=&  \hat A_{v_\alpha}^{h’h}\left|\Phi_{\gamma_{\alpha1}}(g_{\alpha1}) \Phi_{\gamma_{\alpha2}}(g_{\alpha2}) \Phi_{\gamma_{23}}(g_{23})   \Phi_{\gamma_{13}}(g_{13})\Phi_\alpha(u)\right\rangle.
\end{eqnarray*}
This gives the proof of the first equation \eqref{a40} at the base-point $v=v_\alpha$,
\begin{eqnarray*}
   \hat A_{v_\alpha}^{h’}\hat A_{v_\alpha}^h=  \hat A_{v_\alpha}^{h’h}.
\end{eqnarray*}
This result is satisfied for any vertices of the above graph. The second equation \eqref{a40} is proved  as follows. The vertex transfoms at different vertices $ v_\alpha$ and $ v_2$ are given by
\begin{eqnarray*}
    \hat A_{v_\alpha}^{h}\hat A_{v_2}^{h’} \left|\Phi_{\gamma_{\alpha1}}(g_{\alpha1}) \Phi_{\gamma_{\alpha2}}(g_{\alpha2}) \Phi_{\gamma_{23}}(g_{23})   \Phi_{\gamma_{13}}(g_{13}) \Phi_\alpha(u)\right\rangle &=&  |h\Phi_{\gamma_{\alpha1}}(g_{\alpha1}) hh’\Phi_{\gamma_{\alpha2}}(g_{\alpha2}) (h’)^{-1}\cr&&\Phi_{\gamma_{23}}(g_{23})   \Phi_{\gamma_{13}}(g_{13}) h\rhd\Phi_\alpha(u)\rangle.
\end{eqnarray*}
On the other hand, we have

\begin{eqnarray*}
  \hat A_{v_2}^{h’} \hat A_{v_\alpha}^{h} \left|\Phi_{\gamma_{\alpha1}}(g_{\alpha1}) \Phi_{\gamma_{\alpha2}}(g_{\alpha2}) \Phi_{\gamma_{23}}(g_{23})   \Phi_{\gamma_{13}}(g_{13}) \Phi_\alpha(u)\right\rangle &=&  |h\Phi_{\gamma_{\alpha1}}(g_{\alpha1}) hh’\Phi_{\gamma_{\alpha2}}(g_{\alpha2}) (h’)^{-1}\cr&&\Phi_{\gamma_{23}}(g_{23})   \Phi_{\gamma_{13}}(g_{13}) h\rhd\Phi_\alpha(u)\rangle.
\end{eqnarray*}
From the latter two equations, we deduce that at the different vertices of the graph, the vertex transforms  commute such that
\begin{eqnarray*}
  \hat A_{v_\alpha}^{h}\hat A_{v_2}^{h’}=   \hat A_{v_2}^{h’} \hat A_{v_\alpha}^{h}\implies [\hat A_{v_\alpha}^{h},\hat A_{v_2}^{h’}] =0
\end{eqnarray*}

\subsection{Proof of Lemma \eqref{lem2}}
We still consider the Hilbert space spanned by vectors of the above graph defined as
$$\mathcal{H}(M,L,\mathcal{G}(\mathcal{X})):= \{\mbox{span}\left|\Phi_{\gamma_{\alpha1}}(g_{\alpha1})\otimes \Phi_{\gamma_{\alpha2}}(g_{\alpha2})\otimes \Phi_{\gamma_{23}}(g_{23}) \otimes  \Phi_{\gamma_{13}}(g_{13})\otimes \Phi_\alpha(u)\right\rangle \}.$$
We consider the edges transforms $\hat A_{\gamma_{13}}^e$ and $\hat A_{\gamma_{13}}^{e'} $
on the edges $\gamma_{13}$ are given by
\begin{eqnarray*}
    \hat A_{\gamma_{13}}^e \left|\Phi_{\gamma_{\alpha1}}(g_{\alpha1}) \Phi_{\gamma_{\alpha2}}(g_{\alpha2}) \Phi_{\gamma_{23}}(g_{23})   \Phi_{\gamma_{13}}(g_{13}) \Phi_\alpha(u)\right\rangle &=&  |\Phi_{\gamma_{\alpha1}}(g_{\alpha1}) \Phi_{\gamma_{\alpha2}}(g_{\alpha2}) \Phi_{\gamma_{23}}(g_{23}) \cr&& \partial(e)\Phi_{\gamma_{13}}(g_{13}) \Phi_\alpha(u)[p^+(v_\alpha\rightarrow s(\gamma_{13}))\rhd e^{-1}] \rangle,\\
  \hat A_{\gamma_{13}}^{e'}  \hat A_{\gamma_{13}}^e \left|\Phi_{\gamma_{\alpha1}}(g_{\alpha1}) \Phi_{\gamma_{\alpha2}}(g_{\alpha2}) \Phi_{\gamma_{23}}(g_{23})   \Phi_{\gamma_{13}}(g_{13}) \Phi_\alpha(u)\right\rangle &=& |\Phi_{\gamma_{\alpha1}}(g_{\alpha1}) \Phi_{\gamma_{\alpha2}}(g_{\alpha2}) \Phi_{\gamma_{23}}(g_{23}) \cr&& \partial(e'e)\Phi_{\gamma_{13}}(g_{13}) \Phi_\alpha(u)[p^+(v_\alpha\rightarrow s(\gamma_{13}))\rhd  {e'}^{-1}]
 \cr&&  [p^+(v_\alpha\rightarrow s(\gamma_{13}))\rhd  e^{-1}] \rangle,\\
 &=& \hat A_{\gamma_{13}}^{e'e}|\Phi_{\gamma_{\alpha1}}(g_{\alpha1}) \Phi_{\gamma_{\alpha2}}(g_{\alpha2}) \Phi_{\gamma_{23}}(g_{23}) \cr&&   \Phi_{\gamma_{13}}(g_{13}) \Phi_\alpha(u)\rangle,
\end{eqnarray*}
where $ p^+(v_\alpha\rightarrow s(\gamma_{13}))= \Phi_{\gamma_{\alpha1}}(g_{\alpha1}) $.  This results of
\begin{eqnarray*}
    \hat A_{\gamma_{13}}^e \hat A_{\gamma_{13}}^{e'}= \hat A_{\gamma_{13}}^{ee'}.
\end{eqnarray*}
\begin{eqnarray*}
    \hat A_{\gamma_{13}}^e \left|\Phi_{\gamma_{\alpha1}}(g_{\alpha1}) \Phi_{\gamma_{\alpha2}}(g_{\alpha2}) \Phi_{\gamma_{23}}(g_{23})   \Phi_{\gamma_{13}}(g_{13}) \Phi_\alpha(u)\right\rangle &=&  |\Phi_{\gamma_{\alpha1}}(g_{\alpha1}) \Phi_{\gamma_{\alpha2}}(g_{\alpha2}) \Phi_{\gamma_{23}}(g_{23}) \cr&& \partial(e)\Phi_{\gamma_{13}}(g_{13}) \Phi_\alpha(u)[p^+(v_\alpha\rightarrow s(\gamma_{13}))\rhd e^{-1}] \rangle,\\
  \hat A_{\gamma_{13}}^{e'}  \hat A_{\gamma_{13}}^e \left|\Phi_{\gamma_{\alpha1}}(g_{\alpha1}) \Phi_{\gamma_{\alpha2}}(g_{\alpha2}) \Phi_{\gamma_{23}}(g_{23})   \Phi_{\gamma_{13}}(g_{13}) \Phi_\alpha(u)\right\rangle &=& |\Phi_{\gamma_{\alpha1}}(g_{\alpha1}) \Phi_{\gamma_{\alpha2}}(g_{\alpha2}) \Phi_{\gamma_{23}}(g_{23}) \cr&& \partial(e'e)\Phi_{\gamma_{13}}(g_{13}) \Phi_\alpha(u)[p^+(v_\alpha\rightarrow s(\gamma_{13}))\rhd  {e'}^{-1}]
 \cr&&  [p^+(v_\alpha\rightarrow s(\gamma_{13}))\rhd  e^{-1}] \rangle,\\
 &=& \hat A_{\gamma_{13}}^{e'e}|\Phi_{\gamma_{\alpha1}}(g_{\alpha1}) \Phi_{\gamma_{\alpha2}}(g_{\alpha2}) \Phi_{\gamma_{23}}(g_{23}) \cr&&   \Phi_{\gamma_{13}}(g_{13}) \Phi_\alpha(u)\rangle,
\end{eqnarray*}
where $ p^+(v_\alpha\rightarrow s(\gamma_{13}))= \Phi_{\gamma_{\alpha1}}(g_{\alpha1})$.  This results of
\begin{eqnarray*}
    \hat A_{\gamma_{13}}^e \hat A_{\gamma_{13}}^{e'}= \hat A_{\gamma_{13}}^{ee'}.
\end{eqnarray*}
The edges transforms  
on the edges $\gamma_{23}$ is given by
\begin{eqnarray*}
    \hat A_{\gamma_{23}}^e \left|\Phi_{\gamma_{\alpha1}}(g_{\alpha1}) \Phi_{\gamma_{\alpha2}}(g_{\alpha2}) \Phi_{\gamma_{23}}(g_{23})   \Phi_{\gamma_{13}}(g_{13}) \Phi_\alpha(u)\right\rangle &=&  |\Phi_{\gamma_{\alpha1}}(g_{\alpha1}) \Phi_{\gamma_{\alpha2}}(g_{\alpha2}) \partial(e)\Phi_{\gamma_{23}}(g_{23}) \cr&& \Phi_{\gamma_{13}}(g_{13}) [p^-(v_\alpha\rightarrow s(\gamma_{23}))\rhd e]\Phi_\alpha(u) \rangle,\\
  \hat A_{\gamma_{23}}^{e'}  \hat A_{\gamma_{23}}^e \left|\Phi_{\gamma_{\alpha1}}(g_{\alpha1}) \Phi_{\gamma_{\alpha2}}(g_{\alpha2}) \Phi_{\gamma_{23}}(g_{23})   \Phi_{\gamma_{13}}(g_{13}) \Phi_\alpha(u)\right\rangle &=& |\Phi_{\gamma_{\alpha1}}(g_{\alpha1}) \Phi_{\gamma_{\alpha2}}(g_{\alpha2}) \partial(e'e)\Phi_{\gamma_{23}}(g_{23}) \cr&& \Phi_{\gamma_{13}}(g_{13}) [p^-(v_\alpha\rightarrow s(\gamma_{13}))\rhd e']\cr&&[p^-(v_\alpha\rightarrow s(\gamma_{13}))\rhd e]\Phi_\alpha(u) \rangle,\\
 &=& \hat A_{\gamma_{23}}^{e'e}|\Phi_{\gamma_{\alpha1}}(g_{\alpha1}) \Phi_{\gamma_{\alpha2}}(g_{\alpha2}) \Phi_{\gamma_{23}}(g_{23}) \cr&&   \Phi_{\gamma_{13}}(g_{13}) \Phi_\alpha(u)\rangle,
\end{eqnarray*}
where $ p^-(v_\alpha\rightarrow s(\gamma_{13}))= \Phi_{\gamma_{\alpha2}}(g_{\alpha 2})$  . This results of
\begin{eqnarray*}
    \hat A_{\gamma_{23}}^e \hat A_{\gamma_{23}}^{e'}= \hat A_{\gamma_{23}}^{ee'}.
\end{eqnarray*}
This result is satisfied for any edges of the above graph. The second equation \eqref{b} is proved  as follows.  Let us consider the edges $\gamma_{\alpha1} $ and $\gamma_{23}$. The action of  edge transforms $\hat A_{\gamma_{\alpha 1}}^e$ and $\hat A_{\gamma_{23}}^{e'}$ on the states $ \left|\Phi_{\gamma_{\alpha1}}(g_{\alpha1}) \Phi_{\gamma_{\alpha2}}(g_{\alpha2}) \Phi_{\gamma_{23}}(g_{23})   \Phi_{\gamma_{13}}(g_{13}) \Phi_\alpha(u)\right\rangle$ are given by
\begin{eqnarray*}
   \hat A_{\gamma_{\alpha 1}}^e \left|\Phi_{\gamma_{\alpha1}}(g_{\alpha1}) \Phi_{\gamma_{\alpha2}}(g_{\alpha2}) \Phi_{\gamma_{23}}(g_{23})   \Phi_{\gamma_{13}}(g_{13}) \Phi_\alpha(u)\right\rangle &=&  |\partial(e)\Phi_{\gamma_{\alpha1}}(g_{\alpha1}) \Phi_{\gamma_{\alpha2}}(g_{\alpha2})\cr&& \Phi_{\gamma_{23}}(g_{23})   \Phi_{\gamma_{13}}(g_{13}) \Phi_\alpha(u)\cr&&[p^+(v_\alpha\rightarrow s(\gamma_{\alpha1}))\rhd e^{-1}]\rangle\\
  \hat A_{\gamma_{23}}^{e'}\hat A_{\gamma_{\alpha 1}}^e  \left|\Phi_{\gamma_{\alpha1}}(g_{\alpha1}) \Phi_{\gamma_{\alpha2}}(g_{\alpha2}) \Phi_{\gamma_{23}}(g_{23})   \Phi_{\gamma_{13}}(g_{13}) \Phi_\alpha(u)\right\rangle &=&  |\partial(e)\Phi_{\gamma_{\alpha1}}(g_{\alpha1}) \Phi_{\gamma_{\alpha2}}(g_{\alpha2})\cr&& \partial (e')\Phi_{\gamma_{23}}(g_{23})   \Phi_{\gamma_{13}}(g_{13})\cr&& [p^-(v_\alpha\rightarrow  s(\gamma_{23}))\rhd e']\Phi_\alpha(u)\cr&&[p^+(v_\alpha\rightarrow  s(\gamma_{\alpha1}))\rhd e^{-1}]\rangle.
\end{eqnarray*}
The action of $   \hat A_{\gamma_{\alpha 1}}^e\hat A_{\gamma_{23}}^{e'} $ reads as follows
\begin{eqnarray*}
  \hat A_{\gamma_{23}}^{e'} \left|\Phi_{\gamma_{\alpha1}}(g_{\alpha1}) \Phi_{\gamma_{\alpha2}}(g_{\alpha2}) \Phi_{\gamma_{23}}(g_{23})   \Phi_{\gamma_{13}}(g_{13}) \Phi_\alpha(u)\right\rangle &=&  |\Phi_{\gamma_{\alpha1}}(g_{\alpha1}) \Phi_{\gamma_{\alpha2}}(g_{\alpha2})\cr&& \partial (e')\Phi_{\gamma_{23}}(g_{23})   \Phi_{\gamma_{13}}(g_{13})\cr&&[p^-(v_\alpha\rightarrow s(\gamma_{23}))\rhd e'] \Phi_\alpha(u)\rangle\\
  \hat A_{\gamma_{\alpha 1}}^e\hat A_{\gamma_{23}}^{e'}  \left|\Phi_{\gamma_{\alpha1}}(g_{\alpha1}) \Phi_{\gamma_{\alpha2}}(g_{\alpha2}) \Phi_{\gamma_{23}}(g_{23})   \Phi_{\gamma_{13}}(g_{13}) \Phi_\alpha(u)\right\rangle &=&  |\partial(e)\Phi_{\gamma_{\alpha1}}(g_{\alpha1}) \Phi_{\gamma_{\alpha2}}(g_{\alpha2})\cr&& \partial (e')\Phi_{\gamma_{23}}(g_{23})   \Phi_{\gamma_{13}}(g_{13}) [p^-(v_\alpha\rightarrow s(\gamma_{23}))\rhd e']\cr&&\Phi_\alpha(u)[p^+(v_\alpha\rightarrow s(\gamma_{\alpha1}))\rhd e^{-1}]\rangle.
\end{eqnarray*}
From the latter two equations, we deduce that, the edges transforms  for different edges  $\gamma_{\alpha1} $ and $\gamma_{23}$ of the graph commute such that
\begin{eqnarray*}\label{b}
\hat A_{\gamma_{\alpha 1}}^e\hat A_{\gamma_{23}}^{e'} = \hat A_{\gamma_{23}}^{e'}\hat A_{\gamma_{\alpha 1}}^e\implies    [\hat A_{\gamma_{\alpha 1}}^e,\hat A_{\gamma_{23}}^{e'} ]= 0, \quad \forall\, e,e'\in   E.
\end{eqnarray*}

\subsection{Proof of  Lemma \eqref{lem3}  }
We  consider the Hilbert space spanned by vector of the above graph Fig \eqref{en61}  defined as
$$\mathcal{H}(M,L,\mathcal{G}(\mathcal{X})):= \{\mbox{span}\left|\Phi_{\gamma_{\alpha1}}(g_{\alpha1})\otimes \Phi_{\gamma_{\alpha2}}(g_{\alpha2})\otimes \Phi_{\gamma_{23}}(g_{23}) \otimes  \Phi_{\gamma_{13}}(g_{13})\otimes \Phi_\alpha(u)\right\rangle \}.$$
Based on this representation, we consider  the vertex $v_\alpha$ as the starting  vertex of the edge $\gamma_{\alpha1}$. We compute $ \hat A_{v_\alpha}^h$ and $\hat A_{\gamma_{\alpha2}}^{e}$. We have
\begin{eqnarray*}
   A_{\gamma_{\alpha2}}^{e}\left|\Phi_{\gamma_{\alpha1}}(g_{\alpha1}) \Phi_{\gamma_{\alpha2}}(g_{\alpha2}) \Phi_{\gamma_{23}}(g_{23})   \Phi_{\gamma_{13}}(g_{13}) \Phi_\alpha(u)\right\rangle &=&  |\Phi_{\gamma_{\alpha1}}(g_{\alpha1})\partial(e) \Phi_{\gamma_{\alpha2}}(g_{\alpha2}) \Phi_{\gamma_{23}}(g_{23})\cr &&  \Phi_{\gamma_{13}}(g_{13}) [p^-(v_\alpha\rightarrow s(\gamma_{\alpha2})\rhd e]\Phi_\alpha(u)\rangle \\
  \hat A_{v_\alpha}^h  A_{\gamma_{\alpha2}}^{e}\left|\Phi_{\gamma_{\alpha1}}(g_{\alpha1}) \Phi_{\gamma_{\alpha2}}(g_{\alpha2}) \Phi_{\gamma_{23}}(g_{23})   \Phi_{\gamma_{13}}(g_{13}) \Phi_\alpha(u)\right\rangle &=&  |h\Phi_{\gamma_{\alpha1}}(g_{\alpha1})\partial(h\rhd e) \Phi_{\gamma_{\alpha2}}(g_{\alpha2}) \Phi_{\gamma_{23}}(g_{23})\cr &&  \Phi_{\gamma_{13}}(g_{13}) [p^-(v_\alpha\rightarrow s(\gamma_{\alpha2}))\rhd e](h\rhd\Phi_\alpha(u))\rangle.
\end{eqnarray*}
The action of $ \hat A_\gamma^{h\rhd e} \hat A_v^h $ reads as follows

\begin{eqnarray*}
  \hat A_{v_\alpha}^h\left|h\Phi_{\gamma_{\alpha1}}(g_{\alpha1}) \Phi_{\gamma_{\alpha2}}(g_{\alpha2}) \Phi_{\gamma_{23}}(g_{23})   \Phi_{\gamma_{13}}(g_{13}) \Phi_\alpha(u)\right\rangle &=&  |h\Phi_{\gamma_{\alpha1}}(g_{\alpha1})h \Phi_{\gamma_{\alpha2}}(g_{\alpha2}) \Phi_{\gamma_{23}}(g_{23})\cr &&  \Phi_{\gamma_{13}}(g_{13}) (h\rhd\Phi_\alpha(u))\rangle \\
   \hat A_\gamma^{h\rhd e} \hat A_v^h\left|\Phi_{\gamma_{\alpha1}}(g_{\alpha1}) \Phi_{\gamma_{\alpha2}}(g_{\alpha2}) \Phi_{\gamma_{23}}(g_{23})   \Phi_{\gamma_{13}}(g_{13}) \Phi_\alpha(u)\right\rangle &=&  |h\Phi_{\gamma_{\alpha1}}(g_{\alpha1})\partial(h\rhd e) \Phi_{\gamma_{\alpha2}}(g_{\alpha2}) \Phi_{\gamma_{23}}(g_{23})\cr &&  \Phi_{\gamma_{13}}(g_{13}) [p^-(v_\alpha\rightarrow s(\gamma_{\alpha2})\rhd e](h\rhd\Phi_\alpha(u))\rangle.
\end{eqnarray*}
This show that  for a vertex $v_\alpha$ which is the starting  vertex of  the edge $\gamma_{\alpha2}$ gives
\begin{eqnarray*}
   \hat A_{\gamma_{\alpha2}}^{h\rhd e} \hat A_v^h=  \hat A_{v_\alpha}^h  A_{\gamma_{\alpha2}}^{e}.
\end{eqnarray*}
This is true for any vertex $v$ which is  the starting vertex of $ \gamma_{\alpha2}   $
\begin{eqnarray*}\label{a}
\hat A_v^h \hat A_\gamma^{e}&=& \hat A_\gamma^{h\rhd e} \hat A_v^h,\quad  \text{if $v$ is the starting vertex of $\gamma$}. %\label{a}\\
% \hat A_v^h \hat A_\gamma^{e}&=& \hat A_\gamma^{e} \hat A_v^h,\quad \quad \, \text{if v is not the starting vertex of $\gamma$.} \label{a2}
\end{eqnarray*}
However for any vertex $v$ which is the starting vertex of $\gamma$, the edge transform and the vertex transform commutes. For example, we consider the vertex transform $ \hat A_{v_\alpha}^h$ at the base point  ${v_\alpha}$  and the edge transform $ \hat A_{\gamma_{23}}^{e}$

\begin{eqnarray*}
  \hat A_{v_\alpha}^h\left|\Phi_{\gamma_{\alpha1}}(g_{\alpha1}) \Phi_{\gamma_{\alpha2}}(g_{\alpha2}) \Phi_{\gamma_{23}}(g_{23})   \Phi_{\gamma_{13}}(g_{13}) \Phi_\alpha(u)\right\rangle &=&  |h\Phi_{\gamma_{\alpha1}}(g_{\alpha1})h \Phi_{\gamma_{\alpha2}}(g_{\alpha2}) \Phi_{\gamma_{23}}(g_{23})\cr &&  \Phi_{\gamma_{13}}(g_{13}) (h\rhd\Phi_\alpha(u))\rangle \\
   A_{\gamma_{23}}^{e} \hat A_{v_\alpha}^h\left|\Phi_{\gamma_{\alpha1}}(g_{\alpha1}) \Phi_{\gamma_{\alpha2}}(g_{\alpha2}) \Phi_{\gamma_{23}}(g_{23})   \Phi_{\gamma_{13}}(g_{13}) \Phi_\alpha(u)\right\rangle &=&  |h\Phi_{\gamma_{\alpha1}}(g_{\alpha1}) h\Phi_{\gamma_{\alpha2}}(g_{\alpha2}) \partial(e)\Phi_{\gamma_{23}}(g_{23})\cr &&  \Phi_{\gamma_{13}}(g_{13}) [p^-(v_\alpha\rightarrow s(\gamma_{23}))\rhd e](h\rhd\Phi_\alpha(u))\rangle,\\
   &=& \hat A_{v_\alpha}^h  A_{\gamma_{23}}^{e}   \left|\Phi_{\gamma_{\alpha1}}(g_{\alpha1})  \Phi_{\gamma_{23}}(g_{23})   \Phi_{\gamma_{13}}(g_{13}) \Phi_\alpha(u)\right\rangle.
\end{eqnarray*}
This results of the  equation
\begin{eqnarray*}
   A_{\gamma_{23}}^{e} \hat A_{v_\alpha}^h= \hat A_{v_\alpha}^h  A_{\gamma_{23}}^{e}.
\end{eqnarray*}
For a vertex $v$ which is not the starting point of $\gamma$, the edge and the vertex transforms commute such that
\begin{eqnarray*}\label{a}
\hat A_v^h \hat A_\gamma^{e}&=& \hat A_\gamma^{e} \hat A_v^h,\quad \quad \, \text{if $v$ is not the starting vertex of $\gamma$.} \label{a2}
\end{eqnarray*}
\subsection*{Proof of  Lemma \eqref{lem4}}
In this section, we  give the proof of   holonomy operators that are  invariant   under the vertex  and the edge transforms such that
\begin{eqnarray*}
[\hat B_{\alpha}, \hat A_v^h]&=&  0  \quad \mbox{and}\quad [\hat B_{\alpha}, \hat A_\gamma^e]=0,\label{b1} \\
{ [\hat B_b, \hat A_v^h] }&=&  0  \quad   \mbox{and}\quad [\hat B_b, \hat A_\gamma^e]=0,\label{b2}.
\end{eqnarray*}
Without loss of generality, let us consider the following basis vector
 \begin{eqnarray*}
\begin{tikzpicture}[thick]
% \node (A0) at (-3,0) {$v$};
\node (A) at (-1.5,0) {$v_\alpha$};
\node (B) at (3.5,0) {$v$};
% \node (C)  at (4.5,0){$*$};
\node (E)  at (3,0){$$};
\node at (0.9,0) {\rotatebox{270}{$\implies$}}node[right=0.99cm]{$\Phi_\alpha(g,e)$};
\path[->] (A) edge [bend left=25] node[above] {$\Phi_{\gamma_s}(g_s)$} (B);
\path[->] (A) edge [bend right=25] node[below] {$\Phi_{\gamma_t}(g_t)$} (B);
% \draw[->] (A0) --node[above=0.01cm] {$\Phi_{\gamma_1}(g_1)$} (A) ;

\end{tikzpicture}      
\end{eqnarray*}
The 1-holonomy and the 2-holonomy of the latter graph are given by
\begin{eqnarray*}
Hol_{v_\alpha}^1&=& \partial \Phi_\alpha(g,e)\Phi_{\gamma_s}(g_s)\Phi_{\gamma_t^{-1}}(g_t^{-1}),\\
Hol_{v_\alpha}^2&=& \Phi_\alpha(g,e).
\end{eqnarray*}
We have the actions of $\hat B_{\alpha}\hat A_v^h $ and $\hat A_v^h \hat B_{\alpha}$ read as  follows
\begin{eqnarray*}
\begin{tikzpicture}[thick]
\node (A) at (-2.5,0) {$v_\alpha$};
\node (B) at (2.5,0) {$v$};
\path[->] (A) edge [bend left=25,""{name=F,}] node[above] {$\Phi_{\gamma_s}(g_s)$} (B);
\path[->] (A) edge [bend right=25, ""{name=D,  }] node[below] {$\Phi_{\gamma_t}(g_t)$} (B);

\node at (0,0) {\rotatebox{270}{$\implies$}} node at (0.7,0) {$\Phi_\alpha(g,e)$};

\node (F)  at (3.5,0){$\xmapsto{\hat B_{\alpha}\hat A_{v_\alpha}^h}  $};
\node (G1)  at (6.1,0){$\delta(h Hol_{v_\alpha}^1h^{-1}, \Phi_{\partial\alpha}(1_G))$};
\node (G)  at (8.1,0){$v_\alpha$};
\node (H)  at (12.3,0){$v$};
\node at (9.8,0) {\rotatebox{270}{$\implies$}} node  at (10.9,0) {$h\rhd\Phi_\alpha(g,e)$};
\path[->] (G) edge [bend left=25] node[above] {$h\Phi_{\gamma_s}(g_s)$} (H);
\path[->] (G) edge [bend right=25] node[below] {$h\Phi_{\gamma_t}(g_t)$} (H);
\end{tikzpicture}      
\end{eqnarray*}
\begin{eqnarray*}
\begin{tikzpicture}[thick]

\node (F)  at (3.5,0){$\xmapsto{\hat B_{\alpha}\hat A_{v_\alpha}^h}  $};
\node (G1)  at (7.1,0){$\delta(h Hol_{v_\alpha}^1h^{-1}, \Phi_{\partial\alpha}(1_G))A_{v_\alpha}^h$};
\node (G)  at (9.9,0){$v_\alpha$};
\node (H)  at (14.3,0){$v$};
\node at (12.,0) {\rotatebox{270}{$\implies$}} node  at (12.9,0) {$\Phi_\alpha(g,e)$};
\path[->] (G) edge [bend left=25] node[above] {$h\Phi_{\gamma_s}(g_s)$} (H);
\path[->] (G) edge [bend right=25] node[below] {$\Phi_{\gamma_t}(g_t)$} (H);
\end{tikzpicture}      
\end{eqnarray*}
\begin{eqnarray*}
\begin{tikzpicture}[thick]

\node (F)  at (3.5,0){$\xmapsto{\hat B_{\alpha}\hat A_{v_\alpha}^h}  $};
\node (G1)  at (7.1,0){$A_{v_\alpha}^h\delta( Hol_{v_\alpha}^1, \Phi_{\partial\alpha}(1_G))$};
\node (G)  at (9.9,0){$v_\alpha$};
\node (H)  at (14.3,0){$v$};
\node at (12.,0) {\rotatebox{270}{$\implies$}} node  at (12.9,0) {$\Phi_\alpha(g,e)$};
\path[->] (G) edge [bend left=25] node[above] {$\Phi_{\gamma_s}(g_s)$} (H);
\path[->] (G) edge [bend right=25] node[below] {$\Phi_{\gamma_t}(g_t)$} (H);
\end{tikzpicture}      
\end{eqnarray*}

\begin{eqnarray*}
\begin{tikzpicture}[thick]

\node (F)  at (3.5,0){$\xmapsto{\hat B_{\alpha}\hat A_{v_\alpha}^h}  $};
\node (G1)  at (7.1,0){$A_{v_\alpha}^h\hat B_{\alpha}$};
\node (G)  at (9.9,0){$v_\alpha$};
\node (H)  at (14.3,0){$v$};
\node at (12.,0) {\rotatebox{270}{$\implies$}} node  at (12.9,0) {$\Phi_\alpha(g,e)$};
\path[->] (G) edge [bend left=25] node[above] {$\Phi_{\gamma_s}(g_s)$} (H);
\path[->] (G) edge [bend right=25] node[below] {$\Phi_{\gamma_t}(g_t)$} (H);
\end{tikzpicture}      
\end{eqnarray*}
 where $A_{v_\alpha}^h Hol_{v_\alpha}^1= h Hol_{v_\alpha}^1 h^{-1}$. Hence, we have
 \begin{eqnarray*}
\hat B_{\alpha}\hat A_{v_\alpha}^h=A_{v_\alpha}^h\hat B_{\alpha}\implies  [\hat A_{v_\alpha},\hat B_{\alpha}]=0.
 \end{eqnarray*}
As before, we compute the  actions of $\hat B_{\alpha}\hat A_{\gamma_s}^e $ and $\hat A_{\gamma_s}^e\hat B_{\alpha}$  as  follows
\begin{eqnarray*}
\begin{tikzpicture}[thick]
\node (A) at (-2.5,0) {$v_\alpha$};
\node (B) at (2.5,0) {$v$};
\path[->] (A) edge [bend left=25,""{name=F,}] node[above] {$\Phi_{\gamma_s}(g_s)$} (B);
\path[->] (A) edge [bend right=25, ""{name=D,  }] node[below] {$\Phi_{\gamma_t}(g_t)$} (B);

\node at (0,0) {\rotatebox{270}{$\implies$}} node at (0.7,0) {$\Phi_\alpha(g,e)$};

\node (F)  at (3.5,0){$\xmapsto{\hat B_{\alpha}\hat A_{\gamma_s}^e}  $};
\node (G1)  at (6.1,0){$\delta( {Hol_{v_\alpha}^1}', \Phi_{\partial\alpha}(1_G))$};
\node (G)  at (8.1,0){$v_\alpha$};
\node (H)  at (12.3,0){$v$};
\node at (9.8,0) {\rotatebox{270}{$\implies$}} node  at (10.9,0) {$\Phi_\alpha(g,e)e^{-1}$};
\path[->] (G) edge [bend left=25] node[above] {$\partial(e)\Phi_{\gamma_s}(g_s)$} (H);
\path[->] (G) edge [bend right=25] node[below] {$\Phi_{\gamma_t}(g_t)$} (H);
\end{tikzpicture}      
\end{eqnarray*}
where
\begin{eqnarray*}
{Hol_{v_\alpha}^1}'&=&\partial\left(\Phi_\alpha(g,e)e^{-1}\right)\partial(e)\Phi_{\gamma_s}(g_s)\Phi_{\gamma_t}(g_t)\\
     &=&  \partial\Phi_\alpha(g,e)\partial(e^{-1})\partial(e)\Phi_{\gamma_s}(g_s)\Phi_{\gamma_t}(g_t)\\
     &=& {Hol_{v_\alpha}^1}.
\end{eqnarray*}
Therefore, we have
\begin{eqnarray*}
\begin{tikzpicture}[thick]
\node (A) at (-2.5,0) {$v_\alpha$};
\node (B) at (2.5,0) {$v$};
\path[->] (A) edge [bend left=25,""{name=F,}] node[above] {$\Phi_{\gamma_s}(g_s)$} (B);
\path[->] (A) edge [bend right=25, ""{name=D,  }] node[below] {$\Phi_{\gamma_t}(g_t)$} (B);

\node at (0,0) {\rotatebox{270}{$\implies$}} node at (0.7,0) {$\Phi_\alpha(g,e)$};

\node (F)  at (3.5,0){$\xmapsto{\hat B_{\alpha}\hat A_{\gamma_s}^e}  $};
\node (G1)  at (6.1,0){$\delta( {Hol_{v_\alpha}^1}, \Phi_{\partial\alpha}(1_G))\hat A_{\gamma_s}^e$};
\node (G)  at (8.1,0){$v_\alpha$};
\node (H)  at (12.3,0){$v$};
\node at (9.8,0) {\rotatebox{270}{$\implies$}} node  at (10.9,0) {$\Phi_\alpha(g,e)$};
\path[->] (G) edge [bend left=25] node[above] {$\Phi_{\gamma_s}(g_s)$} (H);
\path[->] (G) edge [bend right=25] node[below] {$\Phi_{\gamma_t}(g_t)$} (H);
\end{tikzpicture}      
\end{eqnarray*}
\begin{eqnarray*}
\begin{tikzpicture}[thick]

\node (F)  at (3.5,0){$\xmapsto{\hat B_{\alpha}A_{\gamma_s}^e}  $};
\node (G1)  at (7.1,0){$A_{v_\alpha}^h\delta( Hol_{v_\alpha}^1, \Phi_{\partial\alpha}(1_G))$};
\node (G)  at (9.9,0){$v_\alpha$};
\node (H)  at (14.3,0){$v$};
\node at (12.,0) {\rotatebox{270}{$\implies$}} node  at (12.9,0) {$\Phi_\alpha(g,e)$};
\path[->] (G) edge [bend left=25] node[above] {$\Phi_{\gamma_s}(g_s)$} (H);
\path[->] (G) edge [bend right=25] node[below] {$\Phi_{\gamma_t}(g_t)$} (H);
\end{tikzpicture}      
\end{eqnarray*}
\begin{eqnarray*}
\begin{tikzpicture}[thick]

\node (F)  at (3.5,0){$\xmapsto{\hat B_{\alpha}A_{\gamma_s}^e}  $};
\node (G1)  at (7.1,0){$A_{\gamma_s}^e \hat B_{\alpha}  $};
\node (G)  at (9.9,0){$v_\alpha$};
\node (H)  at (14.3,0){$v$};
\node at (12.,0) {\rotatebox{270}{$\implies$}} node  at (12.9,0) {$\Phi_\alpha(g,e)$};
\path[->] (G) edge [bend left=25] node[above] {$\Phi_{\gamma_s}(g_s)$} (H);
\path[->] (G) edge [bend right=25] node[below] {$\Phi_{\gamma_t}(g_t)$} (H);
\end{tikzpicture}      
\end{eqnarray*}
Hence, we have
\begin{eqnarray*}
\hat B_{\alpha}\hat A_{\gamma_s}^e= \hat  A_{\gamma_s}^e\hat B_{\alpha}\implies  [\hat A_{\gamma_s}^e,\hat B_{\alpha}]=0.
\end{eqnarray*}
The actions of $\hat B_b\hat A_v^h $ and $\hat A_v^h \hat B_b$ read as  follows
\begin{eqnarray*}
\begin{tikzpicture}[thick]
\node (A) at (-2.5,0) {$v_\alpha$};
\node (B) at (2.5,0) {$v$};
\path[->] (A) edge [bend left=25,""{name=F,}] node[above] {$\Phi_{\gamma_s}(g_s)$} (B);
\path[->] (A) edge [bend right=25, ""{name=D,  }] node[below] {$\Phi_{\gamma_t}(g_t)$} (B);

\node at (0,0) {\rotatebox{270}{$\implies$}} node at (0.7,0) {$\Phi_\alpha(g,e)$};

\node (F)  at (3.5,0){$\xmapsto{\hat B_b\hat A_{v_\alpha}^h}  $};
\node (G1)  at (6.1,0){$\delta({ Hol_{v_\alpha}^2}', \Phi_{\partial b}(1_G,1_E))$};
\node (G)  at (8.1,0){$v_\alpha$};
\node (H)  at (12.3,0){$v$};
\node at (9.8,0) {\rotatebox{270}{$\implies$}} node  at (10.9,0) {$h\rhd\Phi_\alpha(g,e)$};
\path[->] (G) edge [bend left=25] node[above] {$h\Phi_{\gamma_s}(g_s)$} (H);
\path[->] (G) edge [bend right=25] node[below] {$h\Phi_{\gamma_t}(g_t)$} (H);
\end{tikzpicture}      
\end{eqnarray*}
where
\begin{eqnarray*}
{Hol_{v_\alpha}^2}'=  h\rhd\Phi_\alpha(g,e)= {Hol_{v_\alpha}^2}.
\end{eqnarray*}
Now, we have  
\begin{eqnarray*}
\begin{tikzpicture}[thick]
\node (A) at (-2.5,0) {$v_\alpha$};
\node (B) at (2.5,0) {$v$};
\path[->] (A) edge [bend left=25,""{name=F,}] node[above] {$\Phi_{\gamma_s}(g_s)$} (B);
\path[->] (A) edge [bend right=25, ""{name=D,  }] node[below] {$\Phi_{\gamma_t}(g_t)$} (B);

\node at (0,0) {\rotatebox{270}{$\implies$}} node at (0.7,0) {$\Phi_\alpha(g,e)$};

\node (F)  at (3.5,0){$\xmapsto{\hat B_b\hat A_{v_\alpha}^h}  $};
\node (G1)  at (6.1,0){$\delta({ Hol_{v_\alpha}^2}, \Phi_{\partial b}(1_G,1_E))A_{v_\alpha}^h$};
\node (G)  at (8.4,0){$v_\alpha$};
\node (H)  at (12.3,0){$v$};
\node at (9.8,0) {\rotatebox{270}{$\implies$}} node  at (10.9,0) {$\Phi_\alpha(g,e)$};
\path[->] (G) edge [bend left=25] node[above] {$\Phi_{\gamma_s}(g_s)$} (H);
\path[->] (G) edge [bend right=25] node[below] {$\Phi_{\gamma_t}(g_t)$} (H);
\end{tikzpicture}      
\end{eqnarray*}
\begin{eqnarray*}
\begin{tikzpicture}[thick]

\node (F)  at (3.5,0){$\xmapsto{\hat B_b\hat A_{v_\alpha}^e}  $};
\node (G1)  at (7.1,0){$A_{v_\alpha}^h\delta( Hol_{v_\alpha}^2, \Phi_{\partial b}(1_G,1_E))$};
\node (G)  at (9.9,0){$v_\alpha$};
\node (H)  at (14.3,0){$v$};
\node at (12.,0) {\rotatebox{270}{$\implies$}} node  at (12.9,0) {$\Phi_\alpha(g,e)$};
\path[->] (G) edge [bend left=25] node[above] {$\Phi_{\gamma_s}(g_s)$} (H);
\path[->] (G) edge [bend right=25] node[below] {$\Phi_{\gamma_t}(g_t)$} (H);
\end{tikzpicture}      
\end{eqnarray*}
\begin{eqnarray*}
\begin{tikzpicture}[thick]

\node (F)  at (3.5,0){$\xmapsto{\hat B_b\hat A_{v_\alpha}^e}  $};
\node (G1)  at (7.1,0){$A_{v_\alpha}^h\hat B_b$};
\node (G)  at (9.9,0){$v_\alpha$};
\node (H)  at (14.3,0){$v$};
\node at (12.,0) {\rotatebox{270}{$\implies$}} node  at (12.9,0) {$\Phi_\alpha(g,e)$};
\path[->] (G) edge [bend left=25] node[above] {$\Phi_{\gamma_s}(g_s)$} (H);
\path[->] (G) edge [bend right=25] node[below] {$\Phi_{\gamma_t}(g_t)$} (H);
\end{tikzpicture}      
\end{eqnarray*}
Hence, we have
\begin{eqnarray*}
\hat B_{\alpha}\hat A_{v_\alpha}^h= \hat  A_{\alpha}^h\hat B_{b}\implies  [\hat A_{v_\alpha}^h,\hat B_{b}]=0.
\end{eqnarray*}
Similarly, using the same reasoning  one can show that    $ [\hat A_{\gamma}^e,\hat B_{b}]=0 $

\subsection{Proof of equations \eqref{a1z} and \eqref{a2z}}

Using the  first relation of equations \eqref{a1z} we can verify that $\hat{\mathcal{A}}_v$ is a projector operator
\begin{eqnarray*}
\hat{\mathcal{A}}_v^2&=&\frac{1}{|G|^2}\sum_{h,g\in G}\hat A_v^h \hat A_v^g=\frac{1}{|G|^2}\sum_{h,g\in G} \hat A_v^{hg}= \frac{1}{|G|^2}\sum_{hg=h'\in G} \hat A_v^{h'}= \hat{\mathcal{A}}_v,\\
\hat{\mathcal{A}}_v^\dag&=& \left(\frac{1}{|G|}\sum_{h\in G}\hat {A_v^h}\right)^\dag=\frac{1}{|G|}\sum_{h\in G}\hat {A_v^h}^\dag=\frac{1}{|G|}\sum_{h\in G}\hat A_v^{h^{-1}}=\frac{1}{|G|}\sum_{h^{-1}=h'\in G}\hat A_v^{h'}=\hat{\mathcal{A}}_v.
\end{eqnarray*}
Similarly, using the second relation of equations \eqref{a1z} one obtains $\hat{\mathcal{A}}_\gamma^2=\hat{\mathcal{A}}_\gamma$ and $\hat{\mathcal{A}}_\gamma^\dag=\hat{\mathcal{A}}_\gamma$.
Now, recall  the plaquette term $\hat{\mathcal{B}}_\alpha$ and the blob term $\hat{\mathcal{B}}_b$
\begin{eqnarray*}
\hat B_\alpha   &=&\delta\left( Hol_{v_{\alpha}}^1(M,L,\mathcal{G}(\mathcal{X})) ,\Phi_{\partial\alpha}(1_G)\right)  \\
&=&\delta\left(\partial \Phi_{\alpha_i} (u_i)p^-(v_{\alpha_i}\rightarrow v_{\alpha_i}),\Phi_{\partial\alpha}(1_G)\right), \\
\hat B_b   &=&\delta\left( Hol_{v_{\alpha}}^2(M,L,\mathcal{G}(\mathcal{X})) ,\Phi_{\partial b}(1_G,1_E)\right) \\
&=&\delta\left(\prod_{i}\Phi_{\gamma_i}(g_i)\rhd \left(\Phi_{\alpha_i}(u_i)\right)^{\theta_i},\Phi_{\partial b}(1_G,1_E)\right),
\end{eqnarray*}
where $p^-(v_{\alpha_i}\rightarrow v_{\alpha_i})\in G$ be the oriented paths that aligned  with the orientation of  plaquettes $\alpha_i$ from the base-point $v_{\alpha_i}$ to itself. Now, we have
\begin{eqnarray*}
\hat{\mathcal{B}}_\alpha^2   =\delta\left(\partial \Phi_{\alpha_i} (u_i)p^-(v_{\alpha_i}\rightarrow v_{\alpha_i}),\Phi_{\partial\alpha}(1_G)\right)\delta\left(\partial \Phi_{\alpha_i} (u_i)p^-(v_{\alpha_i}\rightarrow v_{\alpha_i}),\Phi_{\partial\alpha}(1_G)\right)=\hat{\mathcal{B}}_\alpha.   .
\end{eqnarray*}
The transpose of $\hat{\mathcal{B}}_\alpha$ flips not only the orientation of the plaquette but also reverse the orientation of its boundary such that
\begin{eqnarray*}
\hat{\mathcal{B}}_\alpha^\dag   &=& \delta\left(\partial (\Phi_{\alpha_i} (u_i))^{-1}p^+(v_{\alpha_i}\rightarrow v_{\alpha_i}),\Phi_{\partial\alpha}(1_G)\right)= \delta\left(\Phi_{\partial\alpha}(1_G),\partial (\Phi_{\alpha_i} (u_i))p^-(v_{\alpha_i}\rightarrow v_{\alpha_i})\right) = .\hat{\mathcal{B}}_\alpha,
\end{eqnarray*}
so the plaquette energy term is preserved by the its transpose. Similarly, using the same reasoning  one can show that  $\hat{\mathcal{B}}_b^2=\hat{\mathcal{B}}_b $ and  $\hat{\mathcal{B}}_b^\dag=\hat{\mathcal{B}}_b $

\section{Appendix C: Invariances of energy terms}
In this appendix, we provide the proofs of the invariance of the energy terms  ($\hat{\mathcal{A}}_v, \hat{\mathcal{A}}_\gamma, \hat{\mathcal{B}}_\alpha,\hat{\mathcal{B}}_b $) under the transformations $\hat T_i$ required show that the ground state projector $\hat P$ that encode the  quantum information is topological invariant. To do so, we show  that   energy terms are individually   preserved under the edge-flipping transformation $ \hat T_1$, the plaquette-orientation flipping transformation $ \hat T_2$ and the moving base point transformation $ \hat T_3$.

\subsubsection{Flipping the orientation of an edge}

 We denote by $ \hat T_1$ the edge-flipping transformation. We wish to show that this map preserves
each energy terms ($\hat{\mathcal{A}}_v, \hat{\mathcal{A}}_\gamma, \hat{\mathcal{B}}_\alpha,\hat{\mathcal{B}}_b $) individually of the GS projector $\hat P$. 
\begin{itemize}
    \item The vertex transform action on the adjacent edge $\Phi_{\gamma}(g)$ \eqref{t45} and the plaquette $\Phi_{\alpha}(u)$ \eqref {t46} is provided by
\begin{eqnarray}
    \hat A_v^h \Phi_\gamma(g)&=&  
     \begin{cases} h \Phi_\gamma(g)  & \text{if $\gamma$ points away from $v$,} \\
	\Phi_\gamma(g)h^{-1} & \text{if $\gamma$ points towards to $v$,}\\
	\Phi_\gamma(g) & \text{otherwise,}
 \end{cases}\\
  \hat A_v^h \Phi_\alpha(u)&=&  \begin{cases} h \Phi_\alpha(u)  & \text{if $v=v_\alpha$,}\\
	\Phi_\alpha(u) & \text{otherwise.}
 \end{cases}
 \end{eqnarray} 
 We aim to show that the vertex transform $\hat A_v^h$  is invariant under  the edge-flipping transformation  i.e,  $ (\hat T_1^{-1} \hat A_v^h \hat T_1 )| \Phi_\gamma(g) \Phi_\alpha(u)\rangle = \hat A_v^h| \Phi_\gamma(g) \Phi_\alpha(u)\rangle$.   Thus, the   edge-flipping transformation of the  vertex operator $\hat A_v^h$  action on the adjacent edge  $\Phi_{\gamma}(g)$ state  reads as
\begin{eqnarray}
\hat T_1  \Phi_\gamma(g)&=&\Phi_{\gamma^{-1}}(g^{-1}),\\
 \hat A_v^h \hat T_1 \Phi_\gamma(g) &=&  \begin{cases}
   \Phi_{\gamma^{-1}}(g^{-1}) h^{-1} & \text{\quad if $\gamma$  originally pointed away from (now towards) $v$,} \\
        h \Phi_{\gamma^{-1}}(g^{-1}) & \quad\text{if $\gamma$ originally pointed towards $v$,}\\
	\Phi_{\gamma^{-1}}(g^{-1}) & \text{\quad otherwise,}
    \end{cases}\\
\hat T_1^{-1}\hat A_v^h \hat T_1 \Phi_\gamma(g) 
    &=&   \begin{cases} h\Phi_{\gamma}(g)  & \quad \text{if $\gamma$ points away from $v$,}\\
        \Phi_{\gamma}(g)h^{-1} & \quad\ \text{if $\gamma$ points towards to $v$,}\\
	\Phi_{\gamma}(g) & \text{\quad otherwise.} 
    \end{cases}
\end{eqnarray}
We can see that the action of the vertex transform is preserved under the edge-flipping procedure $\hat T_1$, i.e, $ T_1^{-1}\hat A_v^h  \hat T_1 \Phi_{\gamma}(g)= \hat A_v^h\Phi_{\gamma}(g)$.  However, the   edge-flipping transformation  does not affect the adjacent plaquette  $\Phi_{\alpha}(u)$ state i.e,  $ \hat T_1\Phi_{\alpha}(u)= \Phi_{\alpha}(u)$, therefore  we have  $\hat T_1^{-1}\hat A_v^h  \hat T_1\Phi_{\alpha}(u)= \hat A_v^h\Phi_{\alpha}(u)$. Now, having the latter at hand,  we can show that the edge flipping procedure is consistent with the vertex terms $\hat {\mathcal{A}_v}=\frac{1}{|G|}\sum_{h\in G}\hat A_v^h$ through the  following proposition.

\begin{proposition}
    {\it  Let $v\in L^0$ , $\gamma,\in L^1$, $\alpha\in L^2$ and $\hat T_1$ a flipping edge operator. Based on $\hat T_1^{-1}\hat A_v^h  \hat T_1= \hat A_v^h$},  one can easily show that
\begin{eqnarray}
     \hat T_1^{-1}\hat{\mathcal{A}}_v \hat T_1=\hat{\mathcal{A}}_v.
\end{eqnarray}
\end{proposition}
\item Let consider the energy term $\hat {\mathcal{A}}_\gamma =\frac{1}{|E|}\sum_{e\in  E}\hat A_\gamma^e$ for an edge $\gamma$. We show that this  energy term is invariant under the edge-flipping $\hat T_1$ i.e,  $ (\hat T_1^{-1} \hat {\mathcal{A}}_\gamma \hat T_1 )| \Phi_\gamma(g) \Phi_\alpha(u)\rangle = \hat A_\gamma| \Phi_\gamma(g) \Phi_\alpha(u)\rangle$. Initially, recall how the edge transform $\hat A_\gamma^e$ acts on the edge $\gamma.$ Thus, if the edge was originally labeled $\Phi_{\gamma}(g)$, the edge transform operates as follows: $\hat A_{\gamma}^{e} \Phi_\gamma(g)= \partial(e) \Phi_\gamma(g)$. Therefore, we have
\begin{eqnarray}
     \hat T_1^{-1} \hat A_{\gamma}^{e}  \hat T_1\Phi_\gamma(g) &=& \hat T_1^{-1} \hat A_{\gamma}^{e} (\Phi_{\gamma^{-1}}(g^{-1})),\cr
      &=& \hat T_1^{-1}   (\partial(e)^{-1} \Phi_{\gamma^{-1}}(g^{-1}))\cr
       &=& \Phi_{\gamma}(g)\partial(e^{-1})\Phi_{\gamma^{-1}}(g^{-1})\cr
        &=& \partial(\Phi_{\gamma}(g)\rhd e^{-1})\Phi_{\gamma}(g)\cr
      &=& \hat A_{\gamma}^{\Phi_{\gamma}(g)\rhd e^{-1}}\Phi_{\gamma}(g) ,  
\end{eqnarray}
where  we  used the first Peiffer \eqref{P1}  condition  $\partial(\Phi_{\gamma}(g)\rhd e)= \Phi_{\gamma}(g)\partial(e)\Phi_{\gamma^{-1}}(g^{-1})$. We therefore see that, unlike the vertex transforms, the individual edge transforms action on the edges  are not invariant under the edge-flipping procedure $\hat T_1$. However, the  edge-flipping procedure of the edge energy term reads as follows
\begin{eqnarray}
    \hat T_1^{-1} \hat A_{\gamma}  \hat T_1\Phi_\gamma(g) &=& \frac{1}{|E|}\sum_{e\in  E}\hat T_1^{-1}\hat A_\gamma^e\hat T_1\cr
    &=&\frac{1}{|E|}\sum_{e\in  E}\hat A_{\gamma}^{\Phi_{\gamma}(g)\rhd e^{-1}}\Phi_{\gamma}(g),\cr
     &=&\frac{1}{|E|}\sum_{e'=\Phi_{\gamma}(g)\rhd e^{-1}\in E}\hat A_{\gamma}^{e'}\Phi_{\gamma}(g)\cr
     &=&\hat A_{\gamma}\Phi_{\gamma}(g).\label{re}
\end{eqnarray}
Although, the individual edge transform actions on the edges  are not invariant under the edge-flipping procedure. However with equation \eqref{re}, it does not mean that the edge energy term itself is not invariant.

 Let $\gamma\in L^{1}$ and  any $e\in  E$. Let ${p^+(v_\alpha\rightarrow s(\gamma))} \in G $ and ${p^-(v_\alpha\rightarrow s(\gamma))} \in G $ be the paths that aligned and antialigned, respectively, with the orientation of the plaquette $\alpha$ from the base point $v_\alpha$ to the edge. The edge operator $ \hat A_{\gamma}^e$  acting on $ \mathcal{H}(M,L,\mathcal{G}(\mathcal{X}))$ based on the plaquette
$ \alpha$, is recalled as follows \eqref{ta}
\begin{eqnarray}
    \hat A_{\gamma}^{e} \Phi_\alpha(u)=  \begin{cases}  \Phi_\alpha(u)\left(p^+(v_\alpha\rightarrow s(\gamma))\rhd e^{-1} \right)  & \text{ if $\gamma$ is on $\alpha$ and aligned with $\alpha$} \\
\left({p^-(v_\alpha\rightarrow s(\gamma)}\rhd e\right) \Phi_\alpha(u)  & \text{if $\gamma$ is on $\alpha$ and aligned against $\alpha$}\\
\Phi_\alpha(u) & \text{otherwise}\\
\end{cases}.
\end{eqnarray}
Now, to show the invariance of the edge transforms under the edge-flipping process $\hat T_1$, we consider two cases.  The case where a labeled edge $\Phi_\gamma(g)$ is  aligned with the orientation of the labeled plaquette  $\Phi_\alpha(u)$ and the other one where this labeled edge is anti-aligned 
 with this orientation. They are  illustrated by the following graphs in the Fig\eqref {en5}
 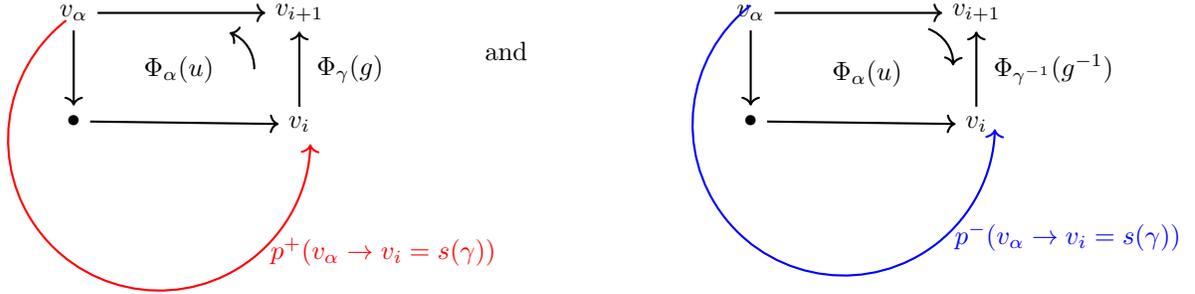
\begin{figure}[H]
\centering
\begin{eqnarray*}
	\begin{tikzpicture}[thick]
		\node (A) at (-1.5,0) {$v_\alpha$};
		\node (B) at (1.5,0) {$v_{i+1}$};
		\node (C) at (4.5,0) [below=of A] {$\bullet$}; 
		\node (D) at (4.5,0) [below=of B] {$v_i$};	
		\draw[->](A)--node [left=0.1cm] {}(C);
		\draw[<-](B)--node [right=0.1cm] {$\Phi_{\gamma}(g)$}(D);
		\draw[->](A)--node [above=0.1cm] {$ $}(B);
		\draw[->](C)--node [below=0.1cm] {$ $}(D); 
        \draw[ thick, ->] (0.9,-0.75) arc (-2:70:0.5);
     %\draw[ thick, ->] (-1.5,0.2) arc (180:-1.70:1.5) node [right,pos=0.7] {$ p^+(v_\alpha\rightarrow v_i)    $};
     \draw[ thick,red, ->] (-1.6,-0.1) arc (-232:-1.8:2.01) node [right,pos=0.8] {$ p^+(v_\alpha\rightarrow v_i=s(\gamma))$};
        
		\node at (-0.1,-0.75){$\Phi_\alpha(u)$};

\node (F)  at (4.25,-0.5){and};
		
		\node (G)  at (7.5,0){$v_\alpha$};
		\node (H)  at (10.5,0){$v_{i+1}$};
		\node (I)  at (6.4,1) [below=of G] {$\bullet$};
		\node (J)  at (10.5,1)[below=of H] {$v_{i}$};

            \draw[->](G)-- node [above=0.1cm] {} (H);
		\draw[->](G)--node [left=0.1cm] {}   (I);
		\draw[<-](H)-- node [right=0.1cm]   {$\Phi_{\gamma^{-1}}(g^{-1})$} (J);
		\draw[->](I)-- node [below=0.1cm] {$ $} (J);
		
		\draw[ thick, <-] (10.19,-0.7) arc (-2:70:0.5);
		\node  at (9.05,-0.8) {$\Phi_{\alpha}(u)$};
\draw[ thick,blue, ->] (7.5,0.1) arc (-232:-1.8:2.01) node [right,pos=0.8] {$ p^-(v_\alpha\rightarrow v_i=s(\gamma))$};
	\end{tikzpicture}	      
\end{eqnarray*}
 \caption{ (Left) The labeled edge $\Phi_\gamma(g)$  aligned with the orientation of the labeled plaquette  $\Phi_\alpha(u)$. (Right) Anti-alignment of  $\Phi_\gamma(g)$ with the orientation of $\Phi_\alpha(u)$. The vertex $v_i= s(\gamma)$ is the source of the edge $\gamma$, $v_{i+1}= t(\gamma)$ is its target. }\label{en5} 
\end{figure}
The path $ p^+(v_\alpha\rightarrow v_i=s(\gamma))$ is the  path that aligned  with the orientation of the plaquette $\alpha$ from the base point $v_\alpha$  and $ p^-(v_\alpha\rightarrow v_i=s(\gamma))$ is the  one that anti-aligned  with the orientation of the plaquette $\alpha$ from its base point $v_\alpha$. The actions of the edge transform reads as follows
\begin{eqnarray}
   \hat A_{\gamma}^{e} \Phi_\alpha(u)= \Phi_\alpha(u)\left(p^+(v_\alpha\rightarrow s(\gamma))\rhd e^{-1} \right)\quad \mbox{and}\quad \hat A_{\gamma}^{e} \Phi_\alpha(u)= \left(p^-(v_\alpha\rightarrow s(\gamma))\rhd e \right)\Phi_\alpha(u)
\end{eqnarray}
These edge-flipping processes  by $\hat T_1$ are illustrated by the  figures \eqref{case1} and \eqref{case2} as follows

 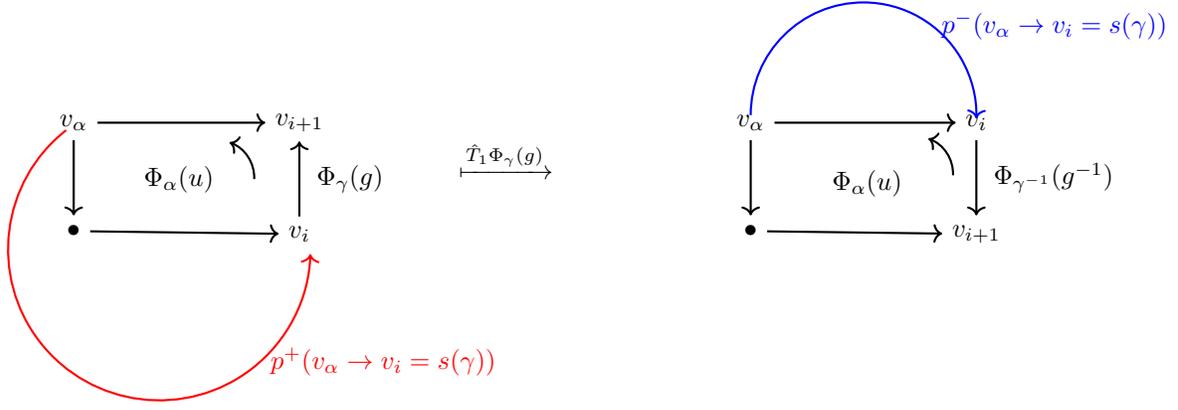
\begin{figure}[H]
\centering
\begin{eqnarray*}
	\begin{tikzpicture}[thick]
		\node (A) at (-1.5,0) {$v_\alpha$};
		\node (B) at (1.5,0) {$v_{i+1}$};
		\node (C) at (4.5,0) [below=of A] {$\bullet$}; 
		\node (D) at (4.5,0) [below=of B] {$v_i$};	
		\draw[->](A)--node [left=0.1cm] {}(C);
		\draw[<-](B)--node [right=0.1cm] {$\Phi_{\gamma}(g)$}(D);
		\draw[->](A)--node [above=0.1cm] {$ $}(B);
		\draw[->](C)--node [below=0.1cm] {$ $}(D); 
        \draw[ thick, ->] (0.9,-0.75) arc (-2:70:0.5);
     %\draw[ thick, ->] (-1.5,0.2) arc (180:-1.70:1.5) node [right,pos=0.7] {$ p^+(v_\alpha\rightarrow v_i)    $};
     \draw[ thick,red, ->] (-1.6,-0.1) arc (-232:-1.8:2.01) node [right,pos=0.8] {$ p^+(v_\alpha\rightarrow v_i=s(\gamma))$};
        
		\node at (-0.1,-0.75){$\Phi_\alpha(u)$};

\node (F)  at (4.25,-0.5){$\xmapsto{\hat T_1 \Phi_{\gamma}(g) }  $} ;
		
		\node (G)  at (7.5,0){$v_\alpha$};
		\node (H)  at (10.5,0){$v_{i}$};
		\node (I)  at (6.4,1) [below=of G] {$\bullet$};
		\node (J)  at (10.5,1)[below=of H] {$v_{i+1}$};

            \draw[->](G)-- node [above=0.1cm] {} (H);
		\draw[->](G)--node [left=0.1cm] {}   (I);
		\draw[->](H)-- node [right=0.1cm]   {$\Phi_{\gamma^{-1}}(g^{-1})$} (J);
		\draw[->](I)-- node [below=0.1cm] {$ $} (J);
		
		\draw[ thick, ->] (10.19,-0.7) arc (-2:70:0.5);
		\node  at (9.05,-0.8) {$\Phi_{\alpha}(u)$};

 \draw[ thick,blue, ->] (7.5,0.1) arc (180:-1.70:1.5) node [right,pos=0.7] {$ p^-(v_\alpha\rightarrow v_i=s(\gamma))  $};
	\end{tikzpicture}	      
\end{eqnarray*}
 \caption{ The labeled flipping edge $ \Phi_{\gamma}(g)  $    }\label{case1} 
\end{figure}
and 
 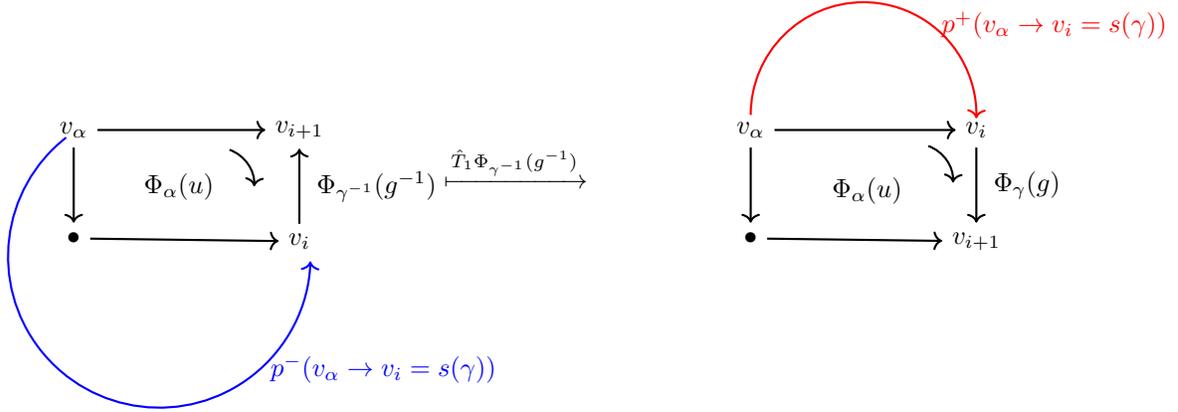
\begin{figure}[H]
\centering
\begin{eqnarray*}
	\begin{tikzpicture}[thick]
		\node (A) at (-1.5,0) {$v_\alpha$};
		\node (B) at (1.5,0) {$v_{i+1}$};
		\node (C) at (4.5,0) [below=of A] {$\bullet$}; 
		\node (D) at (4.5,0) [below=of B] {$v_{i}$};	
		\draw[->](A)--node [left=0.1cm] {}(C);
		\draw[<-](B)--node [right=0.1cm] {$\Phi_{\gamma^{-1}}(g^{-1})$}(D);
		\draw[->](A)--node [above=0.1cm] {$ $}(B);
		\draw[->](C)--node [below=0.1cm] {$ $}(D); 
        \draw[ thick, <-] (0.9,-0.75) arc (-2:70:0.5);
     %\draw[ thick,blue, ->] (-1.5,0.2) arc (180:-1.70:1.5) node [right,pos=0.8] {$ p^-(v_\alpha\rightarrow v_i)    $};
     \draw[ thick,blue, ->] (-1.6,-0.1) arc (-232:-1.8:2.01) node [right,pos=0.8] {$ p^-(v_\alpha\rightarrow v_i=s(\gamma))$};
        
		\node at (-0.1,-0.75){$\Phi_\alpha(u)$};

\node (F)  at (4.38,-0.5){$\xmapsto{\hat T_1 \Phi_{\gamma^{-1}}(g^{-1}) } $};
		
		\node (G)  at (7.5,0){$v_\alpha$};
		\node (H)  at (10.5,0){$v_{i}$};
		\node (I)  at (6.4,1) [below=of G] {$\bullet$};
		\node (J)  at (10.5,1)[below=of H] {$v_{i+1}$};

            \draw[->](G)-- node [above=0.1cm] {} (H);
		\draw[->](G)--node [left=0.1cm] {}   (I);
		\draw[->](H)-- node [right=0.1cm]   {$\Phi_{\gamma}(g)$} (J);
		\draw[->](I)-- node [below=0.1cm] {$ $} (J);
		
		\draw[ thick, <-] (10.19,-0.7) arc (-2:70:0.5);
		\node  at (9.05,-0.8) {$\Phi_{\alpha}(u)$};

% \draw[ thick,blue, ->] (7.5,0.1) arc (180:-1.70:1.5) node [right,pos=0.7] {$ p^-(v_\alpha\rightarrow v_i=s(\gamma))  $};

 \draw[ thick,red, ->] (7.5,0.2) arc (180:-1.70:1.5) node [right,pos=0.7] {$ p^+(v_\alpha\rightarrow v_i=s(\gamma))  $};

	\end{tikzpicture}	      
\end{eqnarray*}
\caption{ The labeled flipping edge $ \Phi_{\gamma^{-1}}(g^{-1})  $    }\label{case2} 
\end{figure}

For the first case \eqref{case1}, we have 
\begin{eqnarray}
   \hat A_{\gamma}^{e} \hat T_1\Phi_\alpha(u)&=& \left({p^-(v_\alpha\rightarrow s(\gamma)})\rhd e\right) \Phi_\alpha(u),\\
  \hat T_1^{-1} \hat A_{\gamma}^{e} \hat T_1\Phi_\alpha(u)&=& \Phi_\alpha(u)\left(p^+(v_\alpha\rightarrow s(\gamma))\rhd e^{-1} \right)\cr
  &=& \hat A_{\gamma}^{e} \Phi_\alpha(u).
\end{eqnarray}
For the second one \eqref{case2}, we have
\begin{eqnarray}
   \hat A_{\gamma}^{e} \hat T_1\Phi_\alpha(u)&=&  \Phi_\alpha(u)\left({p^+(v_\alpha\rightarrow s(\gamma)})\rhd e^{-1}\right),\\
  \hat T_1^{-1} \hat A_{\gamma}^{e} \hat T_1\Phi_\alpha(u)&=& \left(p^-(v_\alpha\rightarrow s(\gamma))\rhd e \right)\Phi_\alpha(u)\cr
  &=& \hat A_{\gamma}^{e} \Phi_\alpha(u).
\end{eqnarray}

\begin{proposition} {\it  Let $\gamma,\in L^1$, $ e\in E$ and $\hat T_1$ a flipping edge operator. Based on the result $  \hat T_1^{-1}\hat A_{\gamma}^{e} \hat T_1\Phi_\alpha(u)= \hat A_{\gamma}^{e} \Phi_\alpha(u) $, we  show that 
\begin{eqnarray}
     \hat T_1^{-1}\hat{\mathcal{A}}_\gamma \hat T_1\Phi_\alpha(u)=\hat{\mathcal{A}}_\gamma\Phi_\alpha(u).
\end{eqnarray}
}
\end{proposition}

\item
Finally,  we consider the plaquette and blob energy terms. Recall that the plaquette term  $\hat B_{\alpha_i}$ and the blob $\hat B_{b_i} $ term are 
\begin{eqnarray}
			\hat B_{\alpha_i} \left| \Phi_{\gamma_i}(g_i)  \Phi_{\alpha_i}(u_i)  \right \rangle &=&\delta\left(	Hol_{v_{\alpha_i}}^1,
            \Phi_{\partial\alpha}(1_G)\right) \left| \Phi_{\gamma_i}(g_i)  \Phi_{\alpha_i}(u_i)  \right \rangle. \\
            \hat B_{b_i}  \left| \Phi_{\gamma_i }(g_i)  \Phi_{\alpha}(u)  \right \rangle  &=&\delta\left(Hol_{v_{\alpha_i}}^2,\Phi_{\partial b_i}(1_G,1_E)\right) \left| \Phi_{\gamma_i}(g_i)  \Phi_{\alpha)i}(u_i)  \right \rangle,
		\end{eqnarray}	
where $ Hol_{v_{\alpha_i}}^1$ and  $ Hol_{v_{b_i}}^2$ are the 1-holonomies  and the 2-holonomies respectively at the base points $v_{\alpha_i}$ and  $v_{b_i}=v_{\alpha_i}$. While, the 1-holonomies   only involve  edges through path elements around the boundaries of  plaquettes, the blob 2-holonomies 
are a product of the plaquette elements around the boundaries of
the blobs. They are expressed as follows
\begin{eqnarray}
    Hol_{v_{\alpha_i}}^1&=&\partial \Phi_{\alpha} (u_i)p^-(v_{\alpha_i}\rightarrow v_{\alpha_i}),\\
    Hol_{v_{\alpha_i}}^2&=& \prod_{i}\Phi_{\gamma_i}(g_i)\rhd \left(\Phi_{\alpha_i}(u_i)\right)^{\theta_i}, 
	\end{eqnarray}	
where $\theta_i$  is $1$ or $-1$ depending on the orientation of plaquettes labeled by $\Phi_{\alpha_i}(u_i)$ and $p^-(v_{\alpha_i}\rightarrow v_{\alpha_i})$ are the paths that anti-aligned with the orientations of the plaquettes $\alpha_i$. The edge flipping  orientations of these holonomies  read as follows
\begin{eqnarray}
  \hat T_1  (Hol_{v_{\alpha_i}}^1)&=&\partial \Phi_{\alpha_i} (u_i)p^+(v_{\alpha_i}\rightarrow v_{\alpha_i}),\\
  \hat T_1  ( Hol_{v_{\alpha}}^2)&=& \prod_{i}\Phi_{\gamma_i^{-1}}(g_i^{-1})\rhd \left(\Phi_{\alpha_i}(u_i)\right)^{\theta_i},
	\end{eqnarray}	
where $p^+(v_{\alpha_i}\rightarrow v_{\alpha_i})$ are the paths that aligned with the orientations of the plaquettes $\alpha_i$.
Consequently, the edge flipping procedure of the plaquette term $\hat B_{\alpha_i}$ and the blob term $\hat  B_{b_i}$ are obtained by
\begin{eqnarray}
 \hat T_1^{-1} \hat B_{\alpha_i} \hat T_1 \left| \Phi_{\gamma_i}(g)  \Phi_{\alpha_i}(u_i)  \right \rangle &=&\delta(\partial \Phi_{\alpha_i} (u_i)p^-(v_{\alpha_i}\rightarrow v_{\alpha_i}), \Phi_{\partial\alpha_i}(1_G))\left| \Phi_{\gamma_i}(g_i)  \Phi_{\alpha_i}(u_i)  \right \rangle,\cr
 &=&\delta\left(	Hol_{v_{\alpha_i}}^1, \Phi_{\partial\alpha_i}(1_G)\right) \left| \Phi_{\gamma_i}(g_i)  \Phi_{\alpha_i}(u_i)  \right \rangle    \cr
 \hat T_1^{-1} \hat B_b\hat T_1 \left| \Phi_{\gamma_i}(g)  \Phi_{\alpha_i}(u_i)  \right \rangle &=& \delta\left(\prod_{i}\Phi_{\gamma_i}(g_i)\rhd \left(\Phi_{\alpha_i}(u_i)\right)^{\theta_i},\Phi_{\partial b_i}(1_G,1_E)\right)\left| \Phi_{\gamma}(g)  \Phi_{\alpha}(u)  \right \rangle.\cr
 &=& \delta\left(	Hol_{v_{\alpha_i}}^2,\Phi_{\partial b_i},\Phi_{\partial b_i}(1_G,1_E)\right) \left| \Phi_{\gamma_i}(g)  \Phi_{\alpha_i}(u_i)  \right \rangle .
	\end{eqnarray}
These show that the plaquette and  blob energy terms are  preserved by the edge-flipping orientation. Now, let  show these results by the following graph \eqref{en50}
\begin{figure}
\centering
\begin{eqnarray*}
	\begin{tikzpicture}[thick]
		\node (A) at (-1.5,0) {$v_\alpha$};
		\node (B) at (1.5,0) {$v_2$};
		\node (C) at (4.5,0) [below=of A] {$v_1$}; 
		\node (D) at (4.5,0)   [below=of B] {$v_3$};	
		
		\draw[->](A)--node [left=0.1cm] {$\Phi_{\gamma_{\alpha 1}}(g_{\alpha1})$}(C);
		\draw[->](B)--node [right=0.1cm] {$\Phi_{\gamma_{23}}(g_{23})$}(D);
		\draw[->](A)--node [above=0.1cm] {$\Phi_{\gamma_{\alpha 2}}(g_{\alpha2})$}(B);
		\draw[->](C)--node [below=0.1cm] {$\Phi_{\gamma_{1 3}}(g_{13})$}(D); 
        \draw[ thick, ->] (0.9,-0.75) arc (-2:70:0.5);
		\node at (-0.1,-0.75){$\Phi_\alpha(u)$};	
	\end{tikzpicture}	      
\end{eqnarray*}
\caption{Configuration graph: $v_\alpha$ is the plaquette base point, $v_i\,(i=1,2,3)$ are the vertices, $\Phi_{\gamma_j}(g_j)$ are the colored edges $\gamma_i$ and   $\Phi_\alpha(u)$ is  the colored plaquette $\alpha_i$ anti-clockwise rotation. }\label{en50}
\end{figure}
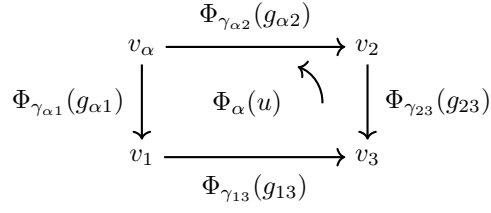

The 1-holonomy $ Hol_{v_{\alpha}}^1 $ and the 2-holonomy are given by
\begin{eqnarray}
    Hol_{v_{\alpha}}^1&=&\partial \Phi_\alpha (u)p^-(v_\alpha\rightarrow v_\alpha)\cr
    &=&\partial \Phi_\alpha (u)\Phi_{\gamma_{ 23}}(g_{23})\Phi_{\gamma_{ \alpha2}}(g_{\alpha2})\Phi_{\gamma_{\alpha 1}}(g_{\alpha1})\Phi_{\gamma_{ 13}^{-1}}(g_{13}^{-1}),\\
    Hol_{v_{\alpha}}^2&=& \Phi_\alpha (u).
\end{eqnarray}
The flipping holonomies $\hat T_1 Hol_{v_{\alpha}}^1 $ and $ \hat T_1 Hol_{v_{\alpha}}^2$ are given by

\begin{eqnarray}
    \hat T_1 Hol_{v_{\alpha}}^1&=& \partial \Phi_\alpha (u)\Phi_{\gamma_{\alpha 1}}(g_{\alpha1})\Phi_{\gamma_{ 13}}(g_{13})\Phi_{\gamma_{ 23}^{-1}}(g_{23}^{-1})\Phi_{\gamma_{ \alpha2}^{-1}}(g_{\alpha2}^{-1}),\\
    \hat T_1 Hol_{v_{\alpha}}^2&=& \Phi_\alpha (u).
\end{eqnarray}
The edge flipping procedures are given by
\begin{eqnarray}
         \hat T_1^{-1} \hat B_\alpha  \hat T_1 &=& \delta\left(\partial \Phi_\alpha (u)\Phi_{\gamma_{ 23}}(g_{23})\Phi_{\gamma_{ \alpha2}}(g_{\alpha2})\Phi_{\gamma_{\alpha 1}}(g_{\alpha1})\Phi_{\gamma_{ 13}^{-1}}(g_{13}^{-1}),\Phi_{\partial\alpha}(1_G)\right) \cr
         &=& \delta\left( Hol_{v_{\alpha}}^1,\Phi_{\partial\alpha}(1_G)\right),\\
            \hat T_1^{-1} \hat B_b  \hat T_1 &=& \Phi_\alpha (u)=\delta\left(	Hol_{v_{\alpha}}^2,\Phi_{\partial b}(1_G,1_E)\right), 
		\end{eqnarray}	   
\end{itemize}
where $  \partial \Phi_\alpha(u)= \Phi_{\gamma_{\alpha 1}}(g_{\alpha1})\Phi_{\gamma_{ 13}}(g_{13})\Phi_{\gamma_{ 23}^{-1}}(g_{23}^{-1})\Phi_{\gamma_{ \alpha2}^{-1}}(g_{\alpha2}^{-1}).$

\subsubsection{Flipping the orientation of a plaquette}
We now consider the analogous procedure where we reverse the orientation of a plaquette. We show in this subsection that the energy terms are invariant under the plaquette  orientation flipping.  We denote this
operation by $ \hat T_2$ for a plaquette $\alpha$ labeled by $\Phi_\alpha(u) $ such that  $ \hat T_2\Phi_\alpha(u)=(\Phi_\alpha(u))^{-1}$.

\begin{itemize}
    \item 
 First consider the vertex
transform $\hat A_v^h$. Apart from the edges, which are unaffected by the orientation of plaquettes, this transform acts only on plaquettes with base point at $v_\alpha$. We have
\begin{eqnarray}
    \hat A_v^h \Phi_\alpha(u)=  \begin{cases} h \rhd \Phi_\alpha(u)  & \text{if $v=v_\alpha$ } \\
	\Phi_\alpha(u) & \text{ if  $v\neq v_\alpha$ }.\\
	\end{cases}
\end{eqnarray}
The plaquette flipping procedure of the vertex transform  is given by
\begin{eqnarray}
    \hat A_v^h \hat T_2\Phi_\alpha(u)&=&  \begin{cases} h\rhd(\Phi_\alpha(u))^{-1}  & \text{if $v=v_\alpha$ } \\
	(\Phi_\alpha(u))^{-1} & \text{ if  $v\neq v_\alpha$ }.\\
	\end{cases}\\
     \hat T_2^{-1}\hat A_v^h \hat T_2\Phi_\alpha(u)&=&  \begin{cases} h\rhd\Phi_\alpha(u)  & \text{if $v=v_\alpha$ } \\
	\Phi_\alpha(u) & \text{ if  $v\neq v_\alpha$ }.\\
	\end{cases}\cr
    &=&  \hat A_v^h\Phi_\alpha(u).
\end{eqnarray}
 This show that   the vertex transform is invariant under the plaquette-flipping procedure. Consequently  the
vertex energy term is also invariant under this plaquette-flipping procedure such as that
\begin{eqnarray}
     \hat T_2^{-1}\hat {\mathcal{A}}_v  \hat T_2=\hat {\mathcal{A}}_v.
\end{eqnarray}
\item Recall the action of the edge transform $\hat A_\gamma^e$
on an adjacent plaquette $\alpha$ labeled by $\Phi_\alpha(u)$ is given by 
	\begin{eqnarray}
    \hat A_{\gamma}^{e} \Phi_\alpha(u)=  \begin{cases}  \Phi_\alpha(u)\left(p^+(v_\alpha\rightarrow s(\gamma))\rhd e^{-1} \right)  & \text{ if $\gamma$ is on $\alpha$ and aligned with $\alpha$} \\
	\left({p^-(v_\alpha\rightarrow s(\gamma))}\rhd e\right) \Phi_\alpha(u)  & \text{if $\gamma$ is on $\alpha$ and aligned against $\alpha$}\\
		\Phi_\alpha(u) & \text{otherwise}.\\
	\end{cases}
\end{eqnarray}
The plaquette flipping procedure of the edge transform  is given by
	\begin{eqnarray}
    \hat A_{\gamma}^{e} \hat T_2\Phi_\alpha(u)&=&  \begin{cases}  \left(p^+(v_\alpha\rightarrow s(\gamma))\rhd e \right)(\Phi_\alpha(u))^{-1}  & \text{ if $\gamma$ is originally  on $\alpha$ and aligned with $\alpha$} \\
	(\Phi_\alpha(u))^{-1}\left({p^-(v_\alpha\rightarrow s(\gamma)   )}\rhd e^{-1}\right)   & \text{if $\gamma$ is originally on $\alpha$ and aligned against $\alpha$}\\
		(\Phi_\alpha(u))^{-1} & \text{otherwise}.\\
	\end{cases}\cr
    \hat T_2^{-1}\hat A_{\gamma}^{e} \hat T_2\Phi_\alpha(u)&=&  \begin{cases}  \Phi_\alpha(u)\left(p^+(v_\alpha\rightarrow s(\gamma))\rhd e^{-1} \right)  & \text{ if $\gamma$ is on $\alpha$ and aligned with $\alpha$} \\
	\left({p^-(v_\alpha\rightarrow s(\gamma))}\rhd e\right) \Phi_\alpha(u)  & \text{if $\gamma$ is on $\alpha$ and aligned against $\alpha$}\\
		\Phi_\alpha(u) & \text{otherwise}.\\
	\end{cases}\\
    &=& \hat A_{\gamma}^{e} \Phi_\alpha(u).
\end{eqnarray}
This  show that the edge transforms are invariant under the plaquette-flipping procedure. This means that the edge energy $ \hat {\mathcal{A}}_\gamma=\frac{1}{|E|}\sum_{e\in E}\hat A_{\gamma}^{e}$ term is invariant under
this procedure i.e 
\begin{eqnarray}
     \hat T_2^{-1}\hat {\mathcal{A}}_\gamma  \hat T_2=\hat {\mathcal{A}}_\gamma.
\end{eqnarray}

\item Finally, we  consider the plaquette term $\hat B_\alpha$ and the blob term $\hat B_b$. Recall their actions on the edge $\gamma$  labeled by $\Phi_\gamma(g)$ and on the plaquette $\alpha$ labeled by $\Phi_{\alpha}(u)$ 
\begin{eqnarray}
			\hat B_{\alpha}\left| \Phi_{\gamma}(g)  \Phi_{\alpha}(u)  \right \rangle  &=&\delta\left(	Hol_{v_{\alpha}}^1,
            \Phi_{\partial\alpha}(1_G)\right)\left| \Phi_{\gamma}(g)  \Phi_{\alpha}(u)  \right \rangle, \\
            \hat B_{b} \left| \Phi_{\gamma}(g)  \Phi_{\alpha}(u)  \right \rangle    &=&\delta\left(Hol_{v_{\alpha}}^2,\Phi_{\partial b}(1_G,1_E)\right)\left| \Phi_{\gamma}(g)  \Phi_{\alpha}(u)  \right \rangle.
		\end{eqnarray}	
If we flip the orientation of the plaquette $\alpha$, then we also reverse the orientation of its boundary, so that $p^-(v_{\alpha_1}\rightarrow v_{\alpha_2}) $ is inverted. Therefore
\begin{eqnarray}
\hat T_2 \left| \Phi_{\gamma}(g)  \Phi_{\alpha}(u)  \right \rangle  &=&  \left| \Phi_{\gamma}(g) (\Phi_\alpha(u))^{-1} \right \rangle     \cr
    \hat B_\alpha \hat T_2 \left| \Phi_{\gamma}(g)  \Phi_{\alpha}(u)  \right \rangle &=& \delta\left(\partial((\Phi_\alpha)^{-1})p^{+} (v_{\alpha}\rightarrow v_{\alpha}), \Phi_{\partial \alpha}(1_G)\right) \left| \Phi_{\gamma}(g) (\Phi_\alpha(u))^{-1} \right \rangle \cr
     \hat T_2^{-1}\hat B_\alpha \hat T_2 \left| \Phi_{\gamma}(g)  \Phi_{\alpha_i}(u_i)  \right) \rangle 
       &=& \delta\left(\Phi_{\partial \alpha}(1_G),\partial(\Phi_\alpha)p^{-}(v_{\alpha}\rightarrow v_{\alpha})\right)\left| \Phi_{\gamma}(g) (\Phi_\alpha(u)) \right \rangle\cr
       &=& \hat B_\alpha\left| \Phi_{\gamma}(g) (\Phi_\alpha(u)) \right \rangle.
\end{eqnarray}
\end{itemize}
For  the blob 2-holonomy energy, we have 
\begin{eqnarray}
      \hat T_2\left| \Phi_{\gamma}(g)  \Phi_{\alpha}(u)  \right \rangle   &=& \left| \Phi_{\gamma}(g) (\Phi_\alpha(u))^{-1} \right \rangle     \cr
      \hat B_b \hat T_2\left| \Phi_{\gamma}(g)  \Phi_{\alpha}(u)  \right \rangle   &=& \delta\left( \prod_{i}\Phi_{\gamma}(g)\rhd \left((\Phi_\alpha(u))^{-1}\right)^{\theta},\Phi_{\partial b}(1_G,1_E)\right)\left| \Phi_{\gamma}(g)  (\Phi_\alpha(u))^{-1} \right \rangle \cr
  \hat T_2^{-1}   \hat B_b \hat T_2\left| \Phi_{\gamma}(g)  \Phi_{\alpha}(u)  \right \rangle   &=& \delta\left( \Phi_{\partial b}(1_G,1_E),\prod_{i}\Phi_{\gamma}(g)\rhd (\Phi_\alpha(u))^{\theta}\right)\left| \Phi_{\gamma}(g)  \Phi_{\alpha}(u)  \right \rangle,\cr
   &=& \hat B_\alpha\left| \Phi_{\gamma}(g) (\Phi_\alpha(u)) \right \rangle.
\end{eqnarray}
This  show that plaquette and the blob energy terms are invariant under the plaquette-flipping procedure
\begin{eqnarray}
    \hat T_2^{-1}\hat {\mathcal{B}}_\alpha  \hat T_2=\hat  {\mathcal{B}}_\alpha \quad \mbox{and}\quad  \hat T_2^{-1}\hat {\mathcal{B}}_b  \hat T_2=\hat  {\mathcal{B}}_b.
\end{eqnarray}

\subsubsection{Moving the base point of a plaquette}
 We finally  consider the  changing of the base point of a plaquette. Considering $ v_{\alpha_1}$ and $ v_{\alpha_2}$, we denote the procedure that moves the base point
of the plaquette $\alpha $ from a vertex $ v_{\alpha_1}$ to a vertex $ v_{\alpha_2}$ by the transformation $\hat T_3 $. This operation changes the labeled plaquette $\Phi_\alpha(u)$ to $p(v_{\alpha_1}\rightarrow  v_{\alpha_2})^{-1}\rhd \Phi_\alpha(u)$,  where $ p(v_{\alpha_1}\rightarrow  v_{\alpha_2})$ is   the path along which we move the base point
\begin{eqnarray}\label{qa}
   \hat T_3\Phi_\alpha(u)= p(v_{\alpha_1}\rightarrow  v_{\alpha_2})^{-1}\rhd \Phi_\alpha(u).
\end{eqnarray}

\begin{itemize}
\item We first consider a vertex transform $ \hat A_v^h$. The vertex transform at a vertex $v$ affects any plaquette whose base point is at the vertex $v_\alpha$. Furthermore, it affects path elements which start or end at the vertex $v_\alpha$. This is relevant because the transformation of the
labeled plaquette  under $ \hat T_3 $  depends on the path element $p (v_{\alpha_1}\rightarrow  v_{\alpha_2})$, which is affected by a vertex transform at $v_{\alpha_1}$ and $v_{\alpha_2}$. There are  two cases to consider.
\begin{itemize}
    \item {\bf First case:}  We consider the vertex transform $\hat A_{v_{\alpha_1}}^h$ at the base point $v_{\alpha_1}$ such that
\begin{eqnarray}
    \hat A_{v_{\alpha_1}}^h \Phi_\alpha(u)=   h \rhd \Phi_\alpha(u).
\end{eqnarray}
The moving base point procedure  from $ v_{\alpha_1}$ to $ v_{\alpha_2}$ for the vertex transform $ \hat A_{v_{\alpha_1}}^h$ is obtained as follows
\begin{eqnarray}
 \hat T_3  \Phi_\alpha(u)&=&  p(v_{\alpha_1}\rightarrow  v_{\alpha_2})^{-1}\rhd \Phi_\alpha(u), \\
 \hat A_{v_{\alpha_1}}^h  \hat T_3  \Phi_\alpha(u)&=&  p(v_{\alpha_1}\rightarrow  v_{\alpha_2})^{-1}\rhd \Phi_\alpha(u),  
\end{eqnarray}
where in the last line, the vertex transform leaves the plaquette element unchanged because the base point of $\alpha$ is no longer at $v_{\alpha_1}$. However, the path element for $ p(v_{\alpha_1}\rightarrow  v_{\alpha_2}) $ has been changed by the action of the vertex transform from its original value of
$ p(v_{\alpha_1}\rightarrow  v_{\alpha_2})$  to $ p(v_{\alpha_1}\rightarrow  v_{\alpha_2})$. This means that when we move the base point back along the path, we pick up a factor of $h\rhd p(v_{\alpha_1}\rightarrow  v_{\alpha_2}) $
acting on the plaquette, rather than just a factor of $ p(v_{\alpha_1}\rightarrow  v_{\alpha_2})$. This gives us
\begin{eqnarray}
 \hat T_3^{{-1}} \hat A_{v_{\alpha_1}}^h  \hat T_3 \Phi_\alpha(u)&=&  hp(v_{\alpha_1}\rightarrow  v_{\alpha_2})  \rhd p(v_{\alpha_1}\rightarrow  v_{\alpha_2})^{-1}\rhd \Phi_\alpha(u), \cr
 &=& h \rhd \Phi_\alpha(u)\cr
  &=&\hat A_{v_{\alpha_1}}^h\Phi_\alpha(u)
\end{eqnarray}
 This  show that the vertex transform at $v_{\alpha_1}$ is invariant under the  moving base point procedure from $v_{\alpha_1}$ to $v_{\alpha_2}$. Consequently  the energy  term $ \hat {\mathcal{A}}_{v_{\alpha_1}}$ is also invariant under this procedure  such that 
 \begin{eqnarray}
  \hat T_3^{{-1}}   \hat {\mathcal{A}}_{v_{\alpha_1}}\hat T_3= \hat {\mathcal{A}}_{v_{\alpha_1}}.
 \end{eqnarray}

\item {\bf Second case:} Now, let’s still consider the  base point at $ v_{\alpha_1}$ and the vertex transform at  $ v_{\alpha_2}$ is given by 
\begin{eqnarray}
 \hat A_{v_{\alpha_2}}^h   \Phi_\alpha(u)= \Phi_\alpha(u).
\end{eqnarray}
The moving base point procedure  from $ v_{\alpha_1}$ to $ v_{\alpha_2}$ for the vertex transform $ \hat A_{v_{\alpha_2}}^h$ is obtained as follows
\begin{eqnarray}
  \hat T_3  \Phi_\alpha(u)&=& p(v_{\alpha_1}\rightarrow  v_{\alpha_2})^{-1}\rhd\Phi_\alpha(u),\\
   \hat A_{v_{\alpha_2}}^h\hat T_3 &=&  h\rhd p(v_{\alpha_1}\rightarrow  v_{\alpha_2})^{-1}\rhd \Phi_\alpha(u),\\
  \hat T_3^{{-1}}   \hat A_{v_{\alpha_2}}^h\hat T_3\Phi_\alpha(u)&=& \left( p(v_{\alpha_1}\rightarrow  v_{\alpha_2})h^{-1}\right)\rhd \left(hp(v_{\alpha_1}\rightarrow  v_{\alpha_2})^{-1}\right)\rhd \Phi_\alpha(u)=\Phi_\alpha(u)\cr
   &=&  \hat A_{v_{\alpha_2}}^h \Phi_\alpha(u).
\end{eqnarray}
\end{itemize}
This means that the vertex energy term at $v_{\alpha_2}$ is also unaffected by the procedure for changing the base points of the plaquette $\alpha$, i.e,
 \begin{eqnarray}
  \hat T_3^{{-1}}  \hat {\mathcal{A}}_{v_{\alpha_2}}\hat T_3 = \hat {\mathcal{A}}_{v_{\alpha_2}}.
 \end{eqnarray}

\item Next, we consider edge energy terms.  Let us  recall the edge transform $\hat A_\gamma^e$ on the adjacent plaquette $ \alpha$ labeled by $\Phi_\alpha(u)$ at the base point $v_{\alpha}$
\begin{eqnarray}
    \hat A_{\gamma}^{e} \Phi_\alpha(u)=  \begin{cases}  \Phi_\alpha(u)\left(p^+(v_\alpha\rightarrow s(\gamma)\rhd e^{-1} \right)  & \text{ if $\gamma$ is on $\alpha$ and aligned with $\alpha$} \\
	\left({p^-(v_\alpha\rightarrow s(\gamma))}\rhd e\right) \Phi_\alpha(u)  & \text{if $\gamma$ is on $\alpha$ and aligned against $\alpha$}\\
		\Phi_\alpha(u) & \text{otherwise}.\\
	\end{cases}
\end{eqnarray}
%Let consider the  following graph to describe how the move of base point %from $v_{\alpha_1}$ 
There are two scenarios of the move plaquette base point procedure with respect to the edge $\phi_\gamma(g)$ on the plaquette. One in which the edge $\phi_\gamma(g)$ is not part of the path taken when moving the plaquette from the base point $v_{\alpha_1}$ to the base point $v_{\alpha_2}$, and the other in which the edge $\Phi_\gamma(g)$ is part of the path taken when moving the base point. For the sake of simplicity, we only consider here the first scenario, and we refer the reader to the paper \cite{42} or details on the second scenario.

Based on the equation \eqref{qa}, the representation $\hat A_{\gamma}^{e} \hat T_3  \Phi_\alpha(u)$ reads as
%The move  plaquette base point procedure of the edge transform reads as follows
\begin{eqnarray}
    %\hat T_3 \Phi_\alpha(u)&=&  p(v_{\alpha_1}\rightarrow  %v_{\alpha_2})^{-1}\rhd\Phi_\alpha(u) \\
\hat A_{\gamma}^{e} \hat T_3  \Phi_\alpha(u)=  \begin{cases}  p(v_{\alpha_1}\rightarrow  v_{\alpha_2})^{-1}\rhd\Phi_\alpha(u)\left(p^+(v_{\alpha_2}\rightarrow s(\gamma))\rhd e^{-1} \right)  & \text{ if $\gamma$ aligned with $\alpha$} \\
	\left({p^-(v_{\alpha_2}\rightarrow s(\gamma))}\rhd e\right) (p(v_{\alpha_1}\rightarrow  v_{\alpha_2})^{-1}\rhd\Phi_\alpha(u)) & \text{if $\gamma$ aligned against $\alpha$}\\
		\Phi_\alpha(u) & \text{otherwise}.\\
	\end{cases}
\end{eqnarray}
 Since the  edge $\Phi_\gamma(g)$ is not include  in  the path along the  move  the  base point, the  plaquette move base point procedure of the edge transform  $\hat T_3^{-1} \hat A_{\gamma}^{e} \hat T_3   \Phi_\alpha(u)$ reads as follows

\begin{eqnarray}\label{q3}
   \hat T_3^{-1} \hat A_{\gamma}^{e} \hat T_3   \Phi_\alpha(u)
   = \hat T_3^{-1}\left( \hat A_{\gamma}^{e} \hat T_3   \Phi_\alpha(u) \right)\cr
   = \begin{cases} p(v_{\alpha_1}\rightarrow  v_{\alpha_2})\rhd\left( p(v_{\alpha_1}\rightarrow  v_{\alpha_2})^{-1}\rhd\Phi_\alpha(u)\right)\left(p^+(v_{\alpha_2}\rightarrow  s(\gamma))\rhd e^{-1} \right) & \text{if $\gamma$ aligned with $\alpha$}\\
	 p(v_{\alpha_1}\rightarrow  v_{\alpha_2})\rhd\left({p^-(v_{\alpha_2}\rightarrow s(\gamma))}\rhd e\right) p(v_{\alpha_1}\rightarrow  v_{\alpha_2})^{-1}\rhd\Phi_\alpha(u)& \text{if $\gamma$ aligned against $\alpha$}\\
		\Phi_\alpha(u) & \text{otherwise}.
	\end{cases}
\end{eqnarray}
The latter equation is reduced  into  
\begin{eqnarray}\label{q333}
   \hat T_3^{-1} \hat A_{\gamma}^{e} \hat T_3   \Phi_\alpha(u)
    = \begin{cases} \Phi_\alpha(u)\left(p(v_{\alpha_1}\rightarrow  v_{\alpha_2})p^+(v_{\alpha_2}\rightarrow s(\gamma))\rhd e^{-1} \right)  & \text{ if $\gamma$ aligned with $\alpha$} \\
	\left(p(v_{\alpha_1}\rightarrow  v_{\alpha_2}){p^-(v_{\alpha_2}\rightarrow s(\gamma))}\rhd e\right) \Phi_\alpha(u) & \text{if $\gamma$ aligned against $\alpha$}\\
		\Phi_\alpha(u) & \text{otherwise}.
	\end{cases}.
\end{eqnarray}
 It is evident that if the edge $\Phi_\gamma(g)$ lines up with the plaquette's $\alpha$ boundary as illustrated  in figure \eqref {en6} as follows 
 
 \begin{figure}
     \centering
\begin{eqnarray*}
	\begin{tikzpicture}[thick]
		\node (A) at (-1.5,0) {$v_{\alpha_1}$};
		\node (B) at (1.5,0) {$v_{i+1}$};
		\node (C) at (4.5,0) [below=of A] {$v_{\alpha_2}$}; 
		\node (D) at (4.5,0) [below=of B] {$v_i$};	
		\draw[->](A)--node [left=0.1cm] {}(C);
		\draw[<-](B)--node [right=0.1cm] {$\Phi_{\gamma}(g)$}(D);
		\draw[->](A)--node [above=0.1cm] {$ $}(B);
		\draw[->](C)--node [below=0.1cm] {$ $}(D); 
        \draw[ thick, ->] (0.9,-0.75) arc (-2:70:0.5);
     %\draw[ thick, ->] (-1.5,0.2) arc (180:-1.70:1.5) node [right,pos=0.7] {$ p^+(v_\alpha\rightarrow v_i)    $};
     \draw[ thick,red, ->] (-1.6,-0.1) arc (-232:-1.8:2.01) node [right,pos=0.8] {$ p^+(v_{\alpha_1}\rightarrow v_i=s(\gamma))$};
     \draw[ thick,blue, ->] (-1.5,-1.5) arc (-245:59.9:3.2) node [right,pos=0.8] {$ p^+(v_{\alpha_2}\rightarrow v_i=s(\gamma))$};
        \node at (-0.1,-0.75){$\Phi_\alpha(u)$};
   % \draw[ thick,blue, ->](-1.6,-0.1) arc (-240:2.5:0.6) node [right,pos=0.8] {$ p^+(v_{\alpha_2}\rightarrow v_i=s(\gamma))$};
        \node at (-0.1,-0.75){$\Phi_\alpha(u)$};

	\end{tikzpicture}	      
\end{eqnarray*}
\caption{In this figure, we consider the case where the paths $ p^+(v_{\alpha_1}\rightarrow v_i=s(\gamma))$ and  $ p^+(v_{\alpha_2}\rightarrow v_i=s(\gamma))$  aligned with the orientation of the plaquette}\label{en6}

\end{figure}
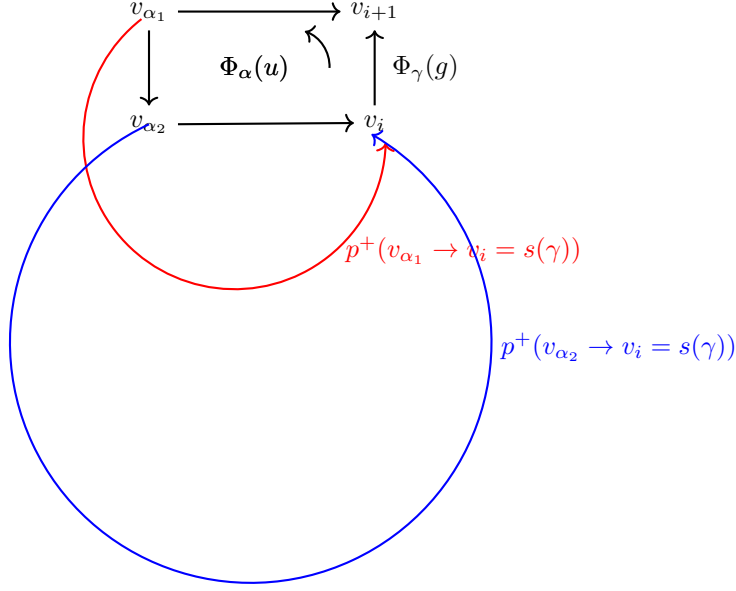

We have: 
 \begin{eqnarray}\label{q1}
     p(v_{\alpha_1}\rightarrow v_{\alpha_2})p^+(v_{\alpha_2}\rightarrow s(\gamma))&=&
     p^+(v_{\alpha_1}\rightarrow s(\gamma))\implies\cr p^+(v_{\alpha_2}\rightarrow s(\gamma))&=&p^+(v_{\alpha_1}\rightarrow v_{i})  p(v_{\alpha_1}\rightarrow v_{\alpha_2})^{-1}.\label{q111}
 \end{eqnarray}
However, if the edge $\Phi_\gamma(g)$ is  anti-aligned with the boundary of the plaquette $\alpha$ as illustrated by the figure \eqref{en90}.
\begin{figure}
\centering
\begin{eqnarray*}
	\begin{tikzpicture}[thick]
		\node (A) at (-1.5,0) {$v_{\alpha_1}$};
		\node (B) at (1.5,0) {$v_{i+1}$};
		\node (C) at (4.5,0) [below=of A] {$v_{\alpha_2}$}; 
		\node (D) at (4.5,0) [below=of B] {$v_i$};	
		\draw[->](A)--node [left=0.1cm] {}(C);
		\draw[<-](B)--node [right=0.1cm] {$\Phi_{\gamma}(g)$}(D);
		\draw[->](A)--node [above=0.1cm] {$ $}(B);
		\draw[->](C)--node [below=0.1cm] {$ $}(D); 
        \draw[ thick, <-] (0.9,-0.75) arc (-2:70:0.5);
     %\draw[ thick, ->] (-1.5,0.2) arc (180:-1.70:1.5) node [right,pos=0.7] {$ p^+(v_\alpha\rightarrow v_i)    $};
     \draw[ thick,red, ->] (-1.6,-0.1) arc (-232:-1.8:2.01) node [right,pos=0.8] {$ p^-(v_{\alpha_1}\rightarrow v_i=s(\gamma))$};
     \draw[ thick,blue, ->] (-1.5,-1.5) arc (-245:59.9:3.2) node [right,pos=0.8] {$ p^-(v_{\alpha_2}\rightarrow v_i=s(\gamma))$};
        \node at (-0.1,-0.75){$\Phi_\alpha(u)$};
   % \draw[ thick,blue, ->](-1.6,-0.1) arc (-240:2.5:0.6) node [right,pos=0.8] {$ p^+(v_{\alpha_2}\rightarrow v_i=s(\gamma))$};
        \node at (-0.1,-0.75){$\Phi_\alpha(u)$};
	\end{tikzpicture}	      
\end{eqnarray*}
\caption{We consider the case where the paths $ p^-(v_{\alpha_1}\rightarrow v_i=s(\gamma))$ and  $ p^-(v_{\alpha_2}\rightarrow v_i=s(\gamma))$ anti-aligned with the orientation of the plaquette $\alpha$}\label{en90}

\end{figure}
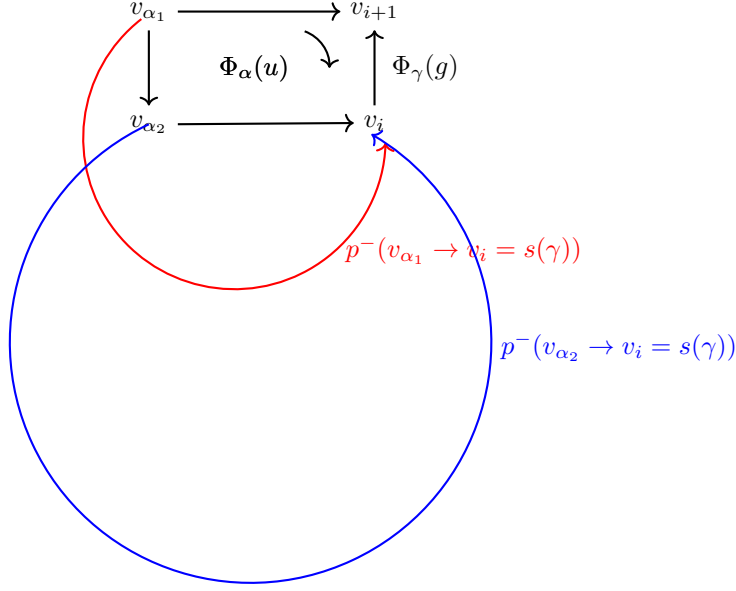
We have
\begin{eqnarray}\label{q222}
     p(v_{\alpha_1}\rightarrow v_{\alpha_2})^{-1}p^-(v_{\alpha_1}\rightarrow s(\gamma))=p^-(v_{\alpha_2}\rightarrow s(\gamma)).
 \end{eqnarray}

Inserting equations \eqref{q111} and \eqref{q222} in equation \eqref{q333}, we have 
\begin{eqnarray}\label{q3}
   \hat T_3^{-1} \hat A_{\gamma}^{e} \hat T_3   \Phi_\alpha(u)
    &=& \begin{cases} \Phi_\alpha(u)\left(p^+(v_{\alpha_2}\rightarrow s(\gamma))\rhd e^{-1} \right)  & \text{ if $\gamma$ aligned with $\alpha$} \\
	\left({p^-(v_{\alpha_1}\rightarrow s(\gamma))}\rhd e\right) \Phi_\alpha(u) & \text{if $\gamma$ aligned against $\alpha$}\\
		\Phi_\alpha(u) & \text{otherwise}.
	\end{cases},\cr
    &=& \hat A_{\gamma}^{e}   \Phi_\alpha(u).
\end{eqnarray}
This show that the action of the edge transform is preserved. Consequently,  the energy  term $ \hat {\mathcal{A}}_\gamma$ is also invariant under this procedure  such that 
 \begin{eqnarray}
  \hat T_3^{{-1}}    \hat {\mathcal{A}}_\gamma\hat T_3= \hat{\mathcal{A}}_\gamma.
 \end{eqnarray}

\item Finally, we consider the plaquette and the blob energy terms. Moving the base point $ v_{\alpha_1}$ to $v_{\alpha_2}$ of a plaquette also affects the boundary of that plaquette. Therefore, the plaque 1-holonomy $Hol_{v_{\alpha_1}}^1=\partial \Phi_\alpha (u)p^-(v_{\alpha_1}\rightarrow v_{\alpha_1}) $  becomes
\begin{eqnarray}
   Hol_{v_{\alpha_2}}^1= \left( p(v_{\alpha_1}\rightarrow  v_{\alpha_2})^{-1}\rhd\Phi_\alpha(u)\right) p(v_{\alpha_1}\rightarrow  v_{\alpha_2})^{-1}  p^-(v_{\alpha_1}\rightarrow v_{\alpha_1})    p(v_{\alpha_1}\rightarrow  v_{\alpha_2}).\label{wq}
\end{eqnarray}
Using the Peiffer first condition \eqref{Pf1} such that 
\begin{eqnarray}
\partial \left( p(v_{\alpha_1}\rightarrow  v_{\alpha_2})^{-1}\rhd\Phi_\alpha(u)\right)=p(v_{\alpha_1}\rightarrow  v_{\alpha_2})^{-1}  \partial \Phi_\alpha(u)    p(v_{\alpha_1}\rightarrow  v_{\alpha_2}).
\end{eqnarray}
Equation \eqref{wq} becomes
\begin{eqnarray}
    Hol_{v_{\alpha_2}}^1&=&p(v_{\alpha_1}\rightarrow  v_{\alpha_2})^{-1}\partial \Phi_\alpha (u)p^-(v_{\alpha_1}\rightarrow v_{\alpha_1}) p(v_{\alpha_1}\rightarrow  v_{\alpha_2})\cr
    &=&p(v_{\alpha_1}\rightarrow  v_{\alpha_2})^{-1}\partial \Phi_\alpha (u)  p^-(v_{\alpha_1}\rightarrow v_{\alpha_1})   p(v_{\alpha_1}\rightarrow  v_{\alpha_2})\cr
     &=&p(v_{\alpha_1}\rightarrow  v_{\alpha_2})^{-1}Hol_{v_{\alpha_1}}^1 p(v_{\alpha_1}\rightarrow  v_{\alpha_2}).
\end{eqnarray}
We see that the
plaquette holonomy is merely conjugated by a path element, which preserves the identity element. Therefore,  the energy term
(which checks if the plaquette holonomy is equal to the identity) is unaffected by the base-point changing procedure
\begin{eqnarray}
     \hat T_3^{-1}\hat B_\alpha \hat T_3 = \hat B_\alpha.
\end{eqnarray}

The next energy term to consider is the blob term. To show the invariance of this term  under the base-point changing procedure, we consider the following graph \eqref{en100} 
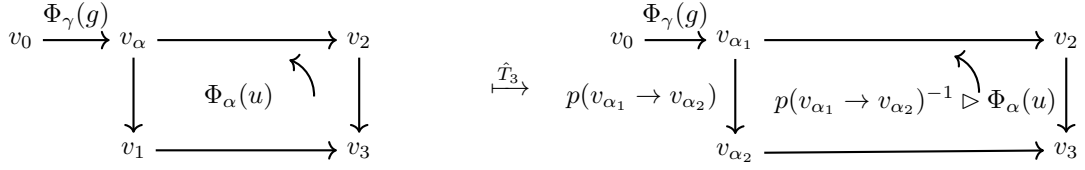
\begin{figure}[H]
\centering
	\begin{tikzpicture}[thick]
    \node (A0) at (-3,0) {$v_0$};
		\node (A) at (-1.5,0) {$v_{\alpha}$};
		\node (B) at (1.5,0) {$v_2$};
		\node (C) at (4.5,0) [below=of A] {$v_1$}; 
		\node (D) at (4.5,0)   [below=of B] {$v_3$};	
		
		\draw[->](A)--node [left=0.1cm] {$ $}(C);
		\draw[->](B)--node [right=0.1cm] {$ $}(D);
		\draw[->](A)--node [above=0.1cm] {$ $}(B);
		\draw[->](C)--node [below=0.1cm] {$ $}(D); 
        \draw[ thick, ->] (0.9,-0.75) arc (-2:70:0.5);
		\node at (-0.1,-0.75){$\Phi_\alpha(u)$};
	\draw[->] (A0) --node[above=0.01cm] {$\Phi_{\gamma}(g)$} (A) ;

		\node (F)  at (3.51,-0.5){$\xmapsto{\hat T_3} $};
        \node (G0)  at (5.,0){$v_{0}$};
		\node (G)  at (6.5,0){$v_{\alpha_1}$};
		\node (H)  at (10.9,0){$v_2$};
		\node (I)  at (6.4,1) [below=of G] {$v_{\alpha_2}$};
		\node (J)  at (10.5,1)[below=of H] {$v_3$};

        %\draw[ thick,blue, ->] (6.5,0) arc (-120:70:0.5) node [right,pos=0.8] {$ p^-(v_\alpha\rightarrow v_i=s(\gamma))$};

        \draw[->] (G0) --node[above=0.01cm] {$\Phi_{\gamma}(g)$} (G);
		
		\draw[->](G)-- node [above=0.1cm] {$ $}                    (H);
		
		\draw[->](G)--node [left=0.1cm] {$  p(v_{\alpha_1}\rightarrow  v_{\alpha_2})$}          (I);
		\draw[->](H)-- node [right=0.1cm] {$ $}        (J);
		\draw[->](I)-- node [below=0.1cm] {$ $}        (J);
		
		\draw[ thick, ->] (9.74,-0.7) arc (-2:70:0.5);
		\node  at (8.9,-0.8) {$ p(v_{\alpha_1}\rightarrow  v_{\alpha_2})^{-1}\rhd \Phi_\alpha(u)$};	
	\end{tikzpicture}
   \caption{Moving the base point of a plaquette 
}\label{en100} 
\end{figure}
The 2-holonomy blob $Hol_{v_{\alpha_1}}^2= \Phi_{\gamma}(g)\rhd \left(\Phi_{\alpha}(u)\right)$ at the base-point $v_\alpha=v_{\alpha_1}$ becomes after the base-point changing 
\begin{eqnarray}
    Hol_{v_{\alpha_2}}^2&=& \left(\Phi_{\gamma}(g) p(v_{\alpha_1}\rightarrow  v_{\alpha_2})\right)\rhd \left(p(v_{\alpha_1}\rightarrow  v_{\alpha_2})^{-1}\rhd(\Phi_{\alpha}(u))\right),\cr
    &=& \Phi_{\gamma}(g)\rhd \left(\Phi_{\alpha}(u)\right),\cr
    &=&   Hol_{v_{\alpha_1}}^2.
\end{eqnarray}
The contribution of the plaquette to the blob 2-holonomy is unchanged by moving the base point of the plaquette. Therefore,  the blob energy term
 is unaffected by the base-point changing procedure
\begin{eqnarray}
     \hat T_3^{-1}\hat B_b \hat T_3= \hat B_b.
\end{eqnarray}

\end{itemize}

\end{document}